\documentclass[oldversion]{aa}

\usepackage{times}
\usepackage{graphicx}
\usepackage{xspace}
\usepackage{epsfig}
\usepackage{natbib}
\usepackage{rotating}
\usepackage{dcolumn}
\usepackage{longtable,lscape}
\usepackage{xcolor}
\usepackage{footmisc,hyperref} 

\newcommand{\gagneM}{{\em Gagn\'eM}\xspace}
\newcommand{\ecat}{{\em eRASS1cat}\xspace}

\def\gsim{\;\lower4pt\hbox{${\buildrel\displaystyle >\over\sim}$}\,}
\def\lsim{\;\lower4pt\hbox{${\buildrel\displaystyle <\over\sim}$}\,}

\begin{document}

\title{A first eROSITA view of ultracool dwarfs\thanks{The full versions of Tables~1 and~2 
are only available in electronic format at the CDS via anonymous ftp to cdsarc.u-strasbg.fr (130.79.128.5)
or via http://cdsweb.u-strasbg.fr/cgi-bin/qcat?J/A+A/}}


\author{B. Stelzer \inst{1,2} \and A. Klutsch \inst{1} \and M. Coffaro \inst{1} \and 
E. Magaudda \inst{1} \and M. Salvato \inst{3}}

\offprints{B. Stelzer}

\institute{
 Institut f\"ur Astronomie \& Astrophysik, Eberhard-Karls-Universit\"at T\"ubingen,
 Sand 1, 72076 T\"ubingen, Germany\label{inst1} \\ \email{B. Stelzer, stelzer@astro.uni-tuebingen.de}
 \and
 INAF - Osservatorio Astronomico di Palermo, Piazza del Parlamento 1, 90134 Palermo,
 Italy\label{inst2} \and
 Max-Planck Institut f\"ur extraterrestrische Physik, Giessenbachstrasse, 85748 Garching, Germany\label{inst3}
}

\titlerunning{A first eROSITA view of ultracool dwarfs}

\date{Received $<$01-May-2021$>$ / Accepted $<$02-June-2021$>$}

\abstract{
We present the first X-ray detections of ultracool dwarfs (UCDs) from the first all-sky survey
of the extended ROentgen Survey with an Imaging Telescope Array ({\em eROSITA}) 
onboard the Russian Spektrum-Roentgen-Gamma (SRG) mission. 
We use three publicly available input catalogs of spectroscopically 
confirmed UCDs and {\em Gaia}-selected UCD candidates that together comprise nearly $20 000$ 
objects. In a careful
source identification procedure we first extracted all X-ray sources from the catalog of
the first survey, eRASS1, that have a UCD or candidate within three times their positional
uncertainty. Then we examined all {\em Gaia} objects in 
the vicinity of these $96$ X-ray sources and we associated them to the most plausible counterpart on the
basis of their spatial separation to the X-ray position and their multiwavelength properties. 
This way we find $40$ UCDs that have a secure identification with an X-ray
source (that is bonafide UCD X-ray emitters) 
and $18$ plausible UCD X-ray emitters for which we consider it likely that the X-ray
source has its origin in the UCD. Twenty-one of the bonafide and plausible X-ray emitting
UCDs have a spectroscopic confirmation, while the others have been selected based on 
{\em Gaia} photometry and we computed spectral types from the $G$-$J$ color. 
The spectral types of the X-ray emitting UCDs and candidates range between M5 and M9.
The distances of the 
eRASS1 UCDs range from $3.5$ to $190$\,pc. The spectroscopically confirmed UCDs at the high
end of the distance distribution are known to be members of nearby star forming regions. The
majority of the UCDs from the eRASS1 sample show a ratio of X-ray to bolometric luminosity
well above the canonical saturation limit of $\log{(L_{\rm x}/L_{\rm bol})} \approx -3$.
For the two most extreme outliers, we verified the hypothesis that these high values are 
due to flaring 
activity through an analysis of the eRASS1 light curve. X-ray spectra could be analyzed
for the two brightest objects in terms of count rate, 
both showing an emission-measure weighted plasma temperature of $\langle kT \rangle = 0.75$\,keV. 
These observations
demonstrate the potential of {\em eROSITA} for advancing our knowledge on the faint coronal
X-ray emission from UCDs by building statistical samples for which the average X-ray brightness,
flares, and coronal temperatures can be derived. 
}


\maketitle

\section{Introduction}\label{sect:intro}

Ultracool dwarfs (UCDs) are defined as objects with spectral type (SpT) M7 and later. 
This group 
spans the hydrogen-burning mass limit \citep[at $M \sim 0.075\,{\rm M_\odot}$;][]{Baraffe02.0} 
and it comprises 
both very low-mass stars and brown dwarfs.
By coincidence, for late-M SpTs ($T_{\rm eff} \sim 2500$\,K), the photospheres become 
effectively neutral, which leads to high electrical resistivity, that is reduced coupling 
between the magnetic field and matter \citep{Mohanty02.1}. This is expected to 
drastically decrease magnetic activity which is the result of magnetic stress that built
up in photospheric fields through their interaction with convective motions. 
The most widely used observational diagnostics of magnetic activity for M dwarfs 
is H$\alpha$ emission from a chromosphere. 
In fact, a strong drop of H$\alpha$ emission is observed at late-M and early-L SpTs 
\citep[e.g.,][]{Gizis00.1, Mohanty03.1, West06.1}.
On the other hand, numerous studies have detected H$\alpha$ emission on UCDs, especially during flares 
 \citep[e.g.,][]{Liebert99.0, Schmidt07.0}. Various possibilities on how late-M and L-type 
 objects may maintain such chromospheric activity have been discussed by \cite{Schmidt15.0}. 
With the advent of the {\em Kepler} mission white-light flares have been observed on
some L dwarfs \citep{Gizis13.0, Paudel20.0}, further demonstrating that the magnetic
activity in such photospherically cool objects shares many similarities with the phenomena
observed on higher-mass late-type stars. Further clues as to the nature of magnetic activity 
of UCDs can be obtained from a study of their outermost atmospheric layer, the corona,
which in active stars is heated to temperatures of several $10^6$\,K leading to radiation
in the X-ray band. 

\cite{Stelzer13.0} show that for M dwarfs, coronal X-ray emission is a more sensitive activity 
diagnostics than H$\alpha$ emission: Many early- to mid-M dwarfs with undetectable H$\alpha$ 
emission have been detected with ROSAT in soft X-rays. However, even for the optimistic 
estimate that the X-ray to bolometric luminosity ($L_{\rm x}/L_{\rm bol}$) remains constant 
throughout the M and L spectral classes, UCDs are faint X-ray emitters as a result of their 
low bolometric luminosity. As a consequence, so far only few UCDs have deep enough X-ray 
observations to constrain their coronal emission \citep{Stelzer06.1}.
In fact, X-ray detections of UCDs have remained rare, with only two  
detections of L dwarfs \citep{Audard07.2, deLuca20.0}. 
On the other hand, the abovementioned 
higher sensitivity to coronal, rather than chromospheric emission of 
the currently available instrumentation, might at least partially make up for the 
weakening of activity signatures at the cool end of the main-sequence. 
Therefore, the sparse sample of 
X-ray detected UCDs is likely also related to the lack of a deep wide-area X-ray survey. 

Studying the emission level and variability in the X-ray band is essential for 
understanding the nature of the outer atmospheres of the objects at the bottom of the 
main-sequence. This is particularly relevant in view of the peculiar behavior of UCDs in 
the radio band where a violation of the G\"udel-Benz relation, an empirical correlation
between radio and X-ray luminosities that holds for all other coronal sources 
\citep{Guedel93.1}, 
is observed \citep{Berger05.1}. Another peculiarity of the radio properties of UCDs is
that for some of them a highly circularly polarized pulsed 
emission component is overlaid on their quiescent, nonvariable gyro-synchrotron emission 
\citep{Burgasser05.1, Hallinan06.1, Hallinan08.1}. 

Despite two decades of observational efforts samples of UCDs with data in all 
relevant wavebands are still sparse. 
\cite{Stelzer12.0} have provided evidence for a distinction of two groups in the 
heterogeneous observational picture of UCD activity: 
A group of objects with X-ray flares and persistent X-ray emission but no 
detected radio emission resembles `typical' active stars, 
and a second group of objects with radio bursts, mostly also with detected quiescent 
radio emission, but no or very weak X-ray emission that dramatically deviates from this 
`canonical' behavior. 
\cite{Williams14.0} argue that 
the origin of the observed division be the magnetic field strength and morphology
rather than rotation.
The puzzling multiwavelength picture that has emerged for this object class
has recently been reviewed by \cite{Pineda17.0}.    

Many observations of magnetic activity on UCDs have occurred serendipitously, and the
most interesting objects were subsequently observed at other wavebands to search 
for connections between the different magnetically induced emissions. 
As a result, the sample of UCDs with sensitive multiwavelength data may not be
representative of the UCDs as a class. 
With the recent launch of the extended ROentgen Survey with an Imaging Telescope Array 
({\em eROSITA}) onboard the Spektrum-R\"ontgen-Gamma mission 
\citep{Predehl21.0}  
we have entered a new era for the study of X-ray activity on UCDs: While the higher 
sensitivity of observatories like {\em XMM-Newton} and {\em Chandra} is required to detect 
the quiescent X-ray emission of all but the most nearby UCDs, the enormous statistical 
samples to be observed with {\em eROSITA} (many thousand objects) can be expected to 
boost the number of detections among UCDs and provide an unprecedented wealth 
of information on their flaring activity.
In this article we carry out the first exploration
for UCD X-ray emission based on the first {\em eROSITA} All-Sky Survey (eRASS1) which was 
completed in summer 2020. 

We introduce our input sample of UCDs and UCD candidates in Sect.~\ref{sect:sample}. 
Then we proceed to the source identification where we match the eRASS1 catalog with
our input samples and subsequently inspect for all matches 
the plausibility that the UCD rather than another
known {\em Gaia} source is the X-ray emitter (Sect.~\ref{sect:identifications_eRASS1}). 
In Sect.~\ref{sect:eRASS1_multilambda} the multiwavelength properties of those UCDs to 
which we have assigned an X-ray source are presented, and in Sect.~\ref{sect:lc_and_spec}
we describe the analysis of X-ray light curves and spectra for the two UCDs
with the largest number of counts in eRASS1. Section~\ref{sect:results} 
comprises our results and a discussion thereof. Finally, in Sect.~\ref{sect:conclusions}
we summarize the findings and draw the conclusions. Appendices are presented for different 
 subsamples of objects that are in the vicinity of the UCDs and that we 
assign as the more likely counterpart of the X-ray source (App.~\ref{app:cpmpairs},
and \ref{app:nonucds}), 
and App.~\ref{app:selection_procedure} holds a short discussion on
the distinction of stellar and extragalactic objects in multiwavelength diagrams based
on {\em eROSITA} data from an observation obtained during the CalPV phase.

\section{Sample}\label{sect:sample}
 
We have compiled a list of UCDs 
to be matched against the 
eRASS1 source catalog.  
Our input list comprises both spectroscopically confirmed UCDs and new candidate UCDs
identified with {\em Gaia}. The known UCDs are 
drawn from the ``list of M6-M9 dwarfs" 
maintained by 
J.\,Gagn\'e\footnote{https://jgagneastro.wordpress.com/list-of-m6-m9-dwarfs/.} 
henceforth referred to as \gagneM,
and the {\sc gucds} catalog from \cite{Smart19.0} which comprises mostly L and T dwarfs. 
The new {\em Gaia} discovered UCD candidates are from \cite{Reyle18.0}. 
For the sake of homogeneity we limit the sample of known UCDs to the objects with
full photometry and astrometry in {\em Gaia}\,DR2, using the selection criteria of
\cite{Reyle18.0} from her definition of the new candidate UCD sample. 

%
%
\begin{figure*}[th]
\begin{center}
\parbox{17cm}{
\parbox{5.7cm}{
\includegraphics[width=5.7cm]{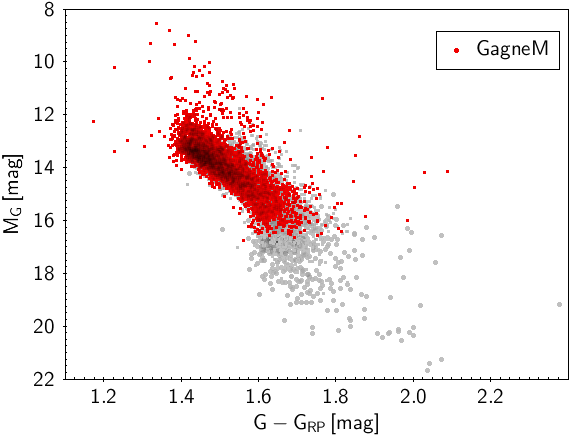} 
}
\parbox{5.7cm}{
\includegraphics[width=5.7cm]{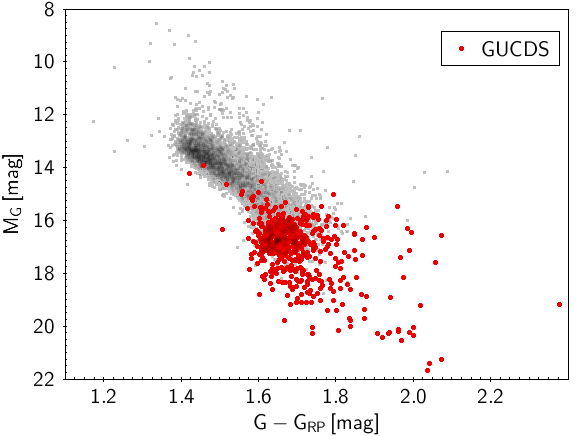}
}
\parbox{5.7cm}{
\includegraphics[width=5.7cm]{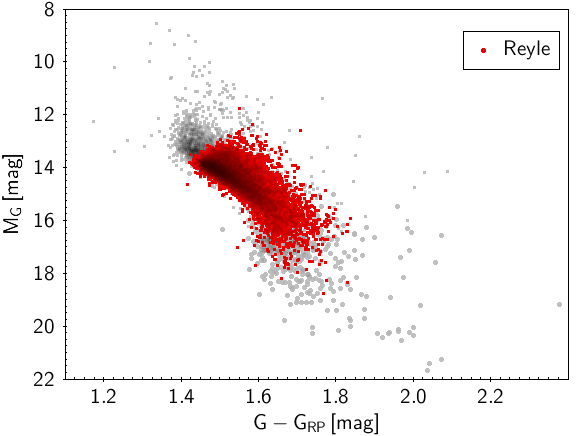}
}
}
\caption{\small {\em Gaia} color-magnitude diagram ($M_{\rm G}$ vs $G - G_{\rm RP}$) 
for the lists of UCDs defined in Sect.~\ref{sect:sample}. In each panel all three master
tables are shown but one of them is high-lighted in red: {\em left} - known late-M dwarfs from 
J.Gagne\'s M6...M9 dwarf archive, 
{\em middle} - known MLTY dwarfs with {\em Gaia}-DR2 data from the {\sc gucds} 
sample \protect\citep{Smart19.0}, 
{\em right} - new {\em Gaia}-discovered UCD candidates 
\protect\citep{Reyle18.0}.} 
\label{fig:MG_GminRP}
\end{center}
\end{figure*}

To perform this selection on the 
\gagneM catalog, 
we first matched 
it
separately against catalogs available from the advanced ADQL 
(Astronomical Data Query Language) interface of the {\em Gaia} archive
\footnote{https://gea.esac.esa.int/archive/.}
in the following way. We used the 2MASS designation given in 
\gagneM 
for the match with the {\it Gaia-DR1 2MASS original valid catalog}. 
This step provides the parameter
{\sc tmass\_oid} which is the 2MASS identifier in the {\em Gaia} catalogs. 
To map the 2MASS identifier to the
{\em Gaia} source number (col. {\sc source\_id}) we matched this output with the 
{\em Gaia\,DR2 2\,MASS best neighbor catalog}.  Subsequently, we extracted all
relevant {\em Gaia}\,DR2 information by matching the output of the last step 
using col. {\sc source\_id} with the full {\em Gaia}\,DR2 catalog. 
As of June 2020 the \gagneM catalog contains $8683$ late-M dwarfs of which 
with the above procedure $8301$ are matched with {\em Gaia}\,DR2. 

For consistency with the \cite{Reyle18.0} catalog 
we required that the objects from \gagneM that we keep in the list 
have full photometry (fluxG, fluxBP, fluxRP $>0$)
and astrometry (pmRA, pmDEC, parallax measured), and that 
the uncertainty of the parallax $\sigma_{\rm \varpi} < 10$\,\%.  
This reduces the \gagneM sample to $4326$ objects, mainly due to the criterion on the
error of the parallax. 

The inspection of the multiband photometry in \gagneM showed that further cleaning of this 
catalog was necessary: a number of objects present SDSS magnitudes much too high
for the $J$ band magnitude. These mismatches in \gagneM could be removed with the exclusion
criteria $z_{\rm SDSS} > 18.0  ~ \&\& ~ J < 15.5$ and $z_{\rm SDSS} > 18.5 ~ \&\& ~ J  < 16.5$. 
Another $130$ objects were disregarded this way. 
An additional $2$ objects were removed because they are 
outliers in the $M_{\rm G}$ vs $G - G_{\rm BP}$ diagram and with difference between the
position in the 2MASS catalog and the position in \gagneM larger than 
$10^{\prime\prime}$ although their proper motion is relatively small. We
have also visually inspected these objects in ESASky\footnote{ESA-Sky is an application to 
visualize and download archived 
astronomical data that is developed at ESAC, Madrid, Spain, by the ESAC Science Data 
Centre (ESDC). It is available at https:/sky.esa.int} confirming the wrong match. 
The final number of objects we consider from \gagneM is, therefore, $4194$. 

The {\sc gucds} catalog was 
extracted by \cite{Smart19.0} primarily from an earlier list of $1885$ UCDs compiled by the
same authors \citep{Smart17.0} 
that is based on the ``list of ultracool dwarfs" 
by J.\,Gagn\'e\footnote{https://jgagneastro.com/list-of-ultracool-dwarfs/.}, 
a catalog that complements \gagneM for cooler spectral types. 
\cite{Smart19.0} also added late-M, L and T dwarfs from some other catalogs 
and defined the {\sc gucds} sample as those objects from 
their list 
that have a match in {\em Gaia}-DR2 with
accurate {\em Gaia} coordinates and astrometry. This list comprises $695$ objects with 
spectral types M8 to T6. 
Applying the criteria on {\em Gaia}\,DR2 photometry and astrometry from \cite{Reyle18.0}  
as described above for \gagneM reduces this catalog to $610$ objects. 
By way of construction of this catalog there is some overlap with \gagneM. 
We, therefore, removed all objects that are in both catalogs from {\sc gucds}
matching the two lists for their {\em Gaia}-DR2 source names. This way our final list
from \cite{Smart19.0} has $560$ entries. 

To summarize, our input catalogs of UCDs to be matched with the {\em eROSITA} source list 
consist of $4194$ objects extracted from the original \gagneM catalog,
$560$ from \cite{Smart19.0} and the full list of $14914$ {\em Gaia}-UCD candidates from 
\cite{Reyle18.0}. 
These lists are henceforth referred to as our `cleaned' samples. 
Figure~\ref{fig:MG_GminRP} shows the $M_{\rm G}$ vs $G - G_{\rm BP}$ diagram for this 
cleaned lists of UCDs.  
The \gagneM and \cite{Reyle18.0} catalogs define a similar area in the {\em Gaia}
color-magnitude diagram because 
the latter one was constructed on the basis of the {\em Gaia} properties of the
former one \citep[see][]{Reyle18.0}. However, \gagneM starts at SpT M5 while \cite{Reyle18.0}
includes only objects with photometric spectral types (SpTs) of M7 or later. 
This explains the extension of
\gagneM at the upper left end of the distribution in Fig.~\ref{fig:MG_GminRP}. 
We also note that the \gagneM catalog comprises a number of late-M dwarfs that are
members or candidate members of young moving groups. These stand
out above the main-sequence in color-magnitude diagrams. While the focus of this work is
to find X-ray emitting field dwarfs, we keep these young objects in the catalog as no
systematic search for their X-ray emission has been performed yet.  

\begin{figure}
\begin{center}
\includegraphics[width=8.5cm]{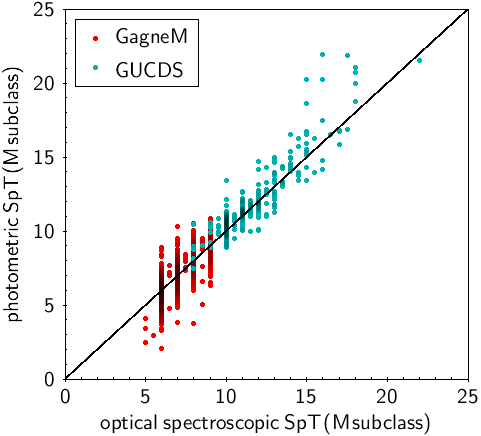} 
\caption{Comparison of spectroscopic and photometric spectral types for the `cleaned' \gagneM
and {\sc gucds} samples. The spectroscopic SpTs are obtained from the original catalogs,
the Gagn\'e dwarf archive and \cite{Smart19.0}, while the photometric SpTs have been computed
using the polynomial relation with $M_{\rm G}$ from \cite{Reyle18.0}. Objects with a flag
indicating youth or uncertain spectroscopic SpT in the two catalogs 
have been omitted, as the $M_{\rm G}-$SpT relation 
fron \cite{Reyle18.0} is valid for the main-sequence.}
\label{fig:sptphot_sptopt}
\end{center}
\end{figure}

The class of UCDs is defined through the spectral type. As described in the previous
paragraphs our samples cover different SpT ranges, and the \gagneM sample comprises also
objects that have slightly earlier SpTs than the canonical boundary
for UCDs, M7. Moreover, for the new UCD candidates from \cite{Reyle18.0} only 
photometric SpTs are available. Since these have been derived on the basis of {\em Gaia}
properties of the known confirmed UCDs, a good agreement between
spectroscopic and photometric SpTs is expected. We have checked this by
computing the photometric SpTs for the cleaned \gagneM and {\sc gucds} samples  
using the polynomial relation with $M_{\rm G}$ given by \cite{Reyle18.0}. The comparison
of the resulting values with the optical spectroscopic values for the SpT are shown
in Fig.~\ref{fig:sptphot_sptopt}. While the  overall agreement is good the scatter in the
relation together with the inclusion of objects with SpT M5 and M6 in \gagneM implies
that not all objects in our lists are genuine UCDs. Given that based on 
Fig.~\ref{fig:sptphot_sptopt} the uncertainties, especially
of the photometric SpTs, amount to a few spectral subclasses we prefer to keep all
objects in our sample even if their spectroscopic SpTs are earlier than M7. 
%
%
%

\section{Identification of UCDs among eRASS1 sources}\label{sect:identifications_eRASS1}

The SRG/{\em eROSITA} data on which we base this work 
are the results of the first All-Sky Survey, eRASS1.
The eRASS1 source catalog is produced at Max Planck Institut f\"ur 
extraterrestrische Physik (MPE) in Garching with the {\em eROSITA} software eSASS;
see \cite{Brunner21.0} 
for a description of the eSASS software. We use the 
catalog version 201008 generated with the data processing version c946\footnote{The 
final eRASS1 catalog will be released in 2022 and presented
in a forthcoming publication.}, and we henceforth call this catalog \ecat.
This catalog comprises all eRASS1 sources in the western half of the sky in
terms of Galactic coordinates, that is $l \geq 180^\circ$. The eRASS data from the other half
of the sky is the property of the Russian {\em eROSITA} consortium and is not available to us. 

We cross-matched each of the three cleaned input catalogs of UCDs and UCD candidates described
in Sect.~\ref{sect:sample} separately with \ecat. For this task we first 
corrected the {\em Gaia}\,DR2 positions (J2015.5) to the expected position at the mean eRASS1 
observing date (mid-March 2020) using the {\em Gaia}\,DR2 proper motions (PM) from our 
cleaned input
catalogs. Then we matched these PM-corrected coordinates 
with the eRASS1 boresight-corrected X-ray positions (cols. {\sc ra\_corr, dec\_corr})
in a radius of 
$30^{\prime\prime}$. Subsequently, we removed all matches for which 
the separation between the PM corrected positions and the eRASS1 positions (henceforth
termed ${\rm sep_{ox}}$) deviate by
$> 3 \times$ the uncertainty in the X-ray position (col.\,{\sc radec\_err} in \ecat). 
This way we obtained $24$ matches in \ecat for our cleaned \gagneM sample with the widest
separation being $24^{\prime\prime}$, and $73$ matches for the sample from \cite{Reyle18.0}
where the largest value of ${\rm sep_{ox}}$ is $20^{\prime\prime}$. 
The latter output catalog comprises one object with joint
Russian-German ownership which was removed from the sample to adhere to the consortium
policies. 
No matches with our cleaned {\sc gucds} sample, derived from the catalog of \cite{Smart19.0},  
fulfilled the separation criterion, that is no objects from this sample are associated
with an eRASS1 X-ray source. 

The corresponding tables of associations between our input catalogs \gagneM and
\cite{Reyle18.0} are henceforth referred to as {\it fullmatch-eRASS1-GagneM} 
and {\it fullmatch-eRASS1-Reyle} 
, respectively.  
We present the most relevant parameters of these catalogs in
Table~\ref{tab:eRASS1}. 
Next to the Gaia {\sc source\_id} (col.~1) 
we provide the boresight corrected eRASS1 positions (cols.~2 and~3), the error on the eRASS1
positions (col.~4), and 
the separation between the proper motion corrected optical positions and the
X-ray positions (${\rm sep_{ox}}$ in col.~5). 
The broad band ($0.2-5.0$\,keV) source count rates and the associated errors are given in 
col.~6.  
In col.~7 we provide the maximum likelihood for the X-ray detection in the broad band. 
In col.~8 we list 
flags for UCDs that are unique {\em Gaia} counterparts to the eRASS1
source (`U'), objects with archival X-ray detections (`X'), and common proper motion pairs 
involving a UCD or UCD candidate (`C').   
In col.~9 we mark all UCDs that after our analysis
are considered `bonafide' or `plausible' counterparts to the eRASS1 sources. For the objects
not marked in col.~9 we discard the UCD as being responsible for the X-ray emission. 
The flags in the last two columns are
based on the detailed examination of alternative possible counterparts as
described in the remainder of this section. 
\begin{table*}
\begin{center}
\caption{Basic X-ray parameters from \ecat for the UCDs from \gagneM and UCD candidates from \cite{Reyle18.0} that are within $3 \times$ the uncertainty of the X-ray position.}
\label{tab:eRASS1}
\begin{tabular}{rrrrrrrcc}
\hline
  \multicolumn{1}{l}{{\em Gaia}\,DR2} &
  \multicolumn{1}{c}{{\sc ra\_corr}} &
  \multicolumn{1}{c}{{\sc dec\_corr}} &
  \multicolumn{1}{c}{delta\_radec} &
  \multicolumn{1}{c}{sep\_ox} &
  \multicolumn{1}{c}{Count rate} &
  \multicolumn{1}{c}{DETML} &
  \multicolumn{1}{c}{flags} &
  \multicolumn{1}{c}{UCD} \\
  \multicolumn{1}{l}{designation} &
  \multicolumn{1}{c}{[deg]} &
  \multicolumn{1}{c}{[deg]} &
  \multicolumn{1}{c}{[$^{\prime\prime}$]} &
  \multicolumn{1}{c}{[$^{\prime\prime}$]} &
  \multicolumn{1}{c}{[cts/s]} &
  \multicolumn{1}{c}{} &
  \multicolumn{1}{c}{} &
  \multicolumn{1}{c}{} \\
\hline
\multicolumn{9}{c}{\gagneM} \\
\hline
  4963614887043956096 & 34.842070 & $-$39.423258 & 4.63 & 2.32 & 0.064 $\pm$ 0.024 & 12.1 &  & P\\
  5084713036143375744 & 54.130367 & $-$26.333831 & 2.61 & 5.16 & 0.164 $\pm$ 0.031 & 80.1 & U & B\\
  4869246209213679872 & 65.587646 & $-$36.102014 & 2.56 & 1.84 & 0.12 $\pm$ 0.023 & 67.6 & U & B\\
  3200303384927512960 & 70.097406 & $-$5.501610 & 4.0 & 5.28 & 0.109 $\pm$ 0.032 & 24.7 & UX & B\\
  3181197137010596608 & 71.715048 & $-$11.280513 & 0.96 & 1.09 & 1.683 $\pm$ 0.128 & 1051.0 & C & B\\
  3164100487113671552 & 114.932856 & 13.080929 & 5.63 & 12.48 & 0.081 $\pm$ 0.038 & 10.2 & C &  \\
  666988221840703232 & 118.101611 & 16.201617 & 2.9 & 3.54 & 0.395 $\pm$ 0.081 & 63.7 & UX & B\\
\hline
\multicolumn{9}{l}{(a) result from counterpart selection: $U$ ... UCD is uniq {\em Gaia} counterpart, $C$ ... common proper motion pair, $X$ ... archival} \\
\multicolumn{9}{l}{X-ray detection} \\
\multicolumn{9}{l}{(b) final counterpart status for the UCD: $B$ ... bona-fide, $P$ ... plausible} \\
\multicolumn{9}{l}{The full table is available in electronic format on the CDS via anonymous ftp to cdsarc.u-strasbg.fr (130.79.128.5) or via}\\
\multicolumn{9}{l}{http://cdsweb.u-strasbg.fr/cgi-bin/qcat?J/A+A/} \\
\end{tabular}
\end{center}
\end{table*}

Given the relatively large {\em eROSITA} positional error in the survey mode, 
the UCDs associated with a source from \ecat 
according to the cross-match described above are not 
necessarily the correct counterparts of the X-ray source. 
We searched for alternative optical/IR counterparts to the
eRASS1 detections identified in the first step with a UCD by reversing the search,
that is we matched each of our two {\it fullmatch-eRASS1} samples  
again with {\em Gaia}\,DR2 
centering the search on the X-ray coordinates ({\sc ra\_corr, dec\_corr}) and
using a match radius of $24^{\prime\prime}$, the maximum value with
a reliable UCD-eRASS1 association in the first step. Then we removed all matches with
${\rm sep_{ox}} > 3$ {\sc radec\_err}. Since the `reverse match' involved no PM correction, in
this step the previously found UCD candidates could be removed if they do not fulfill
the separation criterion. This is, in fact, the case for one object from 
{\it fullmatch-eRASS1-GagneM}, and we added this object again to the list. The resulting
list of potential {\em Gaia}\,DR2 
counterparts comprises now $50$ objects for {\it fullmatch-eRASS1-GagneM} and
$233$ for {\it fullmatch-eRASS1-Reyle}. The task is now to determine for each of the
$96$ X-ray sources whether the UCD or UCD candidate is the most likely {\em Gaia} counterpart. 
%

\subsection{Bonafide eRASS\,1 counterparts}\label{subsect:identifications_eRASS1_bonafide}

We found that 
$12$ X-ray sources in {\it fullmatch-eRASS1-GagneM} 
 and 
$24$ in {\it fullmatch-eRASS1-Reyle} 
have a single {\em Gaia} counterpart, namely
the UCD or UCD candidate. In the following for simplicity we occasionally omit the 
distinction between UCD and UCD candidate, but the reader should keep in mind that the
objects from \gagneM are spectroscopically confirmed while those from \cite{Reyle18.0} are
not. 
In the abovementioned $36$ cases 
the UCD can safely be considered to be the cause for the X-ray emission
as there is no other {\em Gaia} source within $3 \times$ the error of the X-ray position.  
%

To establish the correct counterpart in the remaining cases we took into account several
criteria. First, we carried out a visual inspection of all cases with multiple {\em Gaia}\,DR2
counterparts in ESASky. 
This way we found that among our eRASS1 detections 
$2$ from {\it fullmatch-eRASS1-GagneM}
and 
$12$  
from {\it fullmatch-eRASS1-Reyle} have a common proper motion (CPM) companion;  
one of them is a triple system.  
In all but one cases the comoving companion is the more likely X-ray emitter because 
it is brighter and
closer to the eRASS1 position than the UCD\footnote{The CPM companion
is not in all cases the {\em Gaia} source that is closest to the \ecat position but it is by
far the brightest; see Appendix~\ref{app:cpmpairs}.}. We, therefore, remove
these 
$13$ 
objects from the sample. The one exception is
the UCD {\em Gaia}\,DR2\,3181197137010596608 from our \gagneM catalog that shares a 
similar proper motion
with the source {\em Gaia}\,DR2\,3181197137010596480. These two objects have similar {\em Gaia} 
magnitudes and are a pair of UCDs. Since this pair is very tight (separation of 
$1.6^{\prime\prime}$) we can not infer which of the two is the X-ray source. 
We provide more information on this and all other CPM pairs involving UCDs in 
Appendix~\ref{app:cpmpairs}. 
%
%
%


Another criterion that was useful to assign the X-ray source to the correct {\em Gaia} 
source are 
the X-ray positions in observations with the 
higher spatial resolution instruments {\em XMM-Newton} 
and {\em Chandra}. Such detections are available for 
$7$ of the $24$ objects in {\it fullmatch-eRASS1-GagneM}.
In all these cases the archival X-ray source is spatially associated with the UCD
according to our visual inspection. This relatively high
fraction of objects with archival X-ray data is a result of dedicated studies of 
UCD X-ray emission in the past. 
Two of the corresponding eRASS1 sources have multiple potential {\em Gaia} counterparts,
and we could assign
the UCD as the X-ray emitter thanks to the archival X-ray detection. 
In {\it fullmatch-eRASS1-Reyle} 
only $4$ objects are near an archived {\em XMM-Newton}
or {\em Chandra} source. 
One of these archival X-ray sources is closer to the brighter comoving
companion of a UCD candidate, and this object ({\em Gaia}\,DR2\,5762038930728469888) 
has been disregarded as an X-ray emitter in our previous selection step.
In another case both the UCD candidate ({\em Gaia}\,DR2\,3777108250009273856) 
and its CPM companion are detected with {\em Chandra}. The eRASS1 source is closer to the
CPM companion, and this object has also been removed from the list of UCD
candidates in the previous selection step. 
For the third one, the UCD candidate {\em Gaia}\,DR2\,3902395843353397248, 
both a {\em Chandra} and an {\em XMM-Newton} source are closer to another 
object, {\em Gaia}\,DR2\,3902395813288871936, which has similar optical brightness as the 
UCD candidate. 
This object is a known quasar 
(see Appendix~\ref{app:nonucds}). 
%
%
The last one, Gaia\,DR2\,3754497583659096320, 
has an {\em XMM-Newton} source clearly identified with the UCD candidate and there
is no other {\em Gaia} object in the vicinity that is nearly as bright and close to the 
eRASS1 X-ray position as the UCD candidate. 
The 
{\em XMM-Newton} detection of this UCD was discovered and discussed by \cite{Stelzer20.0}. 
To summarize, with help of the archival X-ray data we 
eliminated one additional UCD candidate from the eRASS1 source list 
{\it fullmatch-eRASS1-Reyle} 
because the X-ray source is identified as an extragalactic object. 

We consider the objects identified above and flagged in col.~8 of 
Table~\ref{tab:eRASS1}
`bonafide' counterparts
to \ecat X-ray sources. To summarize, we have identified the optical counterpart to
$16$ eRASS1 sources from {\it fullmatch-eRASS1-GagneM} 
and $38$ from {\it fullmatch-eRASS1-Reyle}. Among those $54$ `bonafide'
eRASS1 counterparts there are $40$ UCDs 
(flagged in col.~9 of Table~\ref{tab:eRASS1}
with `B'), 
$13$ earlier-type CPM companions to UCDs, and $1$ quasar. In
this way we are more than doubling the number of existing X-ray detections from UCDs. We caution,
however, that $25$ of them, the ones from {\it fullmatch-eRASS1-Reyle}, are 
UCD candidates that still require spectroscopic confirmation. 

\begin{figure*}
\begin{center}
\parbox{18cm}{
\parbox{9.0cm}{
\includegraphics[width=7.5cm]{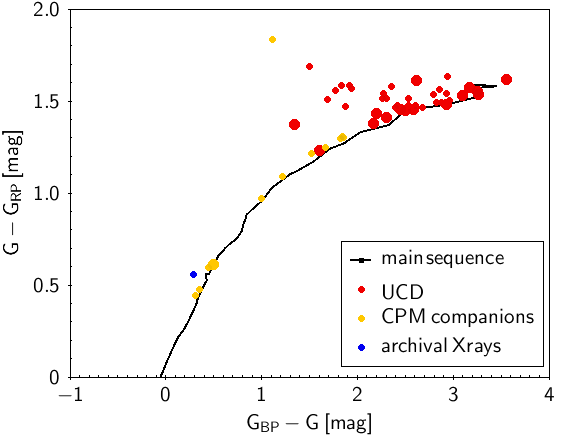}
}
\parbox{9.0cm}{
\includegraphics[width=7.5cm]{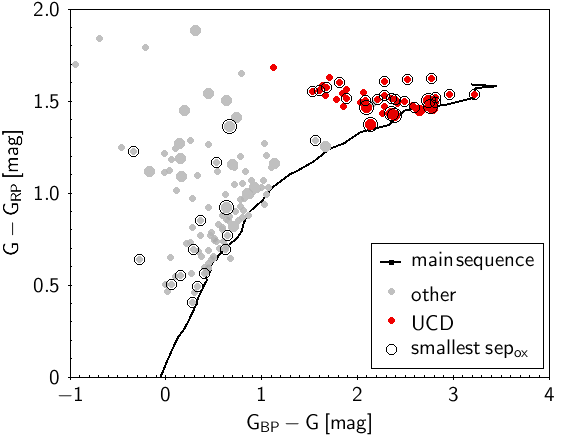}
}
}
\parbox{18cm}{
\parbox{9.0cm}{
\includegraphics[width=7.5cm]{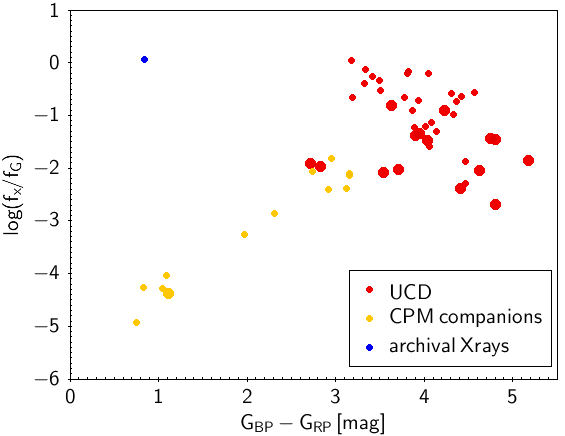}
}
\parbox{9.0cm}{
\includegraphics[width=7.5cm]{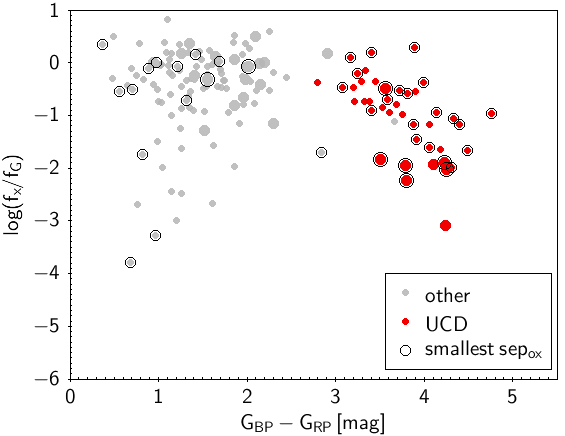}
}
}
\parbox{18cm}{
\parbox{9.0cm}{
\includegraphics[width=7.7cm]{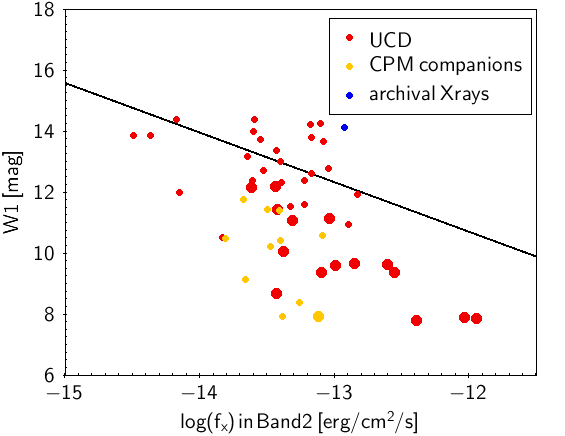}
}
\parbox{9.0cm}{
\includegraphics[width=7.5cm]{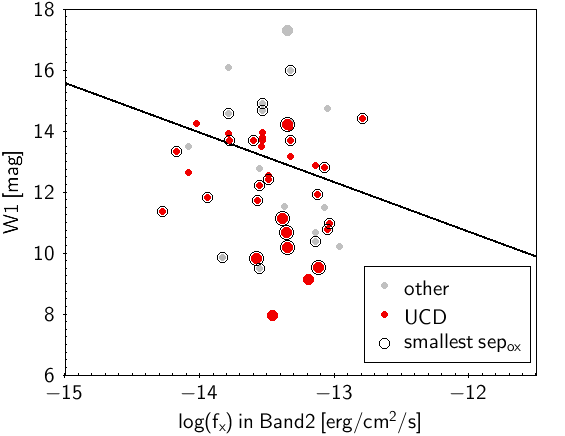}
}
}
\caption{Gaia color-color diagram (top), X-ray to $G$ band flux ratio 
(middle) and $W1$ magnitude vs X-ray flux (bottom) 
for subsamples illustrating the process for the selection of the {\em Gaia}
counterpart to the \ecat source. 
The left panels show the \ecat sources to which
a `bonafide' counterpart has been assigned and the right panels show the possible
counterparts for the remaining sources from \ecat. 
Small circles represent UCD candidates from \cite{Reyle18.0} and
large circles spectroscopically confirmed 
UCDs from \gagneM . UCDs from the input master lists are shown in red,
their CPM companions in yellow, and counterparts assigned based on archival X-ray
detections in blue. All other {\em Gaia} sources within the search box around the \ecat
X-ray positions are marked in gray. In the right panel the brightest {\em Gaia} source
and the one closest to the X-ray position is indicated for each \ecat source.
The black line in the bottom panels is an empirical separator between extragalactic
candidates (above) and stellar candidate (below) introduced by \protect\cite{Salvato18.0}.} 
\label{fig:eRASS1_selectiondiagrams}
\end{center}
\end{figure*}

On the left side of Fig.~\ref{fig:eRASS1_selectiondiagrams} we show 
various parameter combinations for the `bonafide' counterparts to \ecat X-ray sources. 
In particular, we display a {\em Gaia}
color-color diagram  (CCD) and a diagram involving the ratio of X-ray and optical flux,
$f_{\rm x}/f_{\rm G}$. For this latter diagram the X-ray flux is taken from 
\ecat column {\sc ml\_flux\_2}, the eRASS1 energy band `2'. 
It represents the $0.6-2.3$\,keV band and is the energy band
in \ecat that comes closest to the 
$0.5-2.0$\,keV band typically used for studies of X-ray populations. 
The optical flux in our X-ray/optical ratio refers to the {\em Gaia} $G$ band\footnote{The 
$G$ band flux was calculated from the magnitude using the zeropoints, effective wavelength 
and bandwidth provided by the Spanish Virtual Observatory (SVO) filter service at 
http://svo2.cab.inta-csic.es/theory/fps/}. 
The curved path that forms the lower envelope in 
the $G - G_{\rm RP}$ versus $G_{BP} - G$ diagram outlines the stellar main-sequence
as demonstrated by the black line which represents the {\em Gaia} colors from the 
table {\em A Modern Mean Dwarf Stellar Color and Effective 
Temperature Sequence} maintained by E. Mamajek\footnote{\label{note1} This table is available at 
http://www.pas.rochester.edu/$\sim$emamajek}. 
All UCDs show values for the 
X-ray flux above the canonical saturation limit of $\log{(f_{\rm x}/f_{\rm G})} \approx -3$.
We discuss this result in Sect.~\ref{sect:results}. Here we only note that 
in Appendix~\ref{app:selection_procedure} we argue based on the properties of the more than
$20 000$ X-ray sources from the {\it eROSITA Final Equatorial-Depth Survey} 
(eFEDS)\footnote{eFEDS is a $\sim 140$\,sq.deg large area in the southern sky that was
observed with {\em eROSITA} in scanning mode 
during its Calibration \& Performance Verification 
phase with a typical on-source exposure time of $1$\,ksec \citep{Brunner21.0}. 
This is significantly larger than the average exposure during eRASS1, and
the official source catalog comprises $27910$ X-ray detections.} 
that the part of the {\em Gaia} CCD occupied by the UCDs and UCD candidates 
is not populated by any other abundant population of X-ray
emitters, such as extragalactic objects or stars. 
Specifically, while a small number of eFEDS objects classified `galactic' are located in 
the same region as the UCDs the distribution of the `extragalactic' sources are clearly
separated from them (see Fig.~\ref{fig:eFEDS_GminRP_BPminG}). 
In Appendix~\ref{app:selection_procedure} we also examine the other diagrams shown in
Fig.~\ref{fig:eRASS1_selectiondiagrams}, and specifically the one of $W1$ magnitude versus 
the abovementioned Band\,2 X-ray flux (bottom panel in Fig.~\ref{fig:eRASS1_selectiondiagrams}).
In that diagram the black
line denotes the empirical separatrix between `extragalactic' (above) 
and `stellar' (below) X-ray emitters
that  
was defined by \cite{Salvato18.0} from two X-ray samples that bracket the eRASS1 survey 
in terms of X-ray fluxes:
the 2\,RXS \citep{Boller16.0} and XMMSL2 \citep{XMMSL2}
on the low-sensitivity end and the {\em Chandra} COSMOS Legacy Survey \citep{Marchesi16.0}  
on the high-sensitivity
end. Some contamination between the `extragalactic' and `stellar' groups 
was seen in \cite{Salvato18.0}, and might be present in our sample as well. 
In particular, from Fig.~\ref{fig:eRASS1_selectiondiagrams} it can be seen that 
in our sample of `bonafide' eRASS1 sources no confirmed UCDs but some of the UCD candidates
are located in the `nonstellar' part of this diagram. 

%
%
%
%
%
%

\subsection{Plausible eRASS\,1 counterparts}\label{subsect:identifications_eRASS1_plausible}

Putting aside the $54$ identified `bonafide' eRASS1 objects, there are still
$42$ \ecat sources to which we must assign the optical counterpart. Each of these X-ray
sources has at least two {\em Gaia} objects in the search radius.
In the right panels
of Fig.~\ref{fig:eRASS1_selectiondiagrams} the full sample of possible counterparts
to the $42$ still unassigned X-ray sources from {\it fullmatch-eRASS1-GagneM} and
{\it fullmatch-eRASS1-Reyle} is shown. Since these \ecat sources 
have several {\em Gaia} counterparts in the 
search radius they appear more than once in those panels of the figure that involve
{\em Gaia} data. 
Some of the non-UCD {\em Gaia} sources are missing because they have incomplete photometry,
while the UCDs and UCD candidates by definition of our input samples (Sect.~\ref{sect:sample})
are fully characterized with {\em Gaia}.  

The same features described above for the `bonafide'
counterparts are seen, such as the main-sequence path in the CCD and high values of 
$f_{\rm x}/f_{\rm G}$ for the UCDs. In addition a strongly populated area in the upper left of
both diagrams (defined by relatively blue optical color and high $f_{\rm x}/f_{\rm G}$ ratio)
is evident. Only one of the bonafide eRASS1 counterparts is located in that region, the
quasar 2XMM\,J123337.5+073133 (see left side of Fig.~\ref{fig:eRASS1_selectiondiagrams}). 
That region is, therefore, likely dominated by extragalactic sources. 
For the position of
different astrophysical populations in the eRASS1 and {\em Gaia} parameter space 
we refer, again, to  
Appendix\,\ref{app:selection_procedure}. Here we define our criteria for selecting the 
counterpart to each X-ray source from its pool of {\em Gaia} objects. 

To this end, we compare the  
separations between the positions of the {\em Gaia}\,DR2 and the
eRASS1 sources, ${\rm sep_{ox}}$, and the $G$ magnitudes 
of all possible {\em Gaia} counterparts uncovered in our `reversed search' for 
these $42$ eRASS1 sources. The 
{\em Gaia} source with the smallest ${\rm sep_{ox}}$ value for each eRASS1 source is 
highlighted in the viewgraphs 
on the right side of Fig.~\ref{fig:eRASS1_selectiondiagrams} with an annulus. 
For six out of the eight  
still unassigned X-ray sources from {\it fullmatch-eRASS1-GagneM},  
the UCD is the closest 
counterpart to the X-ray position, 
and for the {\it fullmatch-eRASS1-Reyle} sample in $19$ of $34$ cases the UCD candidate has
the smallest ${\rm sep_{ox}}$ value of all {\em Gaia} counterparts. 
The remaining $17$ have a non-UCD as their closest counterpart. 

As a baseline we define as the correct {\em Gaia} counterpart for the
eRASS1 X-ray source the one with the smallest $sep_{\rm ox}$ value. 
$25$ UCDs and UCD candidates fulfill this criterion. 
However, in this work we pursue a conservative approach in which we keep UCDs as 
counterparts to X-ray sources only if there is no other obviously more plausible 
optical counterpart. Therefore, 
we examined all other potential {\em Gaia} counterparts for these $25$ cases. 
This way we identified $7$  
cases where an object with a brighter $G$ magnitude is
in the $3 \times$\,{\sc radec\_err} match radius. 
Their optical/IR properties are presented in 
Appendix~\ref{app:nonucds}. Following the conservative approach described above, 
we consider these brighter {\em Gaia} sources as likely counterpart to the X-ray emitter, 
and we remove the corresponding UCDs from the X-ray sample. 
The other $18$ UCDs (that are both the closest and the brightest {\em Gaia} counterpart
to the X-ray source) are added to the sample investigated 
in this article. We henceforth term these UCD X-ray assignments as `plausible', because 
due to the presence of other {\em Gaia} sources in the 
match radius they 
are somewhat less certain associations than the `bonafide' objects from 
Sect.~\ref{subsect:identifications_eRASS1_bonafide}. 
These UCDs are marked with `P' in col.~9 of Table~\ref{tab:eRASS1}. 

The $17$ non-UCD counterparts with the smallest $sep_{\rm ox}$ among 
all {\em Gaia} counterparts are not of interest to our study. However,
for completeness and future reference we provide optical/IR parameters of 
these other astrophysical X-ray emitters in Appendix~\ref{app:nonucds}.
From a comparison of Fig.~\ref{fig:eRASS1_selectiondiagrams} and 
Fig.\ref{fig:eFEDS_GminRP_BPminG} we can infer that those of them 
that are located in the upper left in the $G - G_{\rm RP}$ vs $G_{\rm BP} - G$ diagram
are likely extragalactic, while for those that are located close to the stellar
main-sequence the nature can not be established based on the {\em Gaia} color-color
diagram.  
Note that for Table~\ref{tab:app_nonUCDs} 
when $G$ and $J$ magnitudes are available we have computed the corresponding 
main-sequence spectral type, but this parameter may be meaningless if the object
is extragalactic.

\subsection{Reliability of the X-ray sources}\label{subsect:identifications_eRASS1_spurious}

The detection threshold for the source detection applied to generate \ecat was
{\sc det\_like\_0} $=6$. At this threshold about $\sim 6.9$\,\% of sources are expected
to be spurious \citep{Brunner21.0}. 
Among the $58$ eRASS1 detected UCDs $4$ have {\sc det\_like\_0} $\leq 6.5$, 
that is statistically none of them is expected to be a spurious detection.

\section{Multiwavelength properties of eRASS1-detected UCDs}\label{sect:eRASS1_multilambda}

%
%

As we anticipated in Sect.~\ref{sect:identifications_eRASS1}
all UCDs that we have identified as X-ray emitters 
are marked in the last column of Table~\ref{tab:eRASS1}.
To summarize, we have associated $21$ spectroscopically confirmed 
UCDs from \gagneM to an X-ray source from \ecat, of which $15$ as bonafide and $6$ as plausible 
counterparts. Among the UCD candidates from \cite{Reyle18.0} 
$37$ have been assigned as eRASS1 X-ray emitter, $25$ as bonafide and $12$ as plausible counterparts. 
 
While the multiwavelength properties
of the non-UCD X-ray emitters are presented in the appendices of this work as
described above, here we
summarize relevant parameters for the 
X-ray emitting UCDs and UCD candidates. 
Table~\ref{tab:multilam_UCDs} 
lists the photometry used in this article, that is the {\em Gaia} $G$ band, 2MASS $J$ band 
and AllWISE  
$W1$ band (cols.~2-4). For the objects from \gagneM we provide the SpT from optical
spectroscopy (col.~5). 
In order to have a common SpT scale for the \gagneM and the \cite{Reyle18.0} sample 
we have computed spectral types from the $G-J$ color using the table maintained 
by E.Mamajek\footref{note1} 
for all bonafide and plausible UCDs and UCD candidates (col.~6).  
This scale has the advantage that it is valid for the full SpT sequence, while the
SpT - $M_{\rm G}$ relation used in \cite{Reyle18.0} has been calibrated only for SpT 
M7 and later. In fact, we find that the few M5 objects in \gagneM would all 
have an early- to mid-M  
SpT assigned when using the SpT - $M_{\rm G}$ calibration from \cite{Reyle18.0}, 
at odds with their spectroscopic values. 

Since the distance is crucial for the determination of luminosities we have checked the
reliability of the values obtained from inverting the parallaxes. To this end we have
matched all bonafide and plausible X-ray emitting UCDs with the catalog of
\cite{BailerJones18.0} 
in which a probabilistic approach was used to infer distances
from {\em Gaia} parallaxes and its standard deviation 
and a model for {\em Gaia}'s length scale in the Galaxy. 
The values $d = 1 / \varpi$, where $\varpi$ is the {\em Gaia}\,DR2 parallax, 
are for all targets in excellent agreement with the distances
obtained by \cite{BailerJones18.0}. 
This is not unexpected since our sample by 
construction comprises only objects with well-constrained parallax 
(see Sect.~\ref{sect:sample}). We conclude that all
bonafide and plausible eRASS1 detections among the UCDs and UCD candidates 
have a reliable distance value. 
In col.~7 of Table~\ref{tab:multilam_UCDs} 
we provide the {\em Gaia} distance extracted from the catalog of
\cite{BailerJones18.0}. 

The last two columns of Table~\ref{tab:multilam_UCDs} hold the X-ray luminosity ($L_{\rm x}$)
and the fractional X-ray luminosity ($L_{\rm x}/L_{\rm bol}$) in logarithmic scale. 
The bolometric luminosity required for the $L_{\rm x}/L_{\rm bol}$ ratio 
has been calculated with the polynomial fit with SpT from \cite{Filippazzo15.0}, and
the spectral types used were those derived from the $G-J$ color.  
The X-ray flux
that together with the distance defines $L_{\rm x}$ was derived from the
$0.2-2.3$\,keV count rates in \ecat by summing the values in cols. {\sc ml\_rate\_1}
and {\sc ml\_rate\_2} and 
applying the rate-to-flux conversion factor $CF_{\rm M, 12}
= (7.37 \pm 0.55) \cdot  10^{-13}\,{\rm erg/cts/cm^2}$. This value was derived from 
fits of a thermal model ({\sc apec}) to eFEDS spectra of early-M dwarfs studied by 
\cite{Magaudda21.0}. 
Note that in \cite{Magaudda21.0} a
slightly higher value is reported which refers to a broader {\em eROSITA} energy band ($0.2-5.0$\,keV),
but the same approach and the same sample of M dwarf {\em eROSITA} spectra was used to derive the
value for the $0.2-2.3$\,keV band (that is the sum of eRASS1 energy bands `1' and `2') 
cited above specifically for our work.  

We recall that in the multiwavelength diagrams that supported the source 
identification (Fig.~\ref{fig:eRASS1_selectiondiagrams}) we
have made use of the X-ray fluxes tabulated in \ecat . 
These fluxes have been calculated for an absorbed power-law 
(with $N_{\rm H} = 10^{20}\,{\rm cm^{-2}}$ and $\Gamma = 1.7$) which is an incorrect 
assumption for coronal X-ray emitters. The use of these fluxes for our source
identification procedure is justified by the fact that many
of the potential {\em Gaia} counterparts to the X-ray sources are nonstellar. 
To get an idea on the effect of the inappropriate use of \ecat fluxes 
on our sample of UCDs we have computed the $CF$ for the power-law model given above 
as the mean of the ratio between the $0.2-2.3$\,keV count rates and the fluxes in the
same band given in \ecat. 
We find for our sample of $58$ UCDs and UCD candidates a mean and standard deviation of 
$CF_{\rm PL} = (9.75 \pm 0.21) \cdot 10^{-13}\,{\rm erg/cts/cm^2}$, that is a $30$\,\% 
higher $CF$ than the value obtained from the M dwarf spectra.

\begin{sidewaystable} 
\begin{center}
\caption{Multi-wavelength properties of the UCDs and UCD candidates identified as bonafide or plausible counterparts to an eRASS1 X-ray source.}
\label{tab:multilam_UCDs}
\begin{tabular}{rrrrllrrrcccc}
\hline
\hline
\noalign{\smallskip}
  \multicolumn{1}{l}{{\em Gaia}\,DR2 designation} &
  \multicolumn{1}{c}{$G$} &
  \multicolumn{1}{c}{$J$} &
  \multicolumn{1}{c}{$W1$} &
  \multicolumn{1}{c}{${\rm SpT_{spec}}$} &
  \multicolumn{1}{c}{${\rm SpT_{G-J}}$} &
  \multicolumn{1}{c}{d} &
  \multicolumn{1}{c}{$\log{L_{\rm x}}$} &
  \multicolumn{1}{c}{$\log{(L_{\rm x}/L_{\rm bol})}$} &
  \multicolumn{1}{c}{Young} &
  \multicolumn{1}{c}{Young} &
  \multicolumn{1}{c}{Binary} &
  \multicolumn{1}{c}{Binary} \\
  \multicolumn{1}{l}{of UCD} &
  \multicolumn{1}{c}{[mag]} &
  \multicolumn{1}{c}{[mag]} &
  \multicolumn{1}{c}{[mag]} &
  \multicolumn{1}{c}{} &
  \multicolumn{1}{c}{} &
  \multicolumn{1}{c}{[pc]} &
  \multicolumn{1}{c}{[erg/s]} &
  \multicolumn{1}{c}{} &
  \multicolumn{1}{c}{flag} &
  \multicolumn{1}{c}{ref} &
  \multicolumn{1}{c}{flag} &
  \multicolumn{1}{c}{ref} \\
\noalign{\smallskip}
\hline
\noalign{\smallskip}
\multicolumn{13}{c}{\gagneM} \\ 
\noalign{\smallskip}
\hline
\noalign{\smallskip}
  659464504288593536  & 15.02 & 11.05 &  9.80  & M6      & M6-M6.5V &  $13.552_{-0.025}^{+0.026}$ & 27.21 & $-$3.33 & \dots & \dots & \dots & \dots \\
\noalign{\smallskip}
  666988221840703232  & 14.77 & 10.88 &  9.61 & M7      & M6-M6.5V &  $18.883_{-0.036}^{+0.037}$ & 28.07 & $-$2.51 & Y & 5,6 & \dots & \dots \\
\noalign{\smallskip}
  703790044252850688  & 12.19 &  8.24 &  6.95 & M6.5    & M6-M6.5V &   $3.580_{-0.002}^{+0.002}$ & 26.83 & $-$3.72 & \dots & \dots & \dots & \dots \\
\noalign{\smallskip}
  908963788782045568  & 19.23 & 15.57 & 14.27 & M6      & M5.5-M6V & $115.884_{-5.625}^{+6.218}$ & 28.94 & $-$1.78 & \dots & \dots & \dots & \dots \\
\noalign{\smallskip}
  3181197137010596608 & 12.22 &  8.14 &  7.11 & M5 + M6 & M7V      &  $18.968_{-0.024}^{+0.024}$ & 28.73 & $-$1.71 & Y & 5,6 & Y & 1,13 \\
\noalign{\smallskip}
  3200303384927512960 & 14.93 & 10.66 &  9.36 & M6.5    & M7-M7.5V &   $9.757_{-0.010}^{+0.010}$ & 26.96 & $-$3.41 & Y & 5,6 & \dots & \dots \\
\noalign{\smallskip}
  3830128624846458752 & 16.55 & 12.33 & 11.04 & M7      & M7-M7.5V &  $20.123_{-0.099}^{+0.100}$ & 27.52 & $-$2.87 & \dots & \dots & \dots & \dots \\
\noalign{\smallskip}
\hline
\multicolumn{13}{l}{References: (1) \cite{Mason01.0}, (5) \cite{Shkolnik09.0}, (6) \cite{Shkolnik12.0}, (13) \cite{Winters19.0}.} \\
\multicolumn{13}{l}{The full table is available in electronic format on the CDS via anonymous ftp to cdsarc.u-strasbg.fr (130.79.128.5) or via http://cdsweb.u-strasbg.fr/cgi-bin/qcat?J/A+A/} \\
\end{tabular}
\end{center}
\end{sidewaystable}

\section{eRASS1 light curves and spectra}\label{sect:lc_and_spec}

As a result of a combination of the weak activity levels of UCDs and the 
limited sensitivity of eRASS1 most X-ray detections of UCDs have 
a small number of counts. Although, as we argued in 
Sect.~\ref{subsect:identifications_eRASS1_spurious}, statistically no spurious detections
are expected in our UCD X-ray emitter sample, 
$25$\,\% of them 
have a detection likelihood {\sc det\_like\_0} $\leq 8$. For these weak sources on
average only $7$ source counts were collected during eRASS1. 

Two sources stick out with a count rate ($>1$\,cts/s) much above the average. These are also
the only two UCDs for which more than $100$\,counts have been collected during eRASS1,
and thus a meaningful spectral and timing analysis is feasible. 
We study the eRASS1 spectra and light curves of these two UCDs 
for which $227$\,counts and $292$\,counts are listed for the $0.2-5.0$\,keV band in 
\ecat, respectively. These values are clear outliers from the trend of low
count numbers in our sample and 
lead one to suspect  
flaring activity. 
One of the two is the
binary UCD {\em Gaia}\,DR2\,3181197137010596608 and {\em Gaia}\,DR2\,3181197137010596480 
(Sect.~\ref{subsect:identifications_eRASS1_bonafide}), also known as WDS\,J04469-1117A and~B
and listed with SpT M5 + M6 in \gagneM \citep[based on measurements presented by ][]{Shkolnik09.0}.  
The other one is {\em Gaia\,DR2}\,5355751581627180288, alias TWA\,22\,AB, which is -- contrary to
its historical assignment to the TW\,Hya association -- a member of the $\beta$\,Pic moving
group \citep{Malo13.0} 
and an astrometric binary \citep{Bonnefoy09.0} with two components of SpT M6. 
Both objects for which we are performing eRASS1 
timing and spectral analysis are, thus, not genuine UCDs. Nevertheless, since the X-ray
properties for mid- to late-M stars are not well known, their X-ray variability 
and plasma temperature are of high interest.

Our analysis of the light curves and spectra of WDS\,J04469-1117AB
and TWA\,22AB is based on the merged events files of all seven telescopes 
for the corresponding sky tiles that comprise our targets.  
We extracted the $0.2-5.0$\,keV band light curves 
and spectra, and all required complementary files 
with the {\sc srctool} task of the 
{\em eROSITA} Science Analysis System (eSASS), version eSASSuser\_201009. 
To this end, we have defined a $40^{\prime\prime}$-radius circular region centered on the
boresight corrected position of the X-ray source associated with the UCD.  
For the background region we chose an annular region with 
the same center and an inner and outer radius of $90^{\prime\prime}$ and $180^{\prime\prime}$, 
respectively. 

We used the {\sc srctool} command with the {\sc regular-} option   
which produces a light curve with regularly spaced bins in which time intervals without data are 
automatically discarded. 
As a result of the sparse sampling in the survey mode the 
light curves are dominated by such data gaps.
During its All-Sky Survey {\em eROSITA} visits a given sky position several times with a typical 
time lapse between one and the next visit of $\sim 4$\,h (corresponding to one full rotation 
of the spacecraft and called the `scan rate'). 
Each such visit is called an eRODay. The average number of eRODays for
a source is six, and it is determined by the combination of the scan rate, the orbital 
speed of {\em eROSITA} ($\sim 1^\circ {\rm / d}$) and the field-of-view of the 
instrument (diameter $1^\circ$). However, the number of eRODays and the total on-source
exposure time depend also on the sky position, with more visits taking place for objects
near the ecliptic poles where the great circles traced by the telescope intersect each other
for longer than the abovementioned average. 
Moreover, the exposure time per eRODay varies 
between one scan and the next one since in subsequent scans 
the source crosses the circular field-of-view at different positions. 

The {\sc regular-} option of {\sc srctool} produces a light curve with bins of user-defined 
length irrespective of the temporal sequence of the data-taking which is different for
each source. Since the time interval from the first to the last eRODay is dominated by
data gaps, an arbitrary choice of bin size is likely to lead to a 
light curve which comprises bins with extremely low exposure time, such as when a bin just
scratches the beginning or end of an eRODay. For faint sources this results in 
bins with very large uncertainties on the count rate. 
To take account of these peculiarities related to the survey mode, we have used a 
trial-and-error approach to adjust 
the bin size and the start time of the binning such as to obtain 
one single bin per visit of the source. 

The eRASS1 spectra of WDS\,J04469-1117AB and TWA\,22AB, 
the spectra of their associated
background regions, and the response
matrix and ancilliary response files were extracted with {\sc srctool}.    
The spectra were then binned to a minimum of $10$ counts per bin, 
and subsequently fitted in the XSPEC environment v\,12.11.1 \citep{Arnaud96.0}.

Details of 
the spectral fitting and the results from the spectral and temporal analysis for these
two objects are presented in Sect.~\ref{subsect:results_xproperties}.

\section{Results on UCDs detected in eRASS1}\label{sect:results}

\subsection{Overall properties of the population}\label{subsect:results_general}

%
%
%
%
%
\begin{figure}[t]
\begin{center}
\includegraphics[width=8cm]{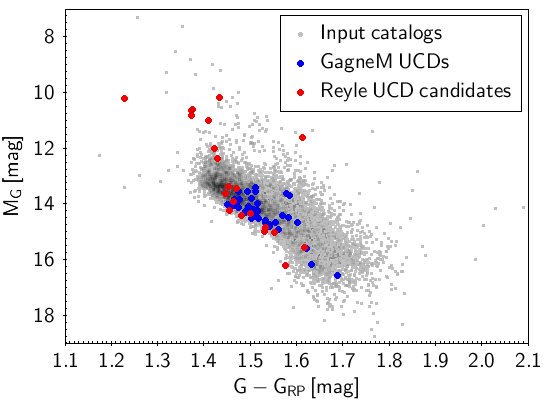}
\caption{{\em Gaia} color-magnitude diagram ($M_{\rm G}$ vs $G - G_{\rm RP}$) 
for the two input lists of UCDs (\gagneM) and UCD candidates \citep{Reyle18.0}  
described in Sect.~\ref{sect:sample}. The subsamples that
our source identification from Sect.~\ref{sect:identifications_eRASS1} 
ascribes as bonafide or plausible counterpart to an eRASS1 X-ray source
are highlighted with larger circles and colors, red for the known UCDs and blue
for the UCD candidates.}
\label{fig:eRASS1_GaiaCMD_all_and_onlyUCDs}
\end{center}
\end{figure}

%
%
%
\begin{figure*}[t]
\begin{center}
\parbox{18cm}{
\parbox{6cm}{
\includegraphics[width=6cm]{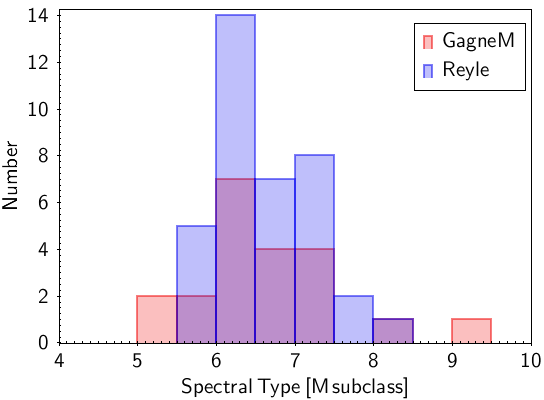}
}
\parbox{6cm}{
\includegraphics[width=6cm]{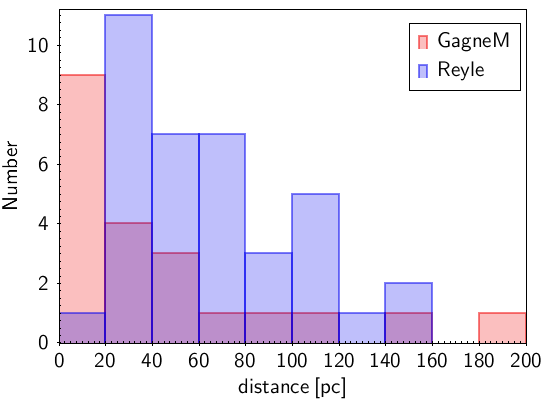}
}
\parbox{6cm}{
\includegraphics[width=6cm]{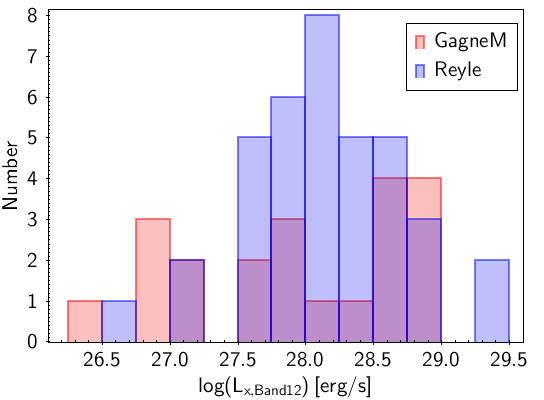}
}
}
\caption{Distribution of spectral types, distances and X-ray luminosities for 
the `bonafide' and `plausible' X-ray emitting 
UCDs and UCD candidates detected in eRASS1. 
SpTs from the \gagneM sample of known UCDs are from optical spectroscopy,
while those from the {\em Gaia} UCD candidates from \cite{Reyle18.0} are photometric
estimates based on $G-J$ color.}
\label{fig:histos_UCDs}
\end{center}
\end{figure*}

Figure~\ref{fig:eRASS1_GaiaCMD_all_and_onlyUCDs} 
shows the X-ray emitting bonafide and plausible UCDs and UCD candidates together with the full
cleaned input samples \gagneM
and \cite{Reyle18.0} 
in the {\em Gaia} color-magnitude diagram. 
Throughout the remainder of this paper, 
the spectroscopically confirmed X-ray emitting UCDs (from \gagneM) are shown in red and the
X-ray emitting UCD candidates \citep[from][]{Reyle18.0} in blue. 

A number of the spectroscopically confirmed UCDs detected in eRASS1 belong to the group of objects
above the main-sequence  (flagged `Young' in Table~\ref{tab:multilam_UCDs}). 
These are late-M stars in young nearby moving groups. 
The strong X-ray emission associated with the youth ($\sim 10 ...100$\,Myr) of 
stars has played a major role in the discovery of such co-moving stellar associations 
based on {\em ROSAT} data 
\citep[e.g.,][]{Kastner97.0, Kastner03.0}.
It is, therefore, not surprising that an disproportionate fraction of these
young objects is detected in eRASS1. In fact, the young objects are among the strongest
X-ray emitters in the sample of eRASS1 UCD detections (as we show at the end of this Section). 
There are only two young objects in the \cite{Reyle18.0}
sample of X-ray emitting UCD candidates 
because that sample was selected from {\em Gaia} data on the basis of the field dwarfs from 
J.Gagn\'e's dwarf archive.  

In the leftmost panel of 
Fig.~\ref{fig:histos_UCDs} 
we show the SpT 
distribution for the combined sample of bonafide and plausible UCDs detected in eRASS1. 
Recall that while the \gagneM 
catalog comprises spectroscopic spectral types, the new UCD candidates
from \cite{Reyle18.0} have only photometric estimates for the SpT. In Sect.~\ref{sect:sample}
we have demonstrated that there is overall good agreement between spectroscopic and
photometric SpTs derived from the absolute {\em Gaia} magnitude, $M_{\rm G}$, 
but in Sect.~\ref{sect:eRASS1_multilambda} we have argued in favor of a calibration with
$G-J$ color which covers the full range of SpTs present in our sample. 
In any case, we caution that in absence of a spectroscopic confirmation the values provided
in \cite{Reyle18.0} must be considered tentative. 
Keeping this in mind, we see in Fig.~\ref{fig:histos_UCDs} that 
the distributions of the number of X-ray detections for
both the known UCDs and the UCD candidates are peaked at SpT M6...M7. 
Note that one UCD, {\em Gaia}\,DR2\,4963614887043956096, has an L4-type companion
\citep{Artigau15.0}. However, since the higher-mass component is the more likely X-ray 
emitter we have assigned a SpT of M6 (the value for the primary) to this object. 
The latest SpT with an eRASS1 detection is M9. This is 
{\em Gaia}\,DR2\,5761985432616501376 (2MASS\,J08533434-0329432) 
with spectroscopic SpT M9 (in the optical) and M8.5 (in the NIR) according to \gagneM. 
No UCDs from the {\sc gucds} sample are detected in eRASS1. 
Therefore, no X-ray detection from any confirmed L dwarf can be reported from our work.

The distance distribution  of the eRASS1-detected UCDs is displayed in 
the middle panel of 
Fig.~\ref{fig:histos_UCDs}. 
It shows a peak at $\sim 30$\,pc and comprises 
values from $3.5$ to $190$\,pc. 
This sample is clearly drawn among the more nearby objects from the input catalogs, \gagneM
and \cite{Reyle18.0}, which is not surprising given the fact that eRASS is a flux-limited
survey. 
The UCD with the largest distance among the eRASS1 detections is CHSM\,17173, a member
of the Cha\,I cloud and likely the youngest in this sample. 

\begin{figure}
\begin{center}
\includegraphics[width=9cm]{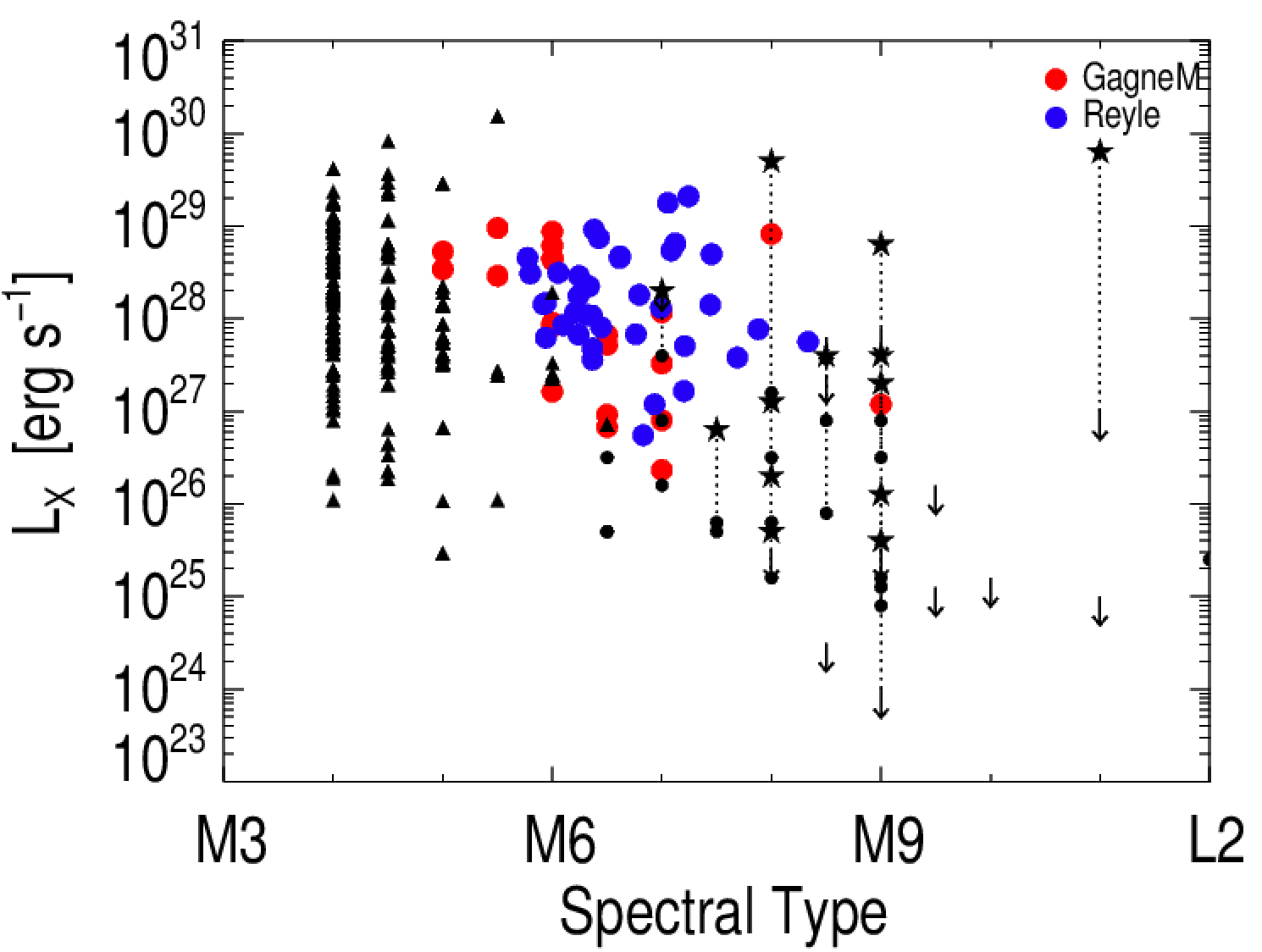}
\includegraphics[width=9cm]{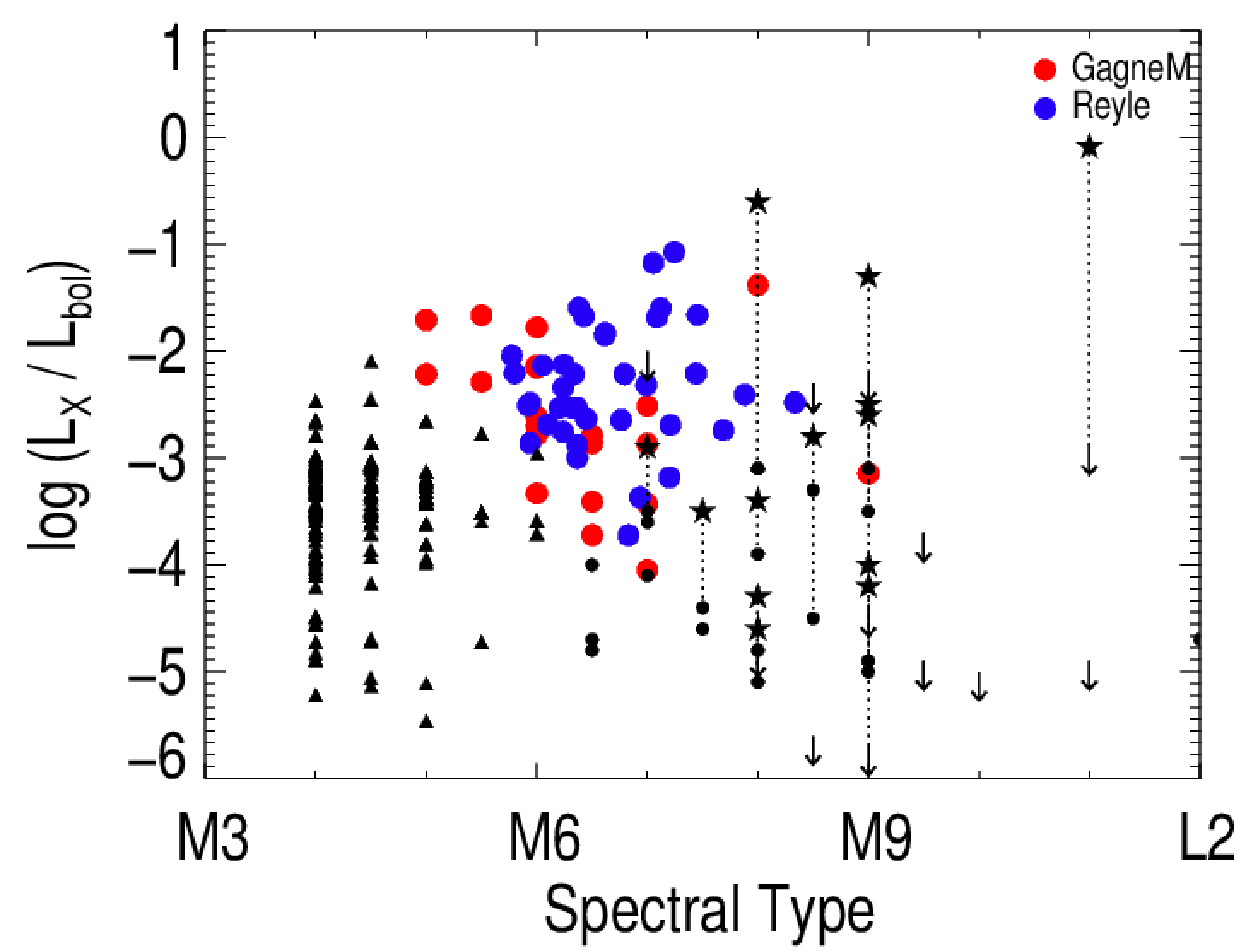}
\caption{X-ray luminosity and its ratio with bolometric luminosity for UCDs and UCD
candidates detected 
in eRASS1 compared to literature samples described in the text of 
Sect.~\ref{subsect:results_general}.}
\label{fig:lxlbol_spt}
\end{center}
\end{figure}

The most important parameters to be delivered from this study are the X-ray luminosities
and the $L_{\rm x}/L_{\rm bol}$ values. 
In Sect.~\ref{sect:eRASS1_multilambda} we have explained how we have computed the 
$L_{\rm x}$ values from 
the tabulated count rates in \ecat and a rate-to-flux conversion factor derived from
{\em eROSITA} observations of early M dwarfs. 
The distribution of the X-ray luminosities computed with this $CF$ and the distances from
\cite{BailerJones18.0} are visualized in the right panel of 
Fig.~\ref{fig:histos_UCDs}. 
The $L_{\rm x}$ values 
are remarkably high, similar to those of mid-M dwarfs. 
This is seen in the top panel of Fig.~\ref{fig:lxlbol_spt} 
where we show the new eRASS1 X-ray detected
UCDs and UCD candidates together with a compilation of X-ray emission from mid-M stars
to L-type UCDs 
from the literature. The literature data for UCDs (black symbols with 
SpT M6 and later in Fig.~\ref{fig:lxlbol_spt})  are from  
\cite{Stelzer12.0}, \cite{Cook14.0}, \cite{Williams14.0} and \cite{deLuca20.0}.  
The earlier-type stars shown as black triangles 
are from \cite{Magaudda20.0} and \cite{Magaudda21.0}.   
While these mid-M dwarfs show X-ray to bolometric luminosity ratios at or below the
canonical saturation limit of $\log{(L_{\rm x}/L_{\rm bol})} \approx -3$, 
for most of the UCDs the X-ray to bolometric luminosity ratio is much higher, 
and about half of them have 
$\log{(L_{\rm x}/L_{\rm bol})}$ values above the upper envelope observed in 
mid-M dwarfs. As the literature data on UCDs shows,  
values as high as $\log{(L_{\rm x}/L_{\rm bol})} \approx -1$ have occasionally
been observed before on UCDs but only during flares (asterisks in Fig.~\ref{fig:lxlbol_spt}).
%
%
%

Since we have seen that the distances are reliable and the probability even for the faintest
objects to be spurious is small, the only remaining possible error source in the
calculation of the X-ray luminosity is the $CF$. 
However, we have shown that even a 
clearly wrong spectral model leads to a difference in the $CF$ on the $30$\,\% level only. 
We may caution 
that many of the UCDs detected in eRASS1 
might have been 
in a high-activity flaring state during the {\em eROSITA} measurement, and therefore
the plasma temperatures could be higher than those of the M dwarfs studied in 
\cite{Magaudda21.0}. In fact, the two UCDs with the highest number of collected
counts for which we analyzed the
eRASS1 light curves both show evidence for flares (see Sect.~\ref{subsect:results_xproperties}).
From their X-ray spectra we derive a mean 
$CF_{\rm UCD,f} = 9.27 \cdot 10^{-13}\,{\rm erg/cts/cm^2}$. This is about $25$\,\%
higher than the value we used and would, therefore, lead to even higher $L_{\rm x}$ values. 
On the other hand, there 
is virtually no observational basis for our assumption that the X-ray temperatures of 
UCDs are similar to those of early-M dwarfs. It is, actually, not implausible to speculate
that the coronae of UCDs are somewhat cooler than those of objects with earlier M spectral 
type given the typical correlation between X-ray luminosity and temperature observed
in GKM stars \citep{Johnstone15.0}. The consequence would be a lower $CF$. 
However, considering that all these uncertainties are on the
percent level, the $CF$ can not be responsible for the very high $\log{(L_{\rm x}/L_{\rm bol})}$
level of the eRASS1 UCD detections. 
The final possibility for the apparent high $\log{(L_{\rm x}/L_{\rm bol})}$ values
we observe is a wrong identification of the UCDs and UCD candidates as an X-ray emitter. 
However, we recall
that more than half of the UCDs that we assigned to an X-ray source 
have no other known {\em Gaia} object within $3 \times$ the uncertainty of the X-ray position;
the $36$ objects with flag `U' in Table~\ref{tab:eRASS1}. 

To summarize, there is no obvious reason to question the observed values of $L_{\rm x}$ and
$\log{(L_{\rm x}/L_{\rm bol})}$ of these X-ray detections. Therefore, the most viable 
conclusion is that many of these objects showed significant flaring activity during 
the eRASS1 observation. To test this hypothesis we have extracted the light curves of the two 
UCD candidates with the most extreme $\log{(L_{\rm x}/L_{\rm bol})}$ values in 
Fig.~\ref{fig:lxlbol_spt} following the scheme explained in Sect.~\ref{sect:lc_and_spec}.
These objects are weak detections that have $\sim 10$ times less counts than the two UCDs
mentioned in Sect.~\ref{sect:lc_and_spec} that we discuss in detail in 
Sect.~\ref{subsect:results_xproperties}. 
Their eRASS1 light curves clearly show that outside a short time interval 
the count rate is consistent with zero, therefore fully confirming the suspicion that
the high activity levels in our sample results from flares. A detailed study of 
X-ray variability in the full sample will be performed in another publication. Here, 
as an illustration for the morphology of eRASS light curves, the {\em eROSITA} 
spectral response and, thus, the potential of {\em eROSITA} for UCD science we discuss only 
the two objects with the largest number of counts in our sample. This analysis is
presented in the next subsection.

\subsection{Information from X-ray light curves and spectra}\label{subsect:results_xproperties}

\begin{figure}
\begin{center}
\includegraphics[width=8.5cm]{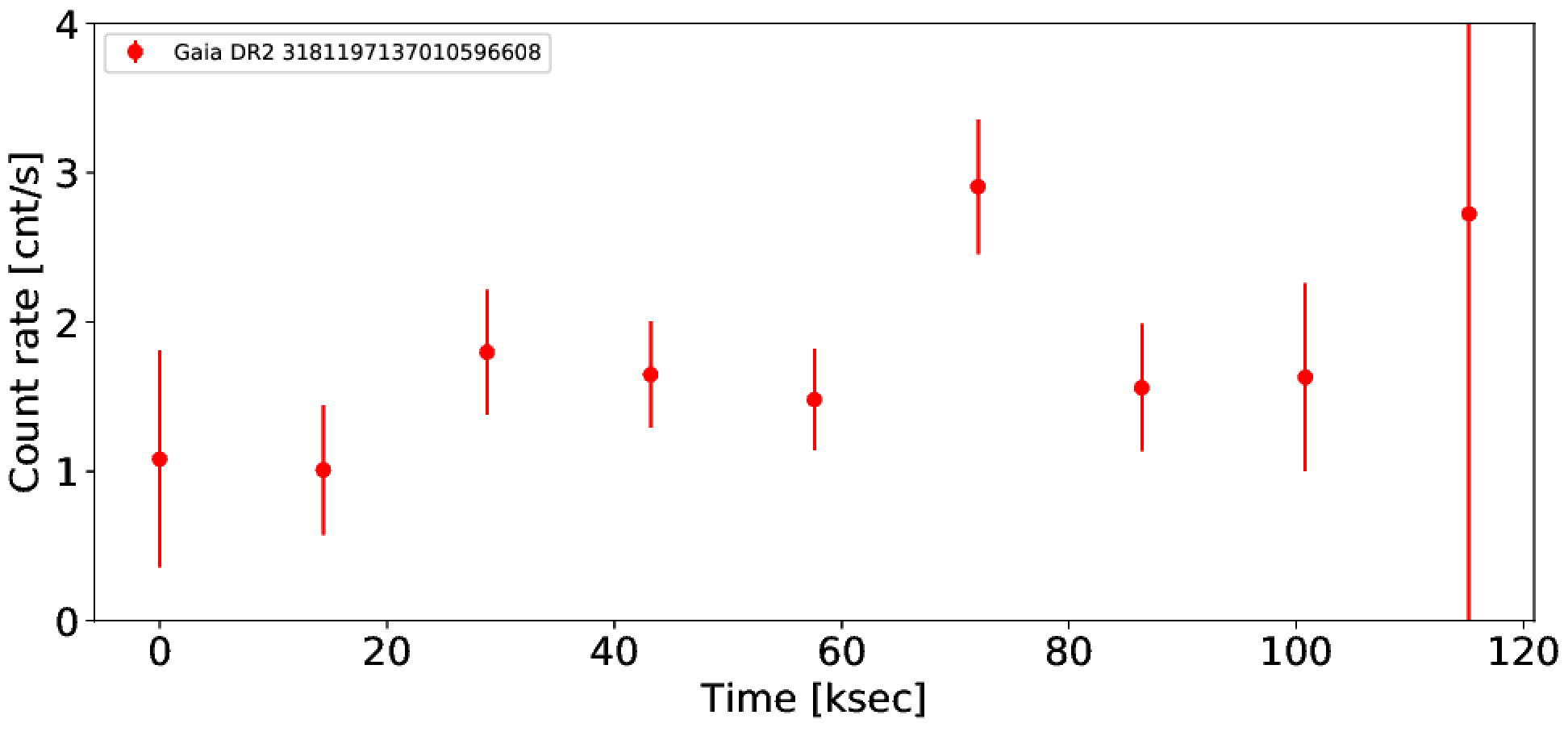}
\includegraphics[width=8.5cm]{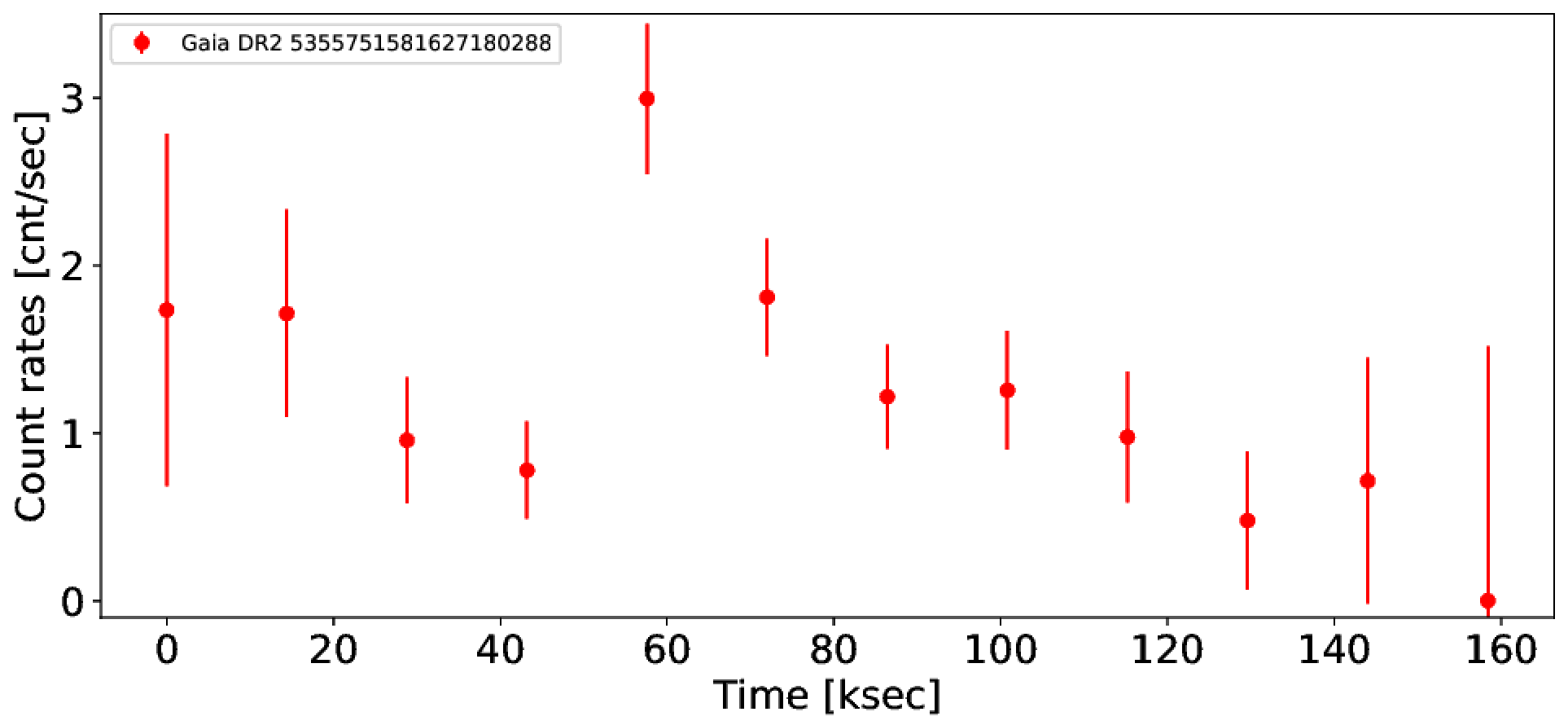}
\caption{eRASS1 light curves of the two UCDs with more than $100$ counts in the \ecat broad band: 
{\em top} - the UCD binary WDS\,J04469-1117AB (bin size is $80$\,s), 
{\em bottom} - TWA\,22 (bin size is $50$\,s).} 
\label{fig:eRASS1_lcs}
\end{center}
\end{figure}

As explained in Sect.~\ref{sect:lc_and_spec} we have extracted 
the eRASS1 light curves and spectra of the two UCDs with the largest number of counts.
The eRASS1 light curves of these two objects, WDS\,J04469-1117AB and TWA\,22, are displayed
in Fig.~\ref{fig:eRASS1_lcs}. Each bin seen in the figure represents one visit of 
the target by {\em eROSITA}, an eRODay. 
%
%
%

Both objects
show clear evidence of flaring. The time lapse between two consecutive eRODays is 
$14.4$\,ksec, corresponding to the $4$\,h interval required for one satellite rotation. 
Flares with a duration on the order of $1$\,h -- as seen before
on UCDs \cite[e.g.,][]{Stelzer06.0} -- thus show up as a single upwards outlying bin. 
This is the case for WDS\,J04469-1117AB. For TWA\,22AB an event with a decay
of at least three and maybe five to six eRODays is seen, that is this flare lasted up to half
a day. As mentioned above a 
more detailed analysis of UCD X-ray variability is deferred to a later work.

The X-ray spectra are fitted with a 
thermal {\sc apec} model. 
The two targets are very close (distance $\lesssim 20$\,pc 
each), and for such small distances the galactic absorption does not play a 
relevant role.
With a one-temperature model these spectra are poorly fitted (high $\chi^2$ values 
and significant residuals). We, therefore, adopted two {\sc apec} components. 
We performed fits with 
different fixed values for the global abundance ranging between $Z = 0.3\,{\rm Z_\odot}$
and $Z = 1.0\,{\rm Z_\odot}$, and we noticed the expected degeneracy between abundance
and emission measure, that is the fits with higher abundance yielded a lower emission
measure. Within the uncertainties, however, all emission measures found for the
abovementioned range of $Z$ are compatible with each other. We, therefore, adopted the
fit with $Z = 0.3\,{\rm Z_\odot}$, a typical value used for modelling stellar corona
observed with poor photon statistics, and we provide the results from the spectral
analysis in Table~\ref{tab:xspec_results}. The EM-weighted mean temperatures are
$\langle kT \rangle = 0.73 \pm 0.11$\,keV for WDS\,J04469-1117AB and 
$\langle kT \rangle = 0.76 \pm 0.10$ for TWA\,22AB.
These mean temperatures are higher than the values derived for 
early-M dwarfs from {\em eROSITA} spectra \citep{Magaudda21.0}. While this might
be surprising at first sight, this is likely to be attributed to a combination of the 
young age, which is $\sim 10$\,Myr for TWA\,22AB \citep{Teixeira09.0}
and $60-300$\,Myr for WDS\,J04469-1117AB \citep{Shkolnik09.0}, 
combined with their evident flaring activity. 
%

\begin{table*}
\centering
\caption{Results from spectral fitting of the two brightest eRASS1-detected UCDs. The abundances were frozen to a global value of $0.3\,{\rm Z_\odot}$.}
\label{tab:xspec_results}
\resizebox{\textwidth}{!}{\begin{tabular}{lcccccc}
\hline
{\em Gaia}\,DR2 & $kT_1$ & $kT_2$ & $\log{EM_1}$ & $\log{EM_2}$ & $f_{\rm x,0.2-2.3\,keV}$ & $\chi^2_{\rm red}$ (dof) \\
designation & [keV]   & [keV]    & [${\rm cm^{-3}}$]  &  [${\rm cm^{-3}}$] &  [$10^{-12}\,{\rm erg \ cm^{-2} s^{-1}}$] & \\ 
\hline
3181197137010596608 & $0.27 \pm 0.08$ & $1.18 \pm 0.20$ & $51.36 \pm 0.41$ & $51.57 \pm 0.30$ & $1.42 \pm 0.11$ & 0.79 (13) \\
5355751581627180288  & $0.28 \pm 0.08$ & $1.23 \pm 0.19$ & $51.20 \pm 0.47$ & $51.51 \pm 0.27$ & $1.05 \pm 0.08$ & 1.32 (14) \\
\hline
\end{tabular}}
\end{table*}


%
\begin{figure}
\begin{center}
\includegraphics[width=8.5cm]{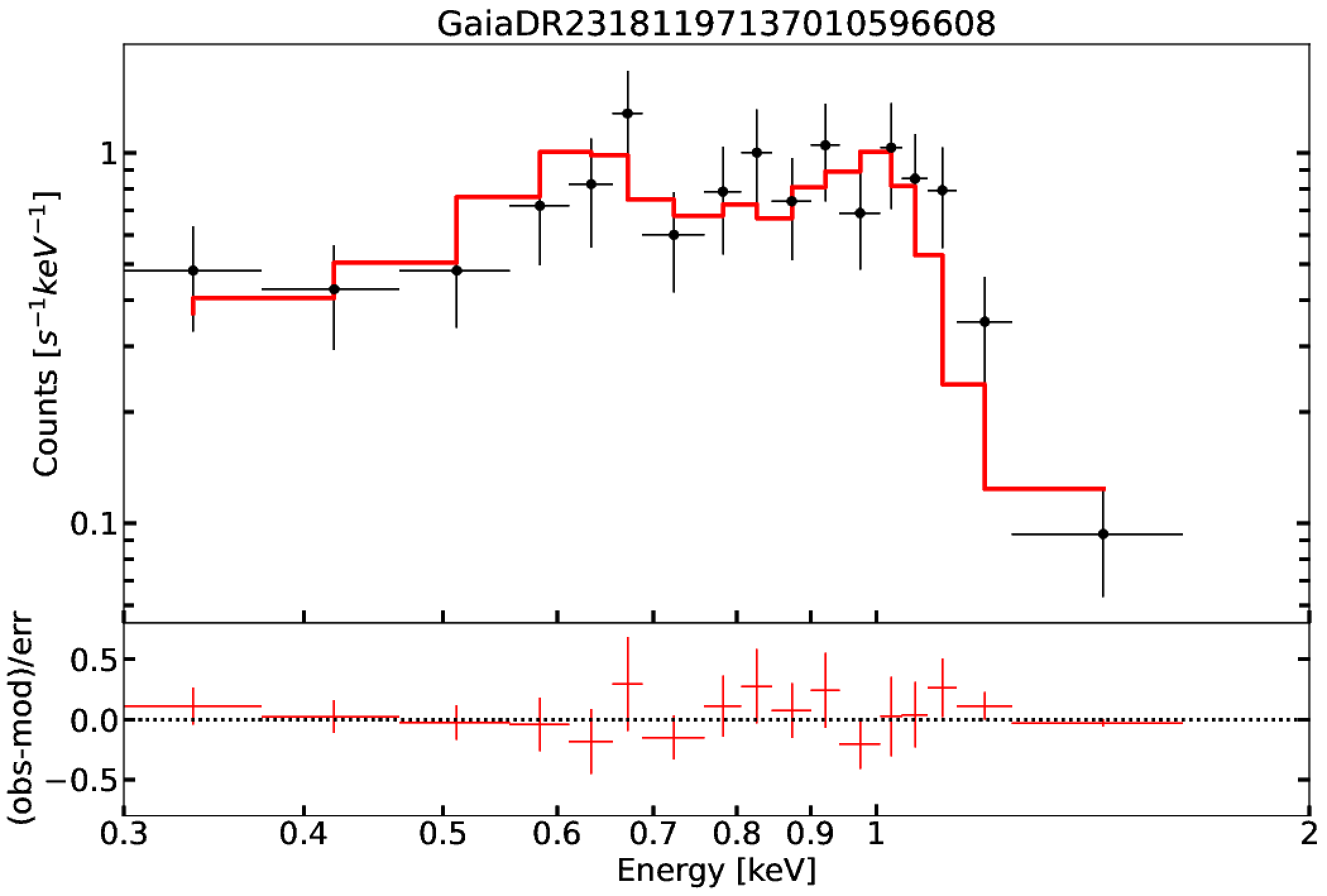}
\includegraphics[width=8.5cm]{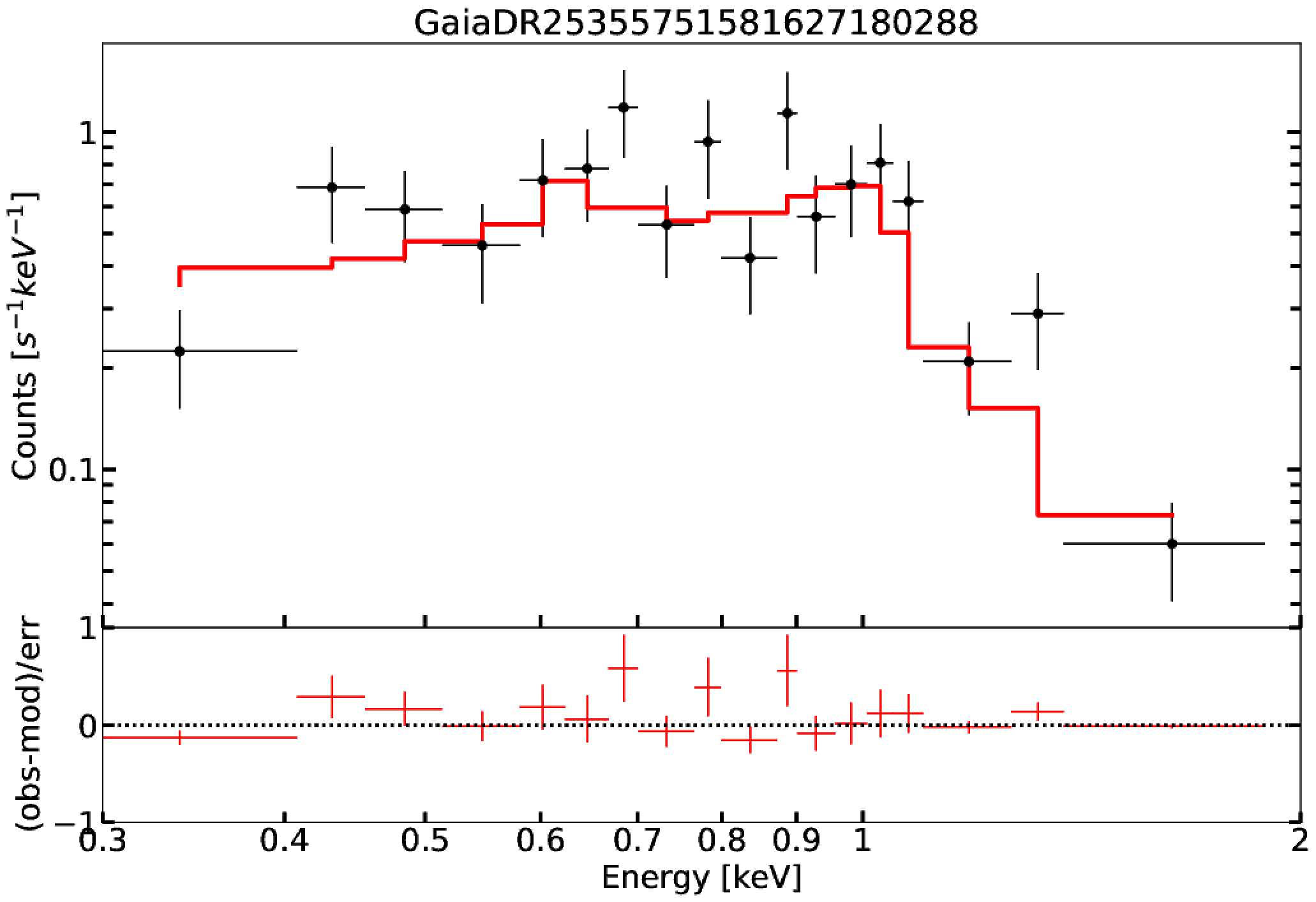}
\caption{eRASS1 spectra of the two UCDs with  more than $100$ counts in the \ecat broad band: 
{\em top} - the UCD binary WDS\,J04469-1117AB, 
{\em bottom} - TWA\,22\,AB. 
Both are shown together with their best-fitting
$2$-T {\sc apec} model and the residuals of the fit.}
\label{fig:eRASS1_spec}
\end{center}
\end{figure}

\begin{figure}
\begin{center}
\includegraphics[width=8.5cm]{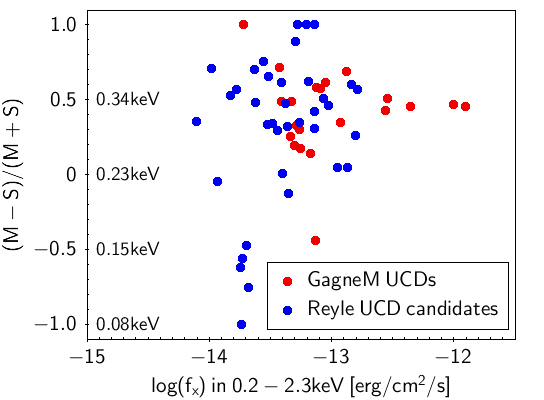}
\caption{eRASS hardness ratios versus flux for UCDs and UCD candidates. 
The flux refers to the $0.2-2.3$\,keV band (the sum of eRASS1 Band\,1 and Band\,2) 
and the hardness ratio
is based on the count rates in the $0.6-2.3$\,keV band (eRASS1 Band\,2, $M$)
and the $0.2-0.6$\,keV band (eRASS1 Band\,1, $S$); see text in 
Sect.~\ref{subsect:results_xproperties} for more details.} 
\label{fig:eRASS1_HR1_fx}
\end{center}
\end{figure}

For the faint sources the only way to study spectral characteristics
is the hardness ratio ($HR$), traditionally defined as 
$(X_{\rm j} - X_{\rm i}) / (X_{\rm j} + X_{\rm i})$, where $X_{\rm j}$ and $X_{\rm i}$
are the count rates in a high ($j$) and a low ($i$) energy band, respectively. 
For convenience we assign letters $S$ (soft), $M$ (medium) and $H$ (hard)
to the standard energy bands of \ecat, Band\,1 ($0.2-0.6$\,keV; $S$),
Band\,2 ($0.6-2.3$\,keV, $M$) and Band\,3 ($2.3-5.0$\,keV, $H$). 
We have examined
the $HR$ involving $H$ and $M$ 
but $50$\,\% of the eRASS1 detected UCDs and UCD candidates have no counts in the hard 
band making this $HR$ of limited use. Fig.~\ref{fig:eRASS1_HR1_fx}
displays the $HR$ defined with $M$ and $S$ 
versus the summed X-ray flux of Band\,1 and Band\,2 ($0.2-2.3$\,keV). 
Side-by-side with the
scale of the $HR$ we provide an estimate of the plasma temperature. 
The conversion between $kT$ and
$HR$ was calculated by \cite{Foster21.0} from simulated 1T-{\sc apec} 
models.
According to their analysis,  
the hardness ratio $(M-S)(M+S)$ 
saturates that is it loses its sensitivity to temperature,
at $HR \approx 0.75$ and $kT \approx 0.5$\,keV. 
This is due to the fact that the bulk of a typical coronal spectrum is comprised in 
the energy range defined by $M$ 
while 
the $S$-band holds only the softest emission. A boundary between different energy bands
at $\sim 1$\,keV would be more appropriate to characterize stellar X-ray sources, 
but this information is not present in the current version of \ecat. 
Figure~\ref{fig:eRASS1_HR1_fx}
shows, however, that the majority of UCDs are captured well by $(M-S)/(M+S)$, that is they
are not saturated.
%
%
%
%
%

A caveat to this $HR1 - kT$ conversion is that {\em eROSITA} spectra of UCDs can
probably not be well described by a $1$-T plasma. This is seen from the spectral analysis we
performed for  
the two UCDs with the largest number of counts, WDS\,J04469-1117\,AB and TWA22\,AB.
Both have
$HR \sim 0.46$, that is according to Fig.~\ref{fig:eRASS1_HR1_fx} we would infer
$kT < 0.35$\,keV. This is,  however, much lower than the mean temperature we have 
computed above from the spectral fits. 
We conclude that $HR1$ is a useful parameter to evaluate the relative
spectral hardness in a sample of {\em eROSITA} spectra but quantitative inferences on 
the plasma temperature must be taken with caution. 

\subsection{Comparison with previous X-ray detections}\label{subsect:results_xarchives}

As described in Sect.~\ref{subsect:identifications_eRASS1_bonafide} 
a total of $9$
UCDs and candidates from {\it fullmatch-eRASS1-GagneM} and {\it fullmatch-eRASS1-Reyle}
had previous {\em XMM-Newton} and/or {\em Chandra} detections. 
For one of them the bulk of the eRASS1 emission was attributed to its nearby CPM
companion which is a separate {\em Chandra} detection, and this object is not considered
in the comparison between eRASS1 and archival X-ray fluxes that we present in this
section. 

For the other $8$ objects we extracted the X-ray fluxes measured in the earlier 
observations from the 4XMM-DR9 \citep{Webb20.0} and CSC\,2.0 \citep{Evans10.0},  
catalogs and list them in Table~\ref{tab:xarchives} together with the eRASS1 fluxes
obtained as explained below. 
The majority of these archival X-ray detections have not been discussed
in previous publications dedicated to UCDs, and a detailed investigation of eventual
literature results would not provide much additional information in our context. 

To enable a comparison with the eRASS1 detections of the same objects we selected the
energy band `9' ($0.5-4.5$\,keV) from 4XMM-DR9. The CSC\,2.0 catalog presents only the 
broad band flux in the $0.5-7.0$\,keV band. However, since UCDs are soft X-ray emitters,
the extension to higher energies should not provide much additional flux. This choice of 
{\em XMM-Newton} and {\em Chandra} energy bands allow for the closest match 
to the energy bands present in \ecat, for which we can provide the flux in the 
$0.6-5.0$\,keV band summing Band\,2 and Band\,3. 
As a side remark we recall from Sect.~\ref{subsect:results_xproperties} 
that for many of the eRASS1 sources associated with UCDs 
there are no counts in Band\,3. 
While there is some contribution from Band\,1 ($0.2-0.6$\,keV)
the hardness ratio discussed in Sect.~\ref{subsect:results_xproperties} 
shows that for most of them the largest number
of counts are collected in Band \,2. This is a combination of the energy dependence of the
effective area of {\em eROSITA} and the peak of the X-ray spectrum of UCDs.

It is important to note that the fluxes in 4XMM-DR9 and CSC\,2.0
refer to a power-law model. In Sect.~\ref{sect:eRASS1_multilambda} we have shown for 
a different energy band that there is a $30$\,\% difference between the fluxes obtained
from a power-law vs a thermal model. Therefore, to examine the variability of the sources
we refer to the power-law fluxes provided in \ecat, even if this is not the appropriate
model for our targets.  
%

\begin{figure}[t]
\begin{center} 
\includegraphics[width=8.5cm]{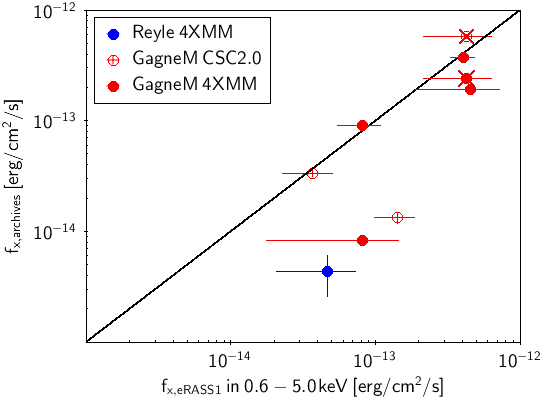}
\caption{X-ray flux from the 4XMM-DR9 and CSC\,2.0 catalogs compared to the \ecat flux
for the $7$ eRASS1 detected UCDs and the one UCD candidate 
with an archival {\em XMM-Newton} and/or {\em Chandra}
detection (one object has both an {\em XMM-Newton} and a {\em Chandra} detection
and is marked with a cross-symbol); 
the values for the archival X-ray fluxes 
are listed in Table~\ref{tab:xarchives}. The black line corresponds to equal
flux with the two compared instruments. To enable the comparison with the archival
fluxes, the {\em eROSITA} fluxes were calculated with a power-law model.}
\label{fig:flux_archives_eRASS1}
\end{center}
\end{figure}

In Fig.~\ref{fig:flux_archives_eRASS1} we compare the archival X-ray fluxes 
to the new measurements obtained during eRASS1, 
for which we adopt the summed 
fluxes {\sc ml\_flux\_2 + ml\_flux\_3} listed in \ecat and their uncertainties. 
For $5$ out of $8$ UCDs the X-ray fluxes
measured with eRASS1 are consistent with the earlier data. For the remaining three, 
two from {\it fullmatch-eRASS1-GagneM} and one from {\it fullmatch-eRASS1-Reyle}, the eRASS1
flux is significantly higher than the {\em XMM-Newton} and {\em Chandra} flux. 
We caution that one of them  
has a low eRASS1 broad band detection likelihood of {\sc det\_like\_0}$ \lesssim 10$. 
In Sect.~\ref{subsect:identifications_eRASS1_spurious}
we have, however, explained that likely none of the UCDs and UCD candidates that we have
identified as eRASS1 X-ray emitters are spurious detections. 
Therefore, the most probable explanation for the enhanced eRASS1 flux is a flare. 
In any case, if these objects had been as faint as
during the archival X-ray observation they would not have been detected during eRASS1.

\begin{table*}\small
\begin{center}
\caption{Archival X-ray fluxes of UCDs and UCD candidates extracted from the 4XMM-DR9 and 
CSC\,2.0 catalogs. These fluxes are based on a power-law model (see text in 
Sect.~\ref{subsect:results_xarchives})}
\label{tab:xarchives}
\begin{tabular}{lrlrlr}
\hline
  \multicolumn{1}{l}{{\em Gaia}\,DR2} &
  \multicolumn{1}{r}{eRASS1 flux} &
  \multicolumn{1}{l}{4XMM} &
  \multicolumn{1}{c}{${\rm flux_{Band\,9}}$} & 
  \multicolumn{1}{l}{CSC\,2.0} &
  \multicolumn{1}{c}{${\rm flux_{aper,b}}$} \\ 
  \multicolumn{1}{l}{designation} &
  \multicolumn{1}{r}{${\rm [10^{-14}\,erg/cm^2/s]}$} &
  \multicolumn{1}{l}{name} &
  \multicolumn{1}{c}{${\rm [10^{-14}\,erg/cm^2/s]}$} &
  \multicolumn{1}{l}{name} &
  \multicolumn{1}{c}{${\rm [10^{-14}\,erg/cm^2/s]}$} \\ 
\hline
  3200303384927512960 & $ 8.10 \pm  2.62$ & J044023.6-053005 & $ 9.10 \pm 0.48$ &  		            &  		    	    \\ 
  666988221840703232  & $45.25 \pm 26.27$ & J075224.1+161210 & $19.15 \pm 1.18$ &      	            &  		   	     \\ 
  703790044252850688  & $40.88 \pm  7.66$ & J082948.1+264624 & $37.42 \pm 0.55$ & 		     	    &  		    	    \\ 
  5761985432616501376 & $14.24 \pm  4.41$ &  		         &  	            & 2CXO J085335.8-032934 & $ 1.35 \pm 0.15$	\\ 
  3875561544817858816 & $42.95 \pm 21.29$ & J102044.0+081422 & $24.10 \pm 1.35$ & 2CXO J102043.8+081421 & $57.94 \pm 2.29$  \\ 
  5201352936175160448 & $ 8.20 \pm  6.43$ & J111022.1-762513 & $ 0.83 \pm 0.07$ &  		            &  	     	    	    \\
  6224387727748521344 & $ 3.71 \pm  1.40$ &  		         &  	            & 2CXO J145637.8-280957 & $ 3.34 \pm 0.21$  \\ 
 3754497583659096320  & $ 4.71 \pm  2.61$ & J103958.3-120357 & $ 0.43 \pm 0.18$ &  		            & 	  	    	   \\ 
\hline\end{tabular}
\end{center}
\end{table*}

%
%
%
%

\section{Summary and conclusions}\label{sect:conclusions}

We have presented the first X-ray observations for ultracool dwarfs from the {\em eROSITA}
all-sky survey. This sample comprises $21$ spectroscopically confirmed UCDs from 
the online catalog of M dwarfs maintained by J.Gagn\'e and
$37$ UCD candidates identified with {\em Gaia} data by \cite{Reyle18.0}. 
Through a careful examination of
alternative optical counterparts to the eRASS1 X-ray sources and an estimate of the
expected number of spurious
detections for weak X-ray emission we have selected this sample from an initial input 
catalog of $96$ UCDs and UCD candidates that are located in the vicinity of an eRASS1
X-ray source. 

In the course of this selection procedure we found that $14$ of these 
objects are in a comoving binary or triple system. This binary fraction of 
$\sim 15$\,\% is similar
to the multiplicity rate of $10.4$\,\% established by \cite{Reyle21.0} 
irrespective of X-ray emission  for the full sample of 
{\em Gaia}-selected UCD candidates, which is 
one of the two parent samples of our study. 

Eight of the $58$ eRASS1-detected UCDs and UCD candidates have a previous X-ray
detection with {\em XMM-Newton} and/or {\em Chandra}. For one of them, the candidate UCD was 
identified as an X-ray emitter
in our recent study for serendipitous {\em XMM-Newton} detections of 
{\em Gaia}-selected UCD candidates \cite{Stelzer20.0}. 
 
With the scope to understand the nature of the eRASS1 X-ray emitters 
for which the optical counterpart is not obviously assigned to the UCD or a companion star  
we have studied three multiwavelength diagrams. To support our source classification 
we have compared the location of the UCDs and
possible alternative {\em Gaia} counterparts in these diagrams
with the source population from the 
large data base provided through the eFEDS fields 
using the systematic classification scheme from \cite{Salvato21.0}.
As a side remark we note here that 
our investigation of this multiwavelength space largely confirms 
the distinction between galactic and extragalactic sources 
provided for the eFEDS X-ray detections 
by \cite{Salvato21.0}, but a small fraction of contaminants might 
be present in each of the two groups. 

Our new eRASS1-detections more than triple 
the number of known X-ray detections in the UCD regime. 
The ratio between the X-ray and bolometric luminosities of most of them are much
higher than the value of $\log{(L_{\rm x}/L_{\rm bol})} \sim -3$ which is the empirical 
upper envelope observed for higher-mass coronally active stars (see e.g., 
\cite{Magaudda21.0}. 
We have found strong evidence that this is the result of flaring activity. In fact, the
faint quiescent X-ray emission levels of UCDs are -- with the exception of the most
nearby ones -- not accessible to eRASS1 because of
its shallow flux limit\footnote{The lowest flux in our sample of eRASS1 detected UCDs
and UCD candidates is 
$\sim 10^{-14}\,{\rm erg/cm^2/s}$ in the $0.2-5.0$\,keV band.}, 
and they are at the limit even for dedicated {\em XMM-Newton} and {\em Chandra} 
observations as previous studies of this subject have shown; see e.g., \cite{Stelzer06.1}. 
Impressive though the increase in X-ray detections we have achieved with eRASS1 is, 
one should be aware that 
we have revealed less than $0.5$\,\% of the targets in our input samples. 

We note that a significant fraction of our new UCD X-ray detections 
do not belong, strictly speaking, 
to this object class as they have SpTs of M5/M6 and/or evidence of young age.  
In fact, $10$ of the $21$ spectroscopically confirmed UCDs with an eRASS1-detection
have been characterized as `young' in the literature (see Table~\ref{tab:multilam_UCDs}).
Our two brightest detections, TWA\,22\,AB and
WDS\,J04469-1117\,AB, belong to this group.
These two objects show both quiescent and flaring X-ray emission in the eRASS1 light curve,
and very similar coronal temperatures of $kT \sim 0.74$\,keV. This is significantly
higher than the temperatures derived from 
analogous observations with {\em eROSITA} for earlier-type M dwarfs \citep{Magaudda21.0}, 
a result that we ascribe to the youth ($\sim 10$\,Myr and $< 300$\,Myr,
respectively) and the variability of these two objects.  
Our detailed analysis of the eRASS1 data for TWA\,22\,AB and
WDS\,J04469-1117\,AB gives a preview on the
potential of {\em eROSITA} for enhancing our knowledge of the spectral and timing properties
of coronal emission at the bottom of the stellar main-sequence.

If we speculate that each of the $58$ UCDs we detected has shown one X-ray flare
during eRASS1 we can estimate the typical flare frequency for UCDs from our overall detection rate. 
Since the detection sensitivity is a strong function of distance we limit this calculation 
to a distance of $50$\,pc. About $10$\,\% of our full input samples, $1858$ objects,
are located in this sky volume, and $33$ of the $58$ X-ray detections are within the
same distance limit. Therefore, under the above hypothesis $\sim 1.7$\,\% of all UCDs
are flaring during the typical timespan between the first and last eRODay of a given
sky position. This coarse estimate does not take into account the strong
dependence of eRASS exposure time on the sky position nor the long data gaps 
inbetween individual eRODays. This value for the flare duty cycle should be verified by a 
systematic analysis of the {\em eROSITA} light curves of all detected UCDs as well as on higher
statistics that will be obtained by combining our results with similar studies
for the other seven all-sky surveys. 

It is finally worth pointing out that while significant efforts are being dedicated to
measuring the X-ray activity levels of UCDs, and the X-ray emission is being studied for 
large samples of early-M dwarfs \cite[e.g.,][]{Stelzer13.0, Magaudda20.0, Magaudda21.0}, 
the regime of mid-M dwarfs (SpT M5 and M6) seems now
the most poorly examined domain. In this vein, the inclusion of such spectral types in
the \gagneM sample is a useful addition. 
The biases of most studies toward stars earlier than M5 is mainly 
because 
optical surveys that are used for target selection 
usually suffer from incompleteness at the cool end;
e.g., the M star studies cited above are based on a  proper-motion survey
\citep{Lepine11.0} and this sample 
was shown by \cite{Stelzer13.0} to be incomplete for SpT later than M4.
Similarly, the earlier {\em ROSAT} study of \cite{Schmitt04.0} included only about a dozen stars with
M5/M6 spectral types. 
Given the uncertainties in the spectral types determined with 
different methods (see e.g., Fig.~\ref{fig:sptphot_sptopt}) a more detailed assessment of 
this observational gap requires a homogeneous treatment of the input samples used 
for the search of X-ray detections. The photometric calibration based on 
{\em Gaia} colors (see Sect.~\ref{subsect:identifications_eRASS1_bonafide}) 
and the increasing number of spectroscopic
studies that have become available for M stars, e.g., through LAMOST \citep{Guo15.0}, 
enable such studies. 
In the near future it should, therefore, be possible to reduce the observational biases and
to obtain a picture of M dwarf activity that is continuous through the whole spectral class.


\appendix

\section{X-ray emitting common proper motion pairs involving UCDs}\label{app:cpmpairs}

\begin{figure*}
\begin{center}
\parbox{17cm}{
\parbox{4.0cm}{\includegraphics[width=4.0cm]{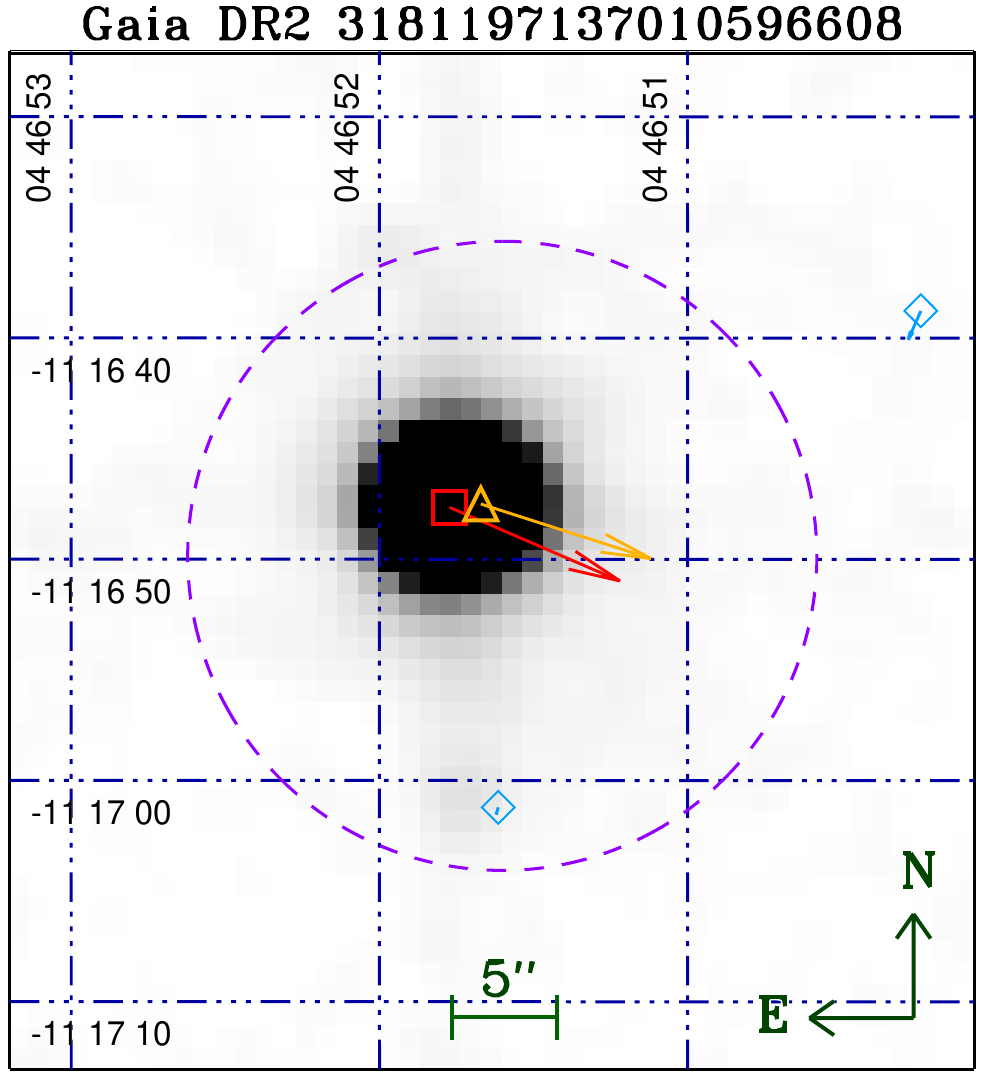}}
\parbox{4.0cm}{\includegraphics[width=4.0cm]{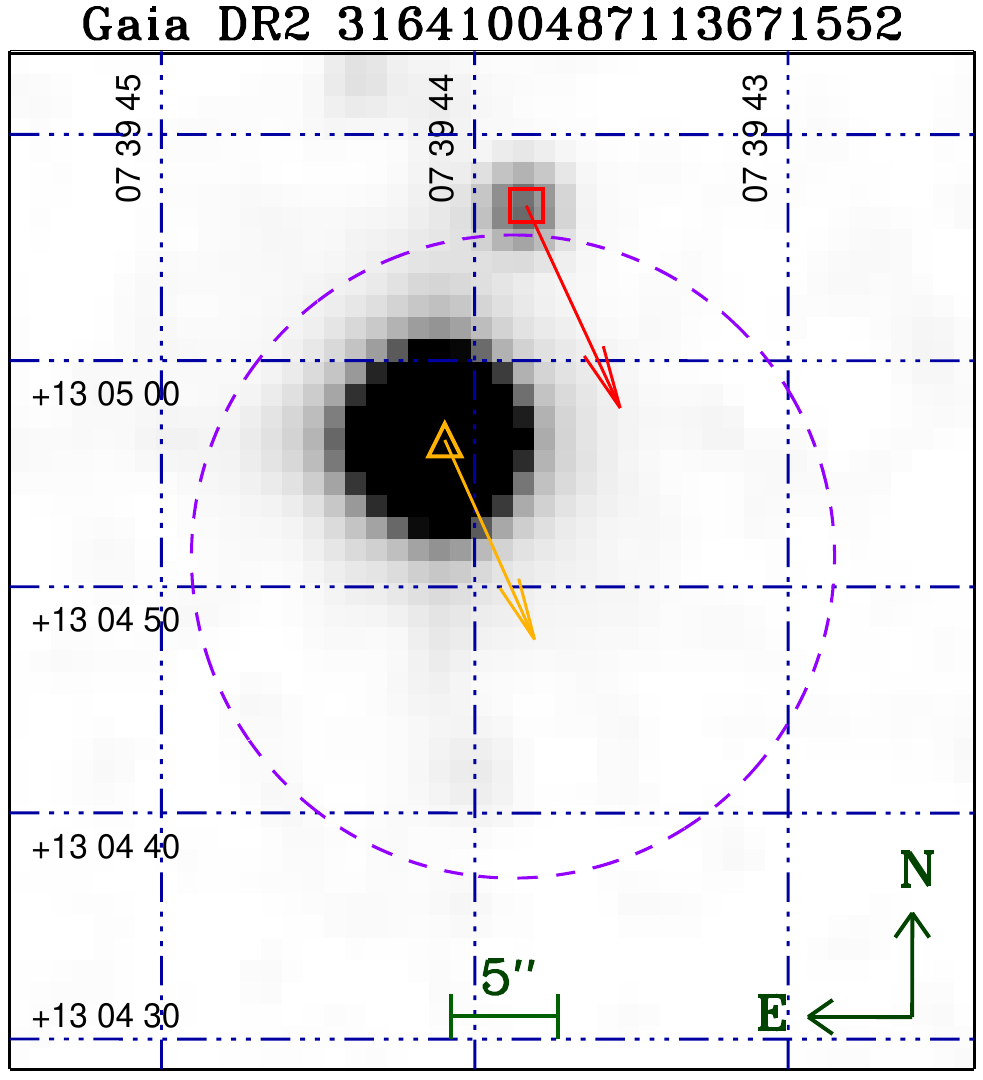}}
\parbox{4.0cm}{\includegraphics[width=4.0cm]{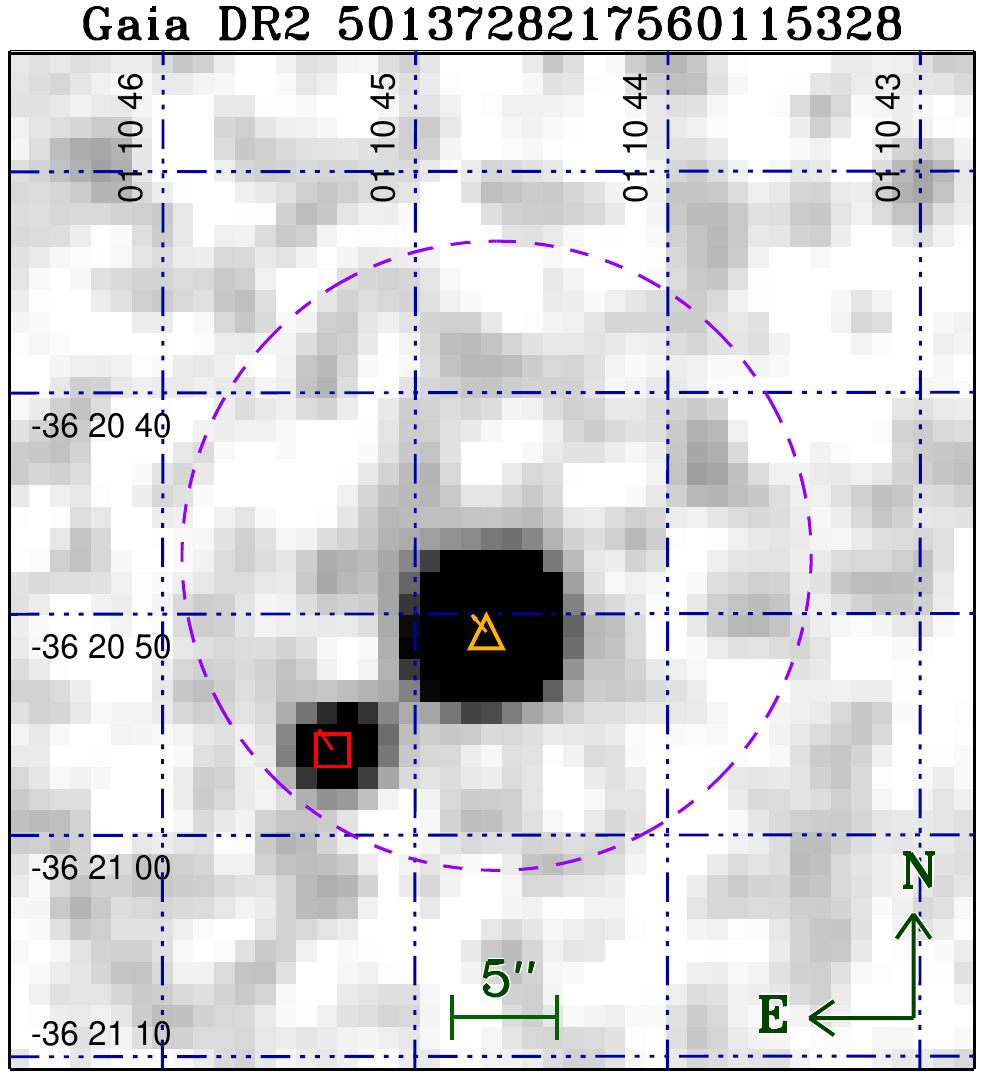}}
\parbox{4.0cm}{\includegraphics[width=4.0cm]{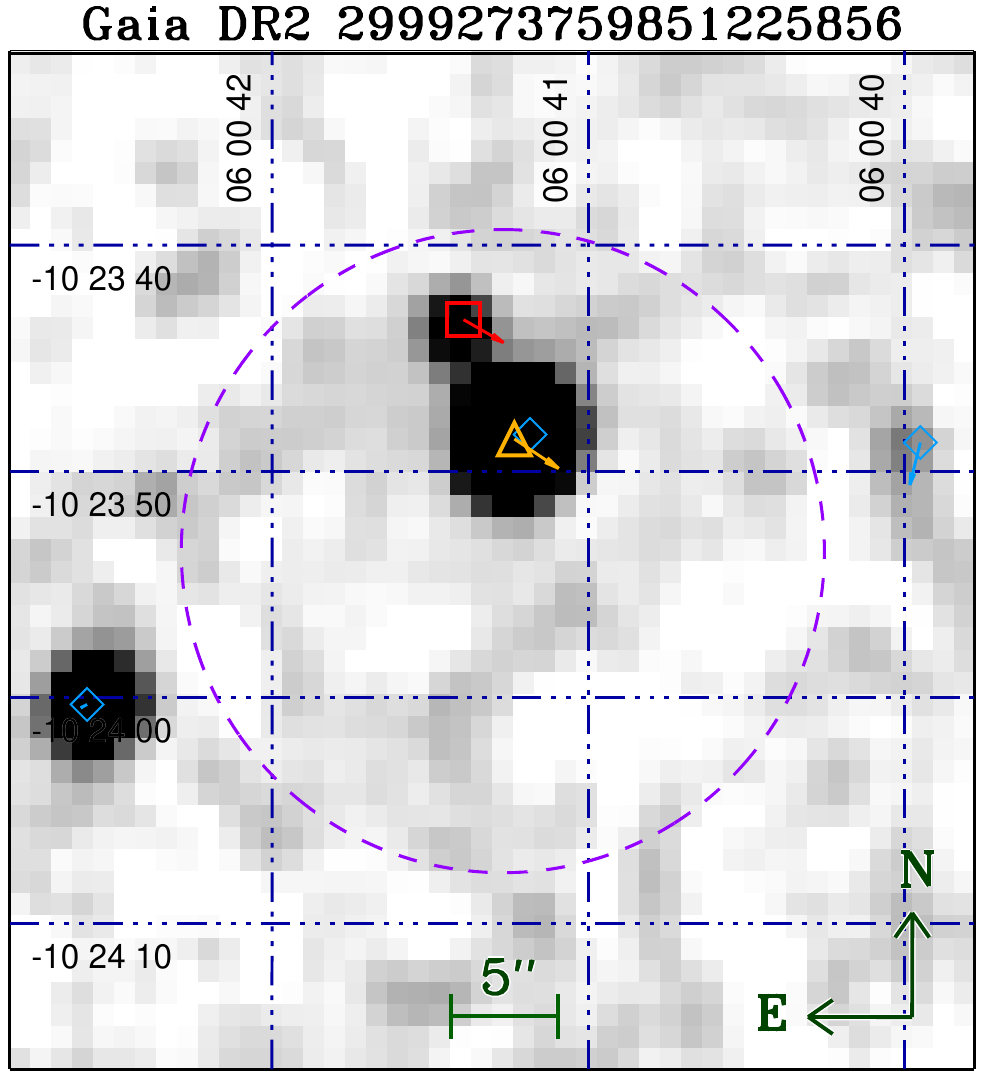}}
}
\parbox{17cm}{
\parbox{4.0cm}{\includegraphics[width=4.0cm]{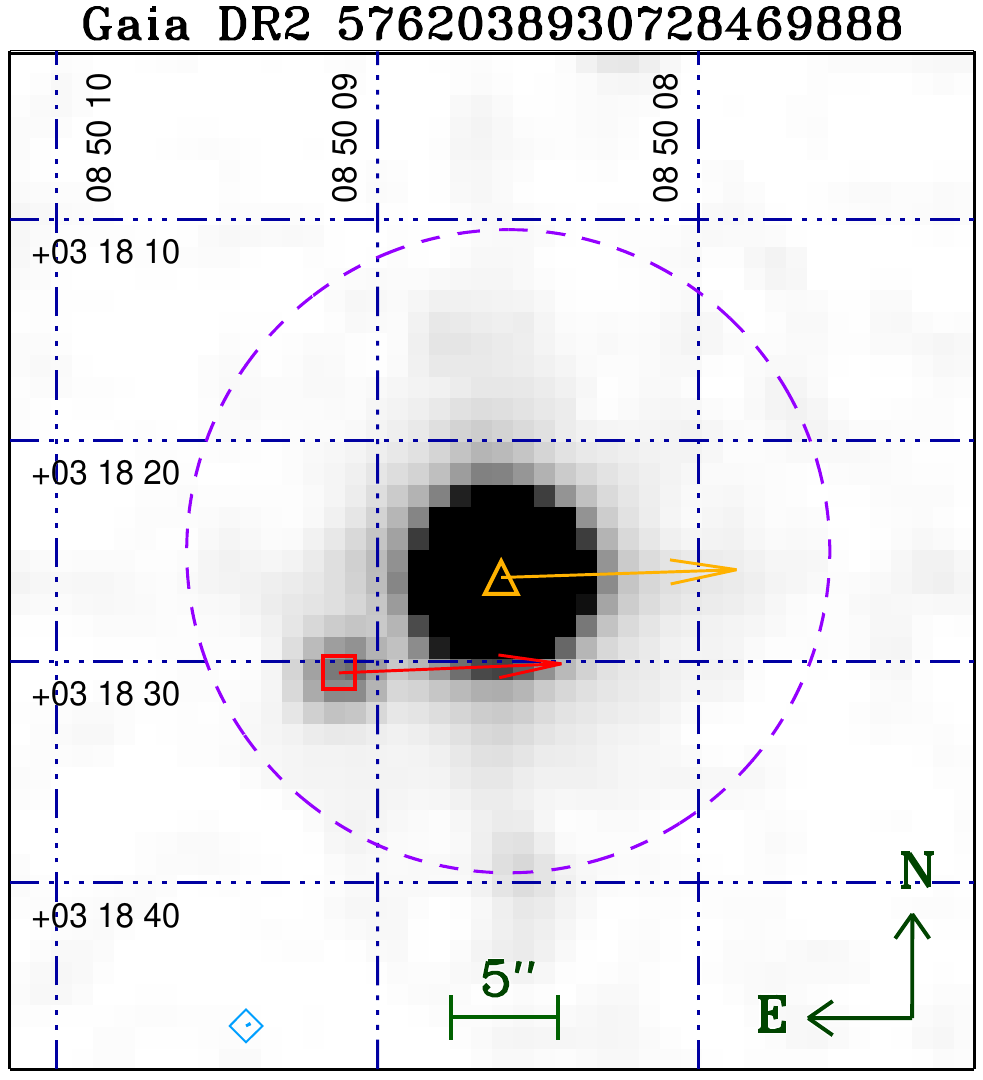}}
\parbox{4.0cm}{\includegraphics[width=4.0cm]{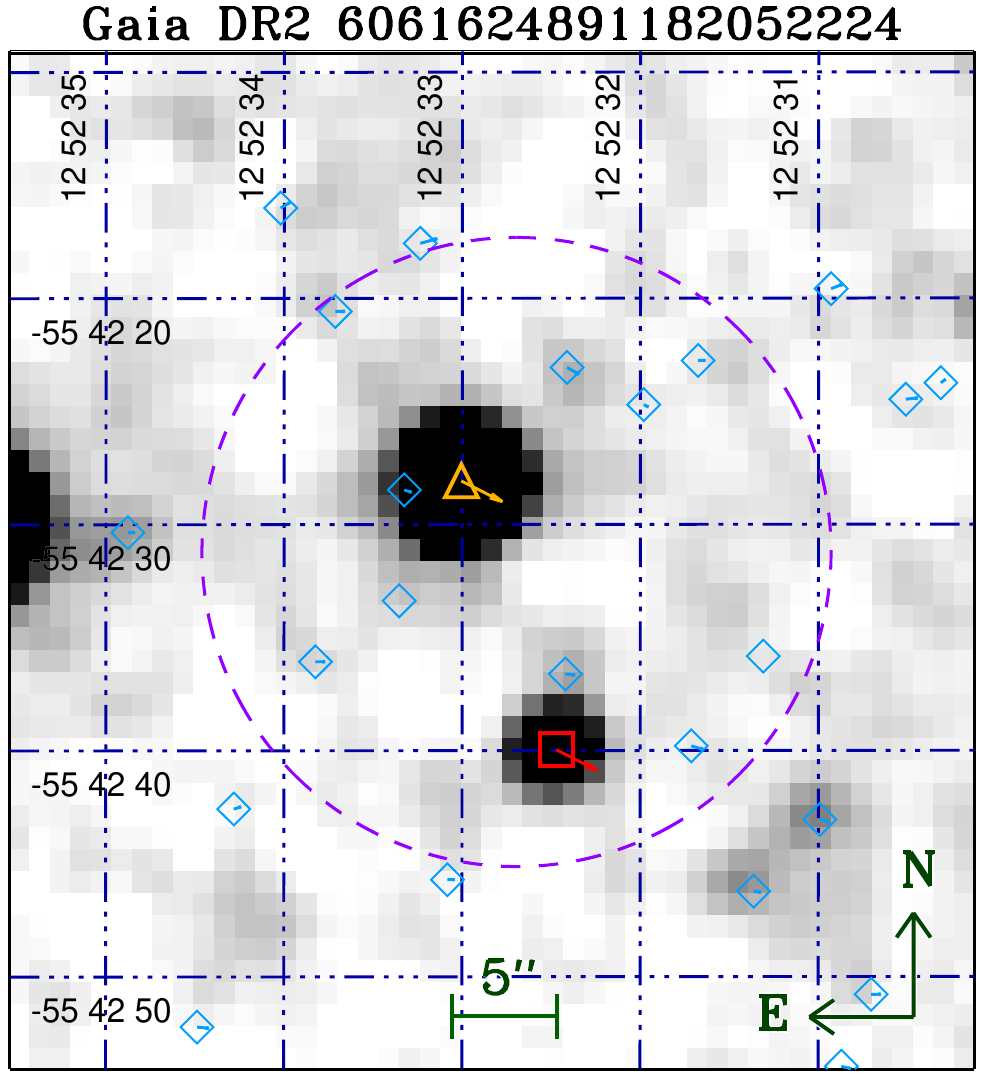}}
\parbox{4.0cm}{\includegraphics[width=4.0cm]{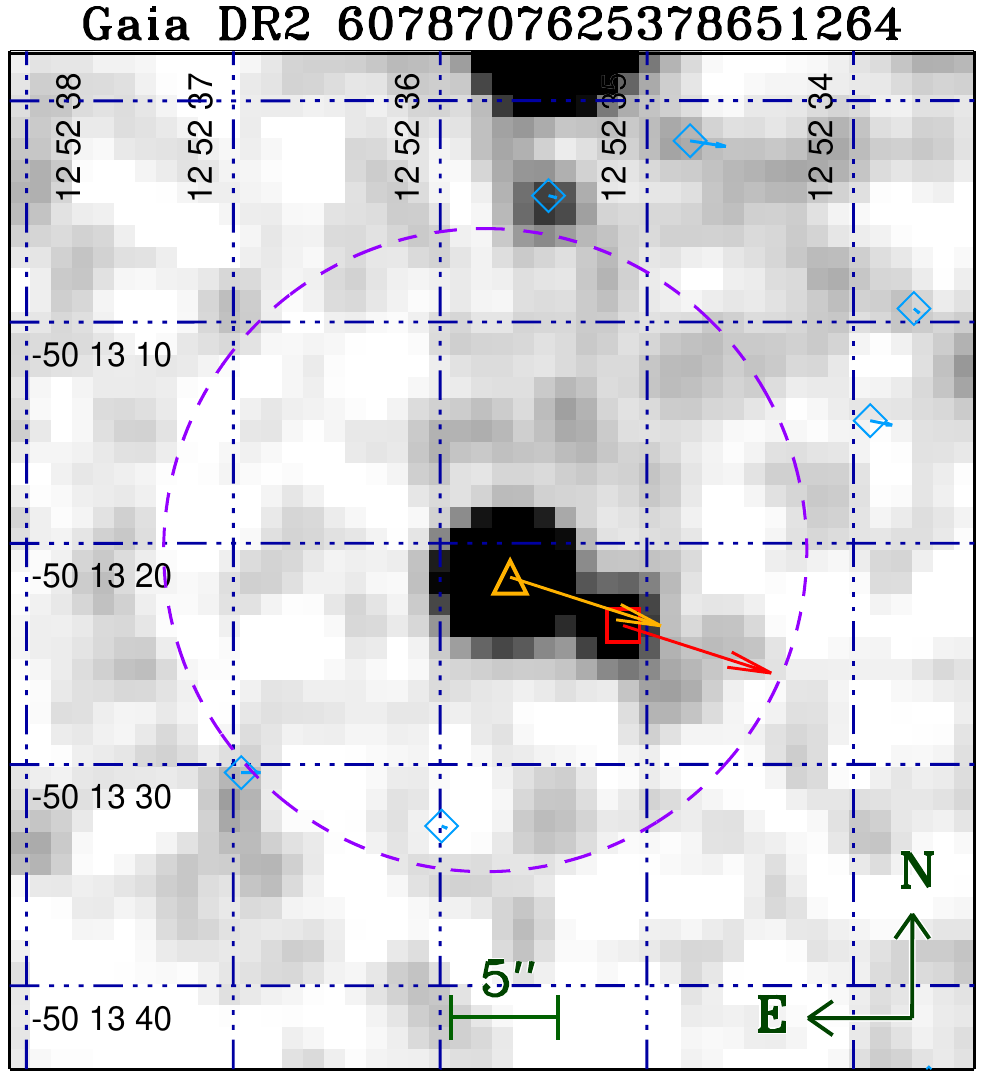}}
\parbox{4.0cm}{\includegraphics[width=4.0cm]{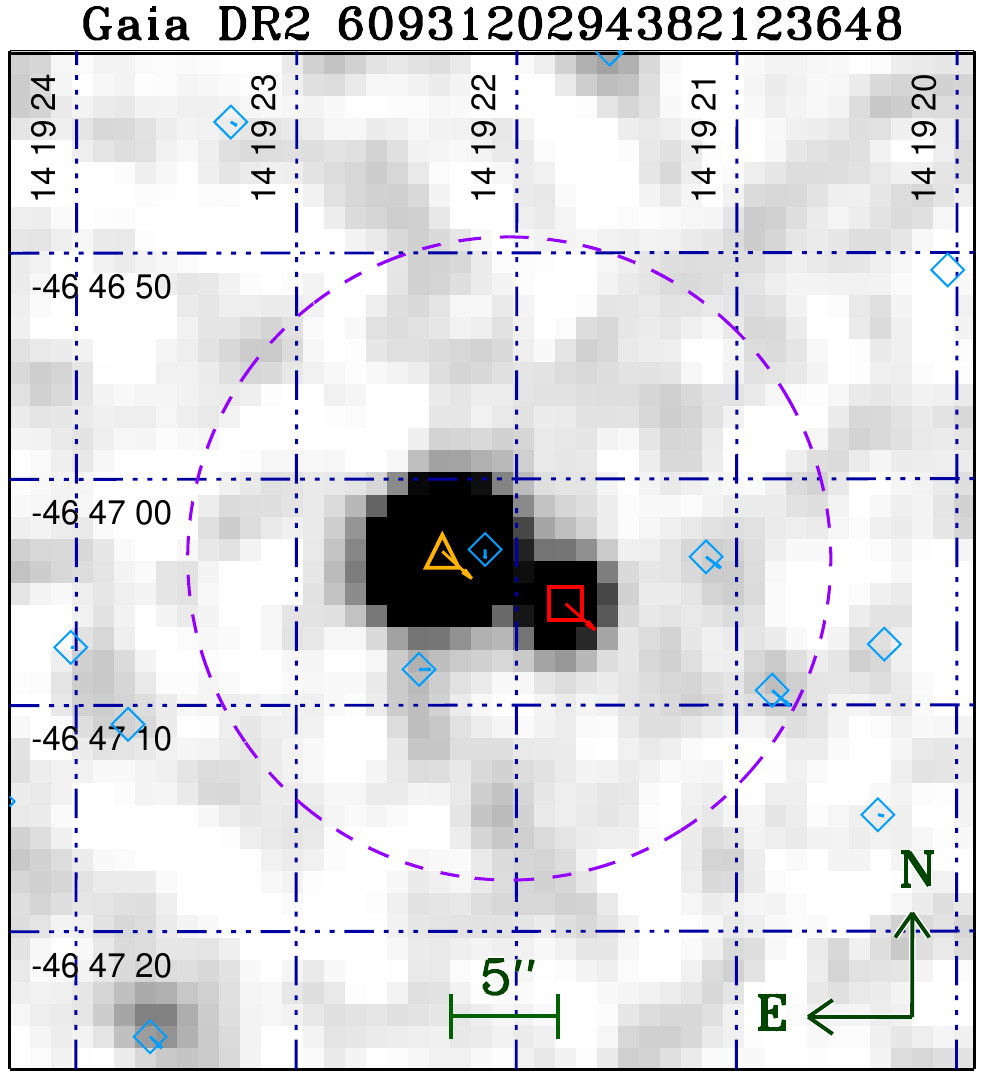}}
}
\parbox{17cm}{
\parbox{4.0cm}{\includegraphics[width=4.0cm]{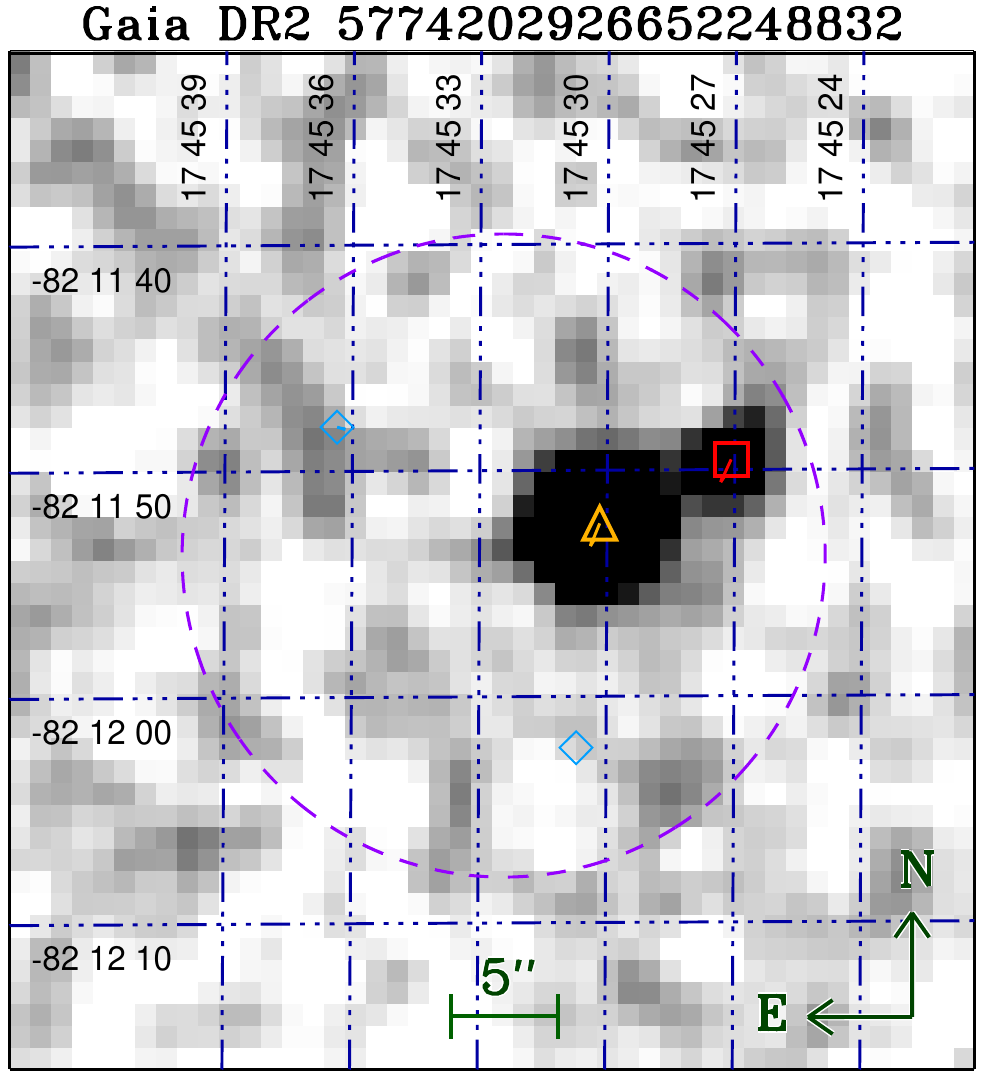}}
\parbox{4.0cm}{\includegraphics[width=4.0cm]{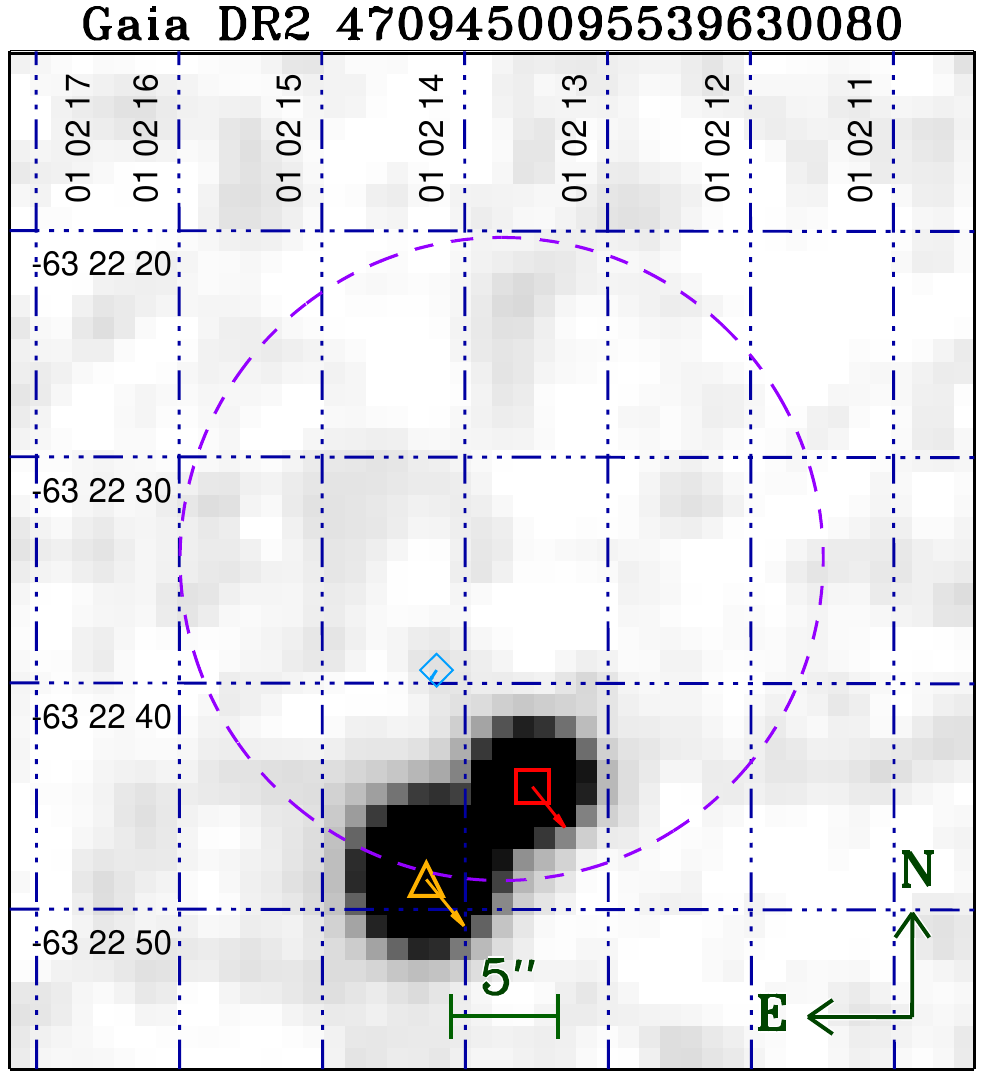}}
\parbox{4.0cm}{\includegraphics[width=4.0cm]{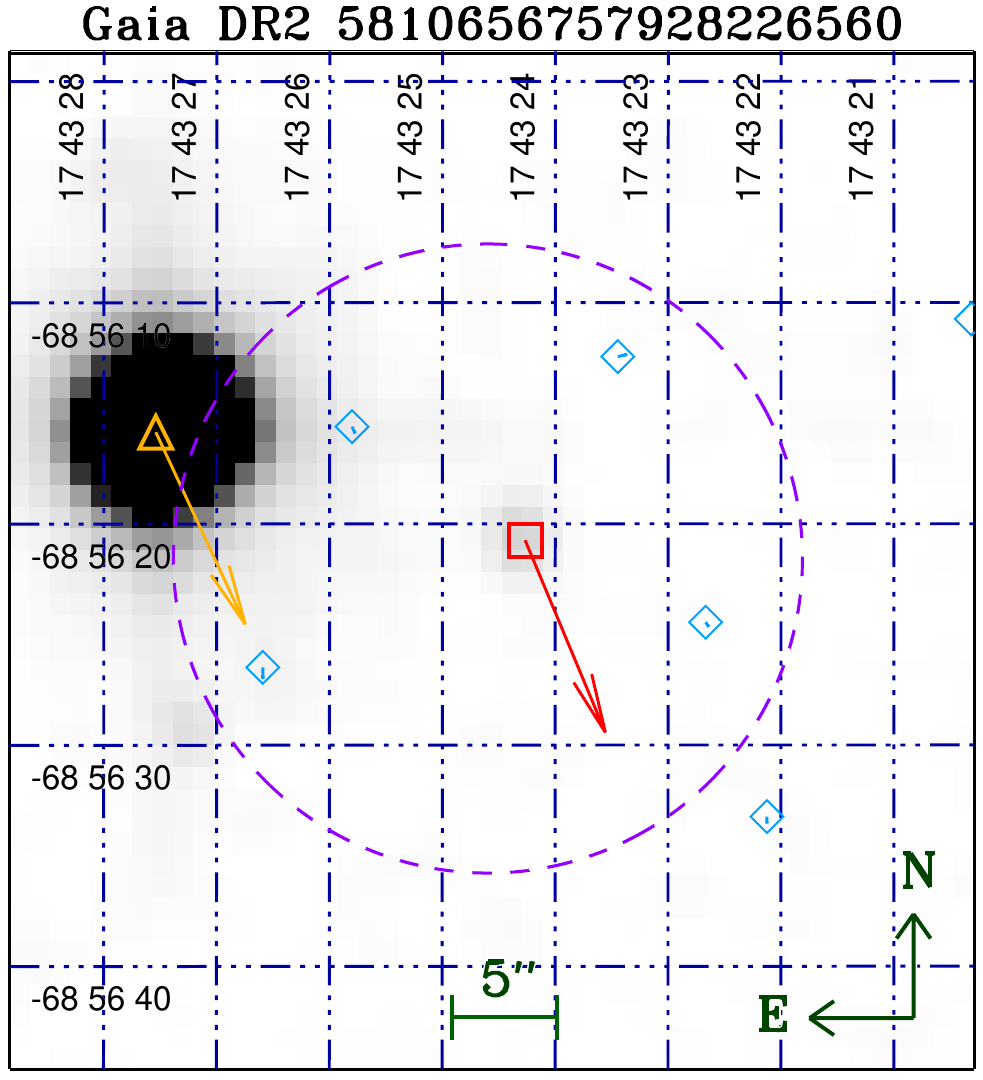}}
}
\parbox{17cm}{
\parbox{4.0cm}{\includegraphics[width=4.0cm]{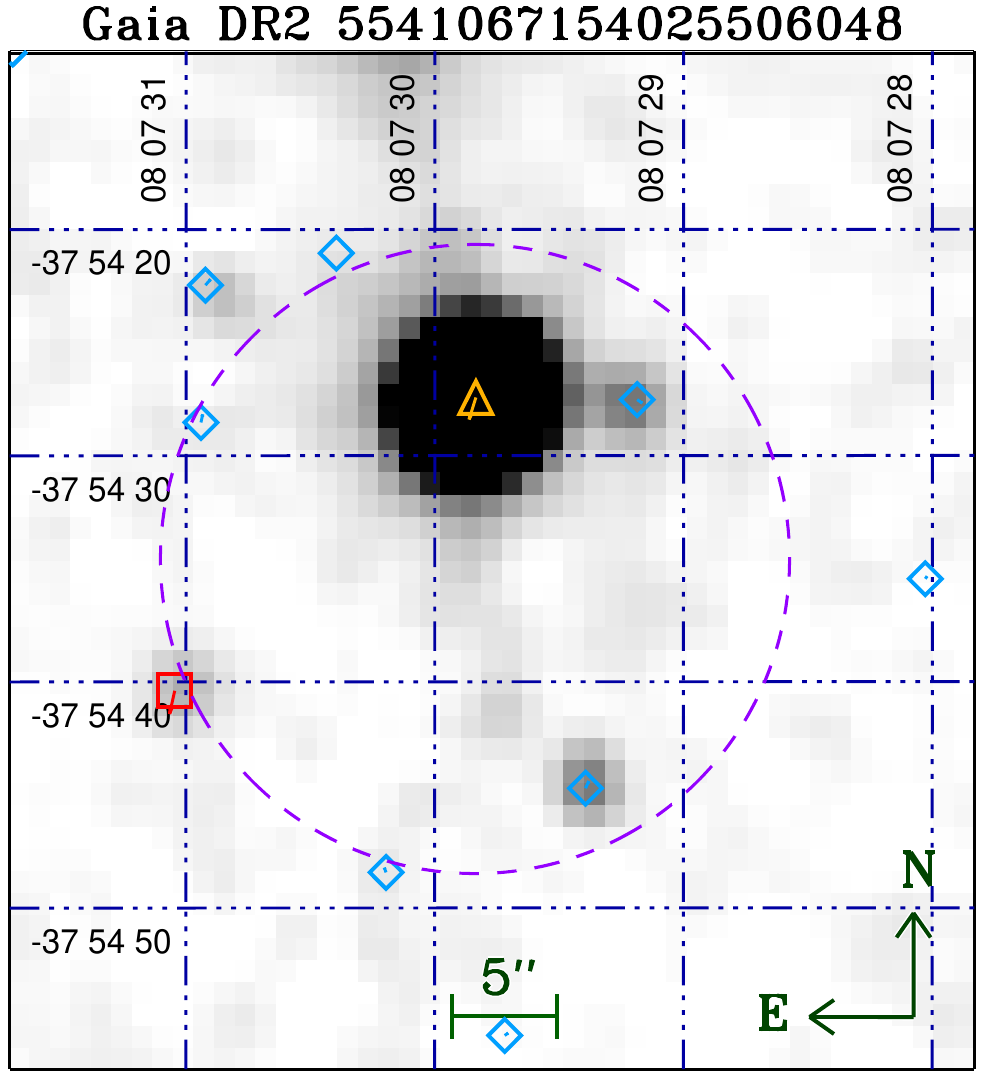}}
\parbox{4.0cm}{\includegraphics[width=4.0cm]{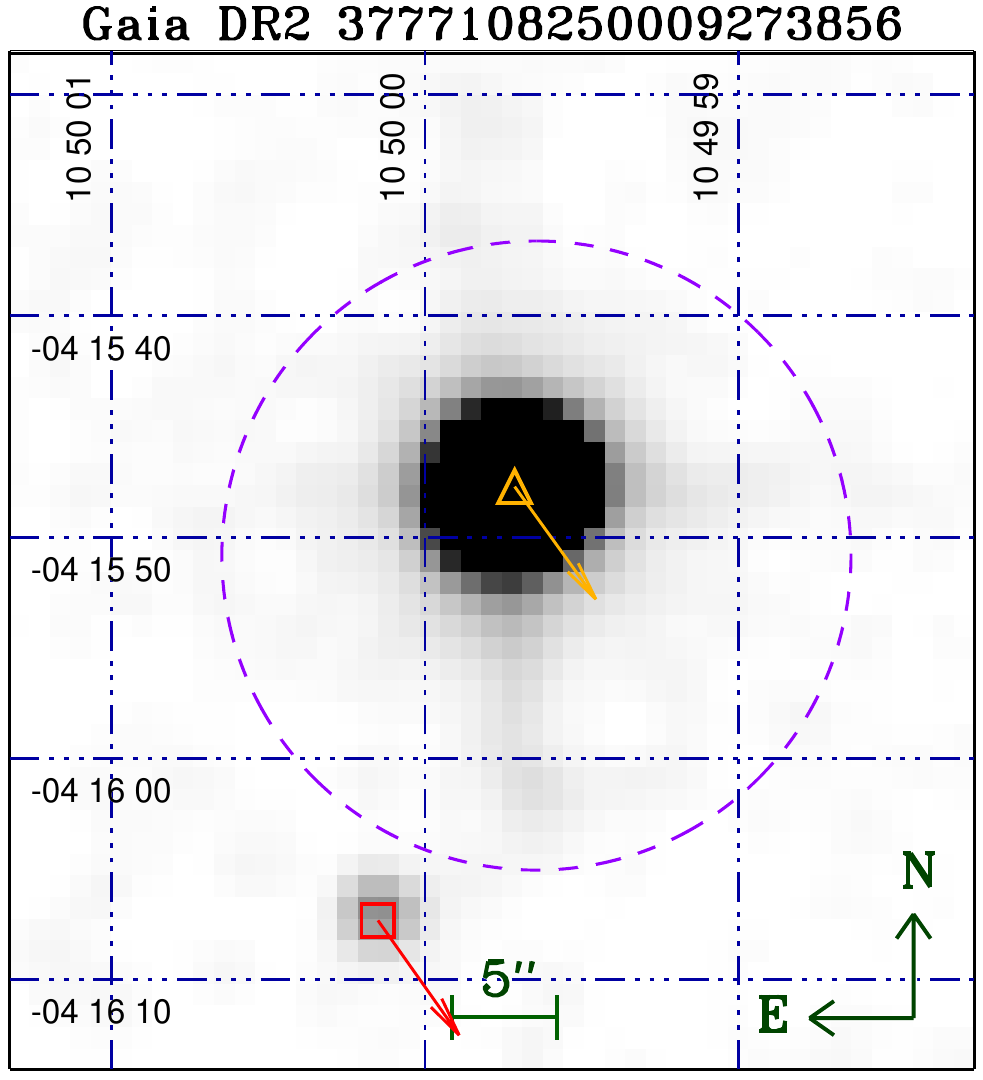}}
\parbox{4.0cm}{\includegraphics[width=4.0cm]{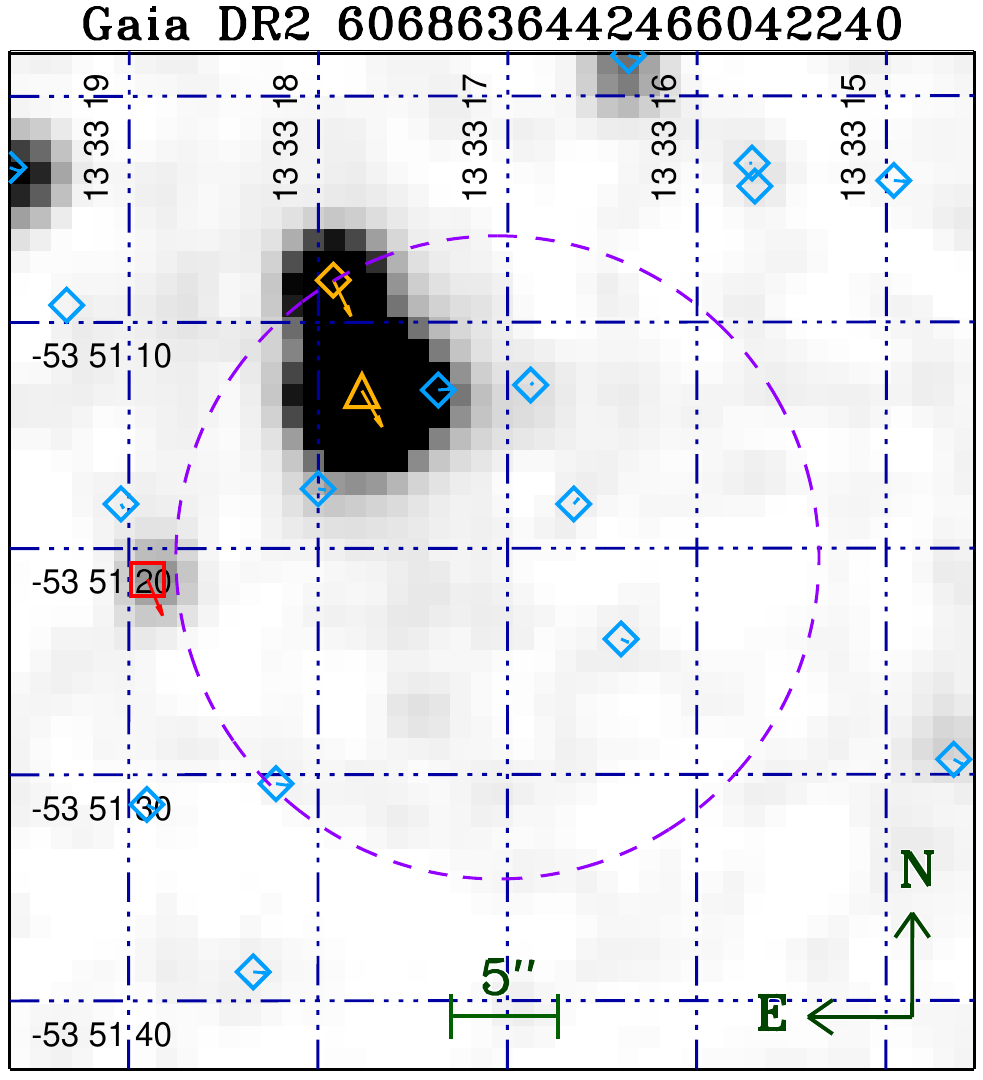}}
}
\caption{
$K_{\rm s}$-band images of UCDs from our input catalogs that are in a CPM binary
or triple system. 
The images are centered on the {\em eROSITA} X-ray source position, and 
a $15^{\prime\prime}$ radius around this position is shown
as magenta dashed circle. The {\em Gaia}\,DR2 source from our input UCD catalogs is represented
by a red square, its CPM companion by an orange triangle and further {\em Gaia}\,DR2 sources
in the field by cyan diamonds. The arrows indicate the PM, and their length and orientation 
are scaled to the {\em Gaia}\,DR2 values. 
Note, that for the actual identification of our UCDs with
the eRASS1 catalog the {\em Gaia}\,DR2 coordinates had been propagated with their proper 
motions to the mean eRASS1 observing data (see Sect.\,\ref{sect:identifications_eRASS1}), 
while here they have been projected backwards to the 2MASS epoch.}
\label{fig:app_cpm_Kband}
\end{center}
\end{figure*}

Here we provide information on CPM binaries involving a UCD or UCD candidate 
from our samples. 
Images of these systems in the 2MASS $K_{\rm s}$ band are shown in 
Fig.~\ref{fig:app_cpm_Kband}. 
In all cases the CPM companions are found as possible {\em Gaia} counterparts 
 in our `reverse' match 
and they are within $3\,\times$\,{\sc radec\_err} of the X-ray position. 
All the comoving companions are also
brighter than the UCD and in all but three cases 
they are closer to the eRASS1 position than the UCD. 
Note that we include here also the UCD pair 
WDS\,J04469-1117\,AB 
which is contrary to the other
objects of this appendix not removed from the 
sample studied in this work. 

Relevant
parameters for the individual systems are summarized in Table~\ref{tab:appA1}. 
We provide for each UCD identified to be in a multiple system (col.\,1) 
the {\em Gaia}\,DR2 identifier
of its comoving companion (col.\,2) and other names for that object (col.\,3). Column\,4 
is the separation between the UCD and its companion. 
Moreover, we 
give for the companion 
its magnitudes in the $G$, $J$ and $W1$ band (cols.\,5$-$7), and its spectral type derived 
from $G-J$ as explained in Sect.~\ref{subsect:identifications_eRASS1_bonafide} (col.~8). 
In col.9 the eRASS1 X-ray fluxes for the 
$0.2-2.3$\,keV band are provided. Since the companions are coronal X-ray sources we did
not take the fluxes from \ecat which are based on a power-law model but we computed them
from the count rates tabulated in \ecat and the conversion factor $CF_{\rm M, 12}$ 
calculated from a
thermal model fit to M dwarf spectra by \cite{Magaudda21.0} as described
in Sect.~\ref{sect:eRASS1_multilambda}.
 Finally, cols.\,10$-$13 comprise flags for indications
for youth / moving group membership and earlier reports on binarity together with the
respective references. 

To summarize, among the 
$13$ pairs comprised of a UCD and a GKM star that are
detected in eRASS1, 
$8$ were previously known as binary stars,   
and $4$ are members of a young moving group but they were unknown as binaries. 
As expected from their brightness in the images 
the companions are of earlier SpT than the UCD, with the exception of 
the abovementioned UCD pair; 
see also Sect.~\ref{subsect:identifications_eRASS1_bonafide}. 
In most cases the companion is a mid-M dwarf. 
Since among late-type, magnetically active stars with earlier SpTs have higher X-ray 
luminosity \citep[e.g.,][]{Schmitt04.0} and in all cases
both components of the CPM system are reasonably close to the X-ray position, 
we associate the X-ray source in 
all these cases with the stellar companion and not with the UCD. These systems have, 
therefore, been removed from our X-ray analysis, again with the exception of the UCD binary. 
Note that the basic X-ray parameters from eRASS1 are found in Table~\ref{tab:eRASS1} 
for these binary systems. 

Table~\ref{tab:appA2} holds the astrometric parameters
from {\em Gaia}\,DR2: 
proper motion in right ascension and declination for both components in the binaries
(cols.2+3 and cols.5+6) and the distance for the companion inferred by 
\cite{BailerJones18.0} from {\em Gaia}\,DR2 data (col.~7). Not unexpectedly, considering
their high PM, all these binaries are nearby with distances $\lesssim 150$\,pc.

%

%
%
%


\begin{sidewaystable}\small
\begin{center}
\caption{Optical/IR parameters of CPM companions to UCDs and UCD candidates 
that have a proper-motion corrected position at the mean eRASS\,1 observing date within $3\,\times$ 
the positional error of an eRASS1 source.}
\label{tab:appA1}
\begin{tabular}{lllrrrrlrcccc}
\hline
  \multicolumn{1}{l}{Gaia\,DR2\,designation} &
  \multicolumn{1}{l}{Gaia\,DR2\,designation} &
  \multicolumn{1}{l}{Other name} &
  \multicolumn{1}{c}{${\rm sep_{UC}}$} &
  \multicolumn{1}{c}{$G$} &
  \multicolumn{1}{c}{$J$} &
  \multicolumn{1}{c}{$W1$} &
  \multicolumn{1}{l}{SpType} &
  \multicolumn{1}{c}{$f_{\rm x,Band\,2}$} &
  \multicolumn{1}{c}{Youth} &
  \multicolumn{1}{c}{Youth} &
  \multicolumn{1}{c}{Binary} &
  \multicolumn{1}{c}{Binary} \\
  \multicolumn{1}{l}{of UCD} &
  \multicolumn{1}{l}{of companion} &
  \multicolumn{1}{l}{of companion} &
  \multicolumn{1}{c}{${\rm [{\prime\prime}]}$} &
  \multicolumn{1}{c}{[mag]} &
  \multicolumn{1}{c}{[mag]} &
  \multicolumn{1}{c}{[mag]} &
  \multicolumn{1}{l}{} &
  \multicolumn{1}{c}{${\rm [erg/cm^2/s]}$} &
  \multicolumn{1}{c}{flag} &
  \multicolumn{1}{c}{ref} &
  \multicolumn{1}{c}{flag} &
  \multicolumn{1}{c}{ref} \\
\hline
  2999273759851225856 & 2999273759850687360 &                  &  5.89 & 16.11 & 12.52 & 11.39 & M5.5$-$M6V &   (4.46 $\pm$ 2.03)$\cdot 10^{-14}$ & Y & 5 & N & \dots \\
  3164100487113671552 & 3164100487115775488 & BD+13 1727       & 11.02 &  9.44 &  8.04 &  7.92 & K2$-$K2.5V &   (5.97 $\pm$ 2.67)$\cdot 10^{-14}$ & N & \dots & Y & 2 \\
  3181197137010596608$^*$ & 3181197137010596480$^*$ & WDS\,J04469-1117\,AB    &  1.52 & 12.22 &  8.14 &  7.84 & M7V        & (123.36 $\pm$ 13.22)$\cdot 10^{-14}$ & N & \dots & Y & 3,4 \\
  3777108250009273856 & 3777111239307270528 & BD$-$03 3000     & 20.68 &  9.58 &  8.25 &  7.93 & K1.5$-$K2V &   (6.07 $\pm$ 2.87)$\cdot 10^{-14}$ & N & \dots & Y & 1,10 \\
  4709450095539630080 & 4709450095539630208 &                  &  6.42 & 14.63 & 11.47 & 10.23 & M4$-$M4.5V &   (4.59 $\pm$ 1.62)$\cdot 10^{-14}$ & N & \dots & Y & 10,11,12 \\
  5013728217560115328 & 5013728221855947264 &                  &  9.05 & 14.0  & 11.45 & 10.49 & M2$-$M2.5V &   (2.73 $\pm$ 1.15)$\cdot 10^{-14}$ & N & \dots & Y & 11,12 \\
  5541067154025506048 & 5541067158327484288 & CD$-$37 4268     & 19.34 &  9.77 &  8.76 &  8.37 & G1$-$G2V   &   (5.54 $\pm$ 1.94)$\cdot 10^{-14}$ & N & \dots & N & \dots \\
  5762038930728469888 & 5762038930729097728 & UCAC4 434-049409 &  8.84 & 10.89 &  8.63 &  7.98 & M0$-$M0.5V &   (9.21 $\pm$ 2.98)$\cdot 10^{-14}$ & N & \dots & Y & 10,12 \\
  5774202926652248832 & 5774202930948040064 &                  &  6.74 & 16.08 & 12.86 & 11.76 & M4$-$M4.5V &   (2.41 $\pm$ 1.10)$\cdot 10^{-14}$ & Y & 7 & N & \dots \\
  5810656757928226560 & 5810656753637344384 & CD$-$68 1857     & 18.18 &  8.58 &  7.62 &  7.89 & G0$-$G1V   &   (3.31 $\pm$ 1.54)$\cdot 10^{-14}$ & N & \dots & Y & 9,10 \\
  6061624891182052224 & 6061625097356185216 &                  & 12.72 & 14.49 & 11.41 & 10.41 & M4$-$M4.5V &   (5.02 $\pm$ 1.45)$\cdot 10^{-14}$ & Y & 6,7 & N & \dots \\
  6068636442466042240 & 6068636438170702848 & TYC 8662-1667-1  & 13.07 & 11.11 &  9.78 &  9.12 & K1.5$-$K2V &   (2.71 $\pm$ 1.21)$\cdot 10^{-14}$ & N & \dots & Y & 12 \\
  6078707625378651264 & 6078707621090084736 &                  &  5.68 & 15.81 & 12.55 & 11.42 & M4.5$-$M5V &   (2.91 $\pm$ 1.14)$\cdot 10^{-14}$ & N & \dots & Y & 10,11 \\
  6093120294382123648 & 6093120298678786944 &                  &  6.21 & 14.56 & 11.59 & 10.58 & M3.5$-$M4V &  (10.45 $\pm$ 2.54)$\cdot 10^{-14}$ & Y & 8 & N & \dots \\
\hline
\multicolumn{13}{l}{$^*$ This pair is a binary UCD that is not resolved with 2\,MASS. Spectral types are M4.9 and M6 and an age range of $60-300$\,Myr was estimated \citep{Shkolnik09.0}} \\
\multicolumn{13}{l}{References:  (1) Washington Double Star Catalog, (2) Cruz et al. (2007), (3) Shkolnik et al. (2009), (4) Shkolnik et al. (2012), (5) Zari et al. (2017), (6) Goldman et al. (2018), (7) Zari et al. (2018),} \\
\multicolumn{13}{l}{(8) Damiani et al. (2019),  (9) Kervella et al. (2019), (10) Hartman \& L{\'e}pine (2020), (11) Sapozhnikov et al. (2020), (12) Tian et al. (2020)} \\
\end{tabular}
\end{center}
\end{sidewaystable}

\begin{sidewaystable}
\begin{center}
\caption{Astrometric parameters of the UCD and UCD candidates and their companions for comoving systems from Table~\ref{tab:appA1}.}
\label{tab:appA2}
\begin{tabular}{rrrrrrrrr}
\hline
  \multicolumn{1}{l}{Gaia\,DR2\,designation} &
  \multicolumn{1}{c}{${\rm pm\_ra_{UCD}}$} &
  \multicolumn{1}{c}{${\rm pm\_dec_{UCD}}$} &
  \multicolumn{1}{l}{Gaia\,DR2\,designation} &
  \multicolumn{1}{c}{${\rm pm\_ra_C}$} &
  \multicolumn{1}{c}{${\rm pm\_dec_C}$} &
  \multicolumn{1}{c}{d} \\
  \multicolumn{1}{l}{of UCD} &
  \multicolumn{1}{c}{[mas/yr]} &
  \multicolumn{1}{c}{[mas/yr]} &
  \multicolumn{1}{l}{of companion} &
  \multicolumn{1}{c}{[mas/yr]} &
  \multicolumn{1}{c}{[mas/yr]} &
  \multicolumn{1}{c}{[pc]} \\
\hline
  2999273759851225856 & $-$32.31 & $-$17.91 & 2999273759850687360 & $-$37.02 & $-$23.02 & 75.96$_{-0.92}^{+0.95}$\\
  3164100487113671552 & $-$77.38 & $-$158.68 & 3164100487115775488 & $-$74.63 & $-$156.23 & 49.58$_{-0.13}^{+0.13}$\\
  3181197137010596608 & $-$142.39 & $-$57.56 & 3181197137010596480 & $-$141.22 & $-$42.79 & 18.9$_{-0.04}^{+0.04}$\\
  3777108250009273856 & $-$66.82 & $-$89.92 & 3777111239307270528 & $-$67.27 & $-$88.25 & 64.53$_{-0.18}^{+0.18}$\\
  4709450095539630080 & $-$26.87 & $-$32.18 & 4709450095539630208 & $-$31.13 & $-$35.96 & 30.19$_{-0.05}^{+0.05}$\\
  5013728217560115328 & 10.92 & 15.35 & 5013728221855947264 & 11.59 & 13.36 & 81.25$_{-0.21}^{+0.21}$\\
  5541067154025506048 & 4.91 & $-$17.43 & 5541067158327484288 & 5.52 & $-$17.31 & 100.86$_{-0.27}^{+0.27}$\\
  5762038930728469888 & $-$181.12 & 7.25 & 5762038930729097728 & $-$193.47 & 6.27 & 36.11$_{-0.29}^{+0.29}$\\
  5774202926652248832 & 8.67 & $-$18.45 & 5774202930948040064 & 7.68 & $-$17.89 & 159.02$_{-1.89}^{+1.93}$\\
  5810656757928226560 & $-$68.17 & $-$151.42 & 5810656753637344384 & $-$74.18 & $-$151.64 & 65.65$_{-1.36}^{+1.42}$\\
  6061624891182052224 & $-$33.99 & $-$15.64 & 6061625097356185216 & $-$34.29 & $-$15.95 & 109.93$_{-0.85}^{+0.86}$\\
  6068636442466042240 & $-$12.69 & $-$27.96 & 6068636438170702848 & $-$15.3 & $-$28.2 & 116.96$_{-0.59}^{+0.6}$\\
  6078707625378651264 & $-$123.32 & $-$38.38 & 6078707621090084736 & $-$124.02 & $-$38.74 & 91.6$_{-0.66}^{+0.67}$\\
  6093120294382123648 & $-$23.56 & $-$20.67 & 6093120298678786944 & $-$22.8 & $-$21.33 & 135.55$_{-1.16}^{+1.18}$\\
\hline\end{tabular}
\end{center}
\end{sidewaystable}

%

\clearpage

\section{eROSITA source populations}\label{app:selection_procedure}

To understand the properties of objects that 
are not bonafide counterparts to an eRASS1 source and that are not
UCDs we examine the {\em eROSITA}, {\em Gaia} and {\em WISE} parameter space.

In Sect.~\ref{subsect:identifications_eRASS1_plausible} we have argued 
for evidence of distinct populations of different astrophysical populations in 
the $G - G_{\rm RP}$ vs $G_{\rm BP} - G$ diagram. To verify this claim 
we make use of the main eFEDS point source catalog 
 (\citeauthor[][\aap ~submitted]{Brunner21.0}). 
We adopt the results from the dedicated source identification procedure described
by \cite{Salvato21.0} that classifies the eFEDS X-ray sources as either galactic or
extragalactic. In the extragalactic category we consider the objects labeled as 
`SECURE EXTRAGALACTIC' and those labeled `LIKELY EXTRAGALACTIC', and analogously 
in the galactic category where we consider the `SECURE GALACTIC' and `LIKELY GALACTIC' ones.  
 We limit our study to the objects that have full {\em Gaia} photometry
($G$, $G_{\rm RB}$ and $G_{\rm RP} > 0$), valid data in the lowest {\em WISE} band ($W1 > 0$)
and reliable counterpart according to a quality flag (CTP\_quality $>= 3$) defined by
\cite{Salvato21.0}.

\begin{figure}
\begin{center}
\includegraphics[width=8.5cm]{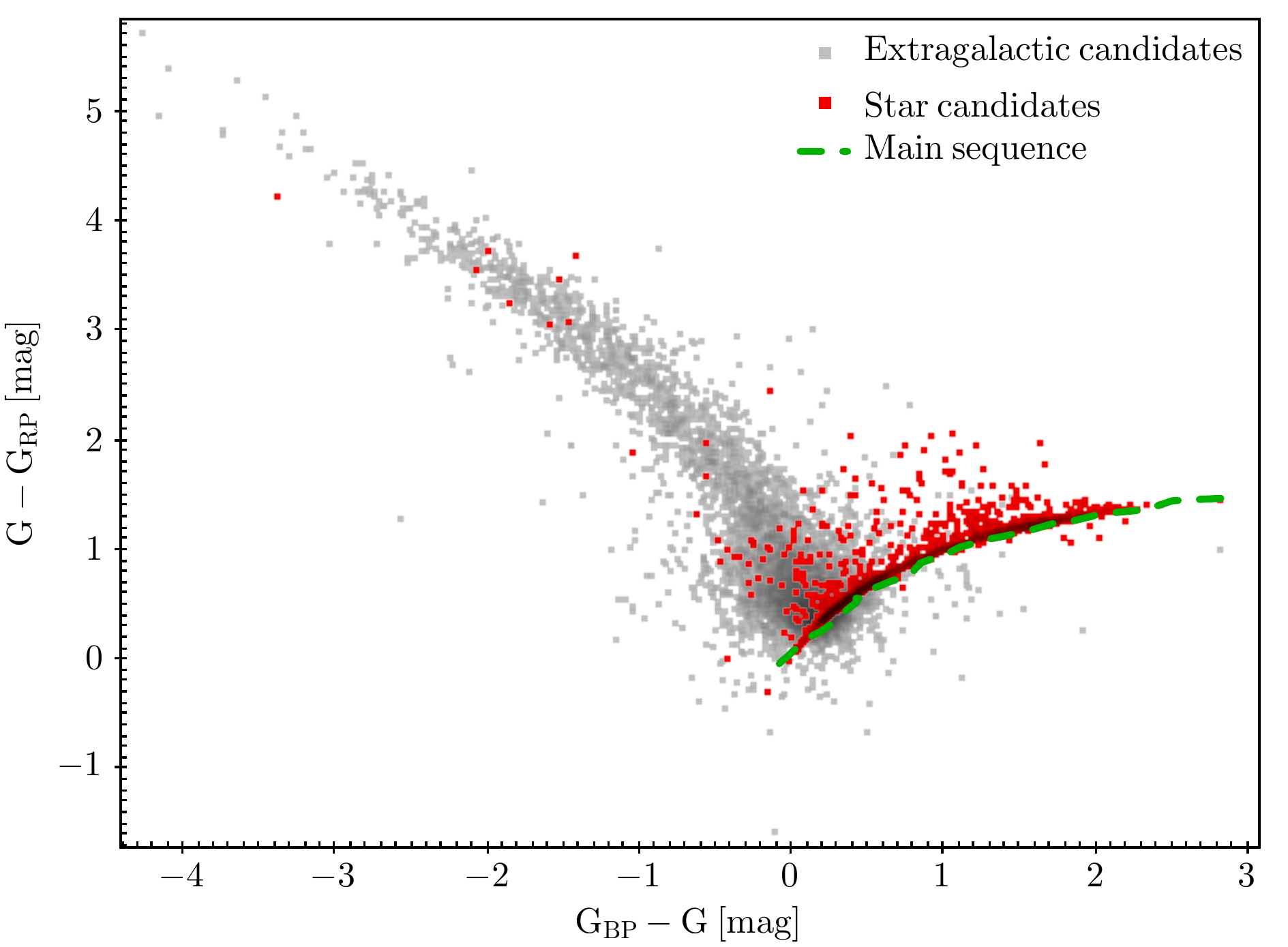}
\caption{{\em Gaia} color-color diagram 
for the two subsamples of eFEDS sources 
identified as extragalactic (gray) and galactic (red) by \protect\cite{Salvato21.0}.
The stellar main-sequence (green) is obtained from the table maintained by E.Mamajek\footref{note1}.}
\label{fig:eFEDS_GminRP_BPminG}
\end{center}
\end{figure}

In Fig.\ref{fig:eFEDS_GminRP_BPminG} we display the two groups of objects in 
the $G - G_{\rm RP}$ vs $G_{\rm BP} - G$ diagram. The distribution of
the `extragalactic' objects in Fig.~\ref{fig:eFEDS_GminRP_BPminG} resembles a combination 
of the `quasar' and `galaxy' CCDs presented by \cite{BailerJones19.0}, with a weak extension 
onto the stellar main-sequence (drawn as a green line). 
The `galactic' objects clearly follow
that sequence, but with a signifiant upwards spread, thus also showing some overlap with
the `extragalactic' population. 
These overlaps might be related to residual contamination of each group with objects from the other
type.
%
%
%

\begin{figure}
\begin{center}
\includegraphics[width=8.5cm]{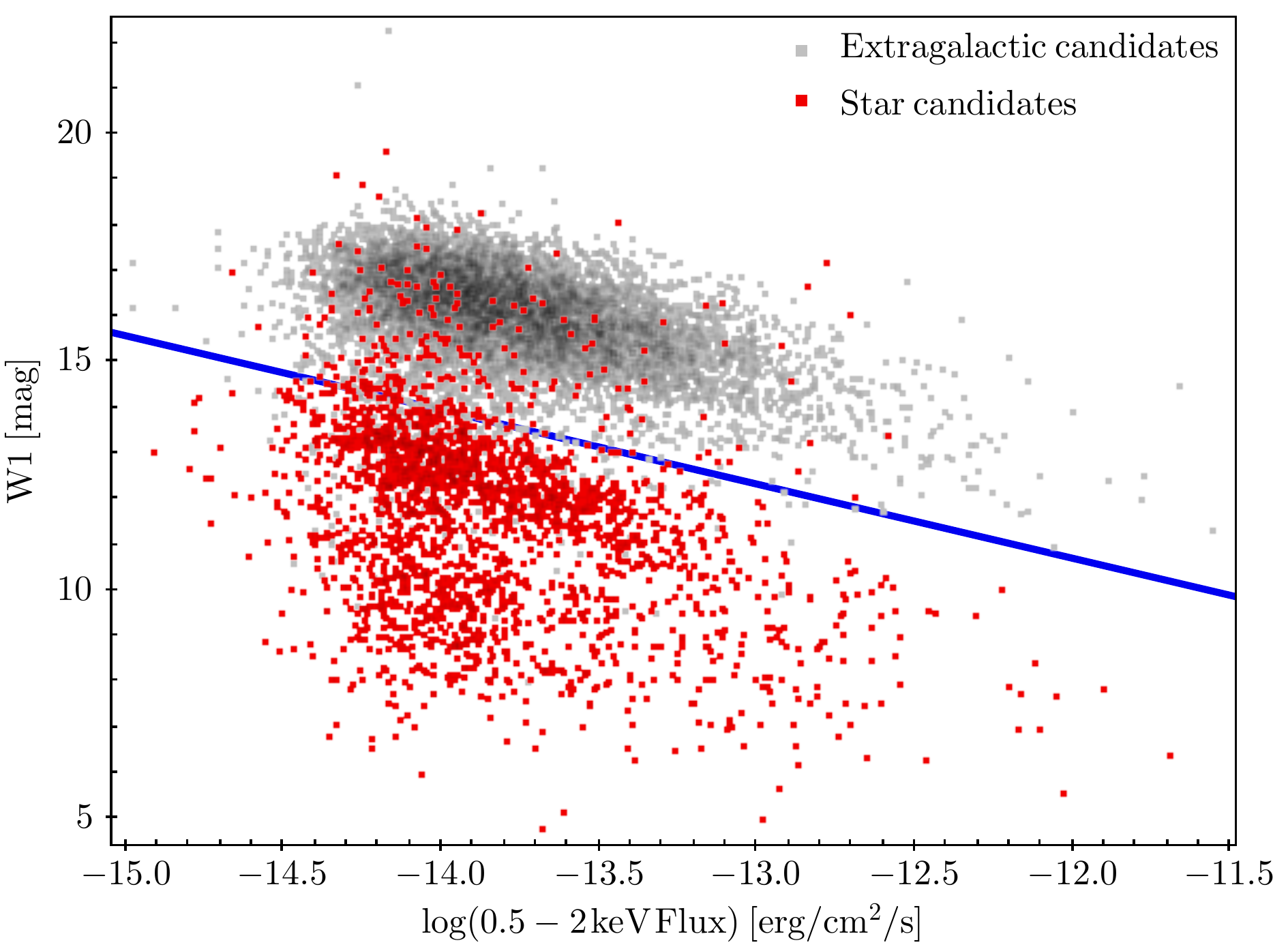}
\caption{WISE $W1$ magnitude versus logarithm of the X-ray flux in 
$0.6-2.0$\,keV for the 
objects from the main eFEDS catalog distinguishing `galactic' from `extragalactic' sources
 as described in the text. 
The line denotes the empirical
separator between stars and other objects as defined by \cite{Salvato18.0} with 
extragalactic sources being located above the line.}
\label{fig:eFEDS_W1_logfx}
\end{center}
\end{figure}

In Fig.~\ref{fig:eFEDS_W1_logfx} we examine for the same two groups of objects the 
$W1$ vs $\log{f_{\rm x}}$ diagram 
which was introduced by \cite{Salvato18.0} to tentatively distinguish extragalactic
from galactic objects 
based on an empirical dividing line shown in blue in Fig.~\ref{fig:eFEDS_W1_logfx}.
Overall this line distinguishes well the two populations 
as already suggested by \cite{Salvato18.0};  
\cite[see][for a discussion of this diagram with relation to 
the eFEDS population]{Salvato21.0}. 

\begin{figure*}
\begin{center}
\parbox{17cm}{
\parbox{8.5cm}{
\includegraphics[width=8.5cm]{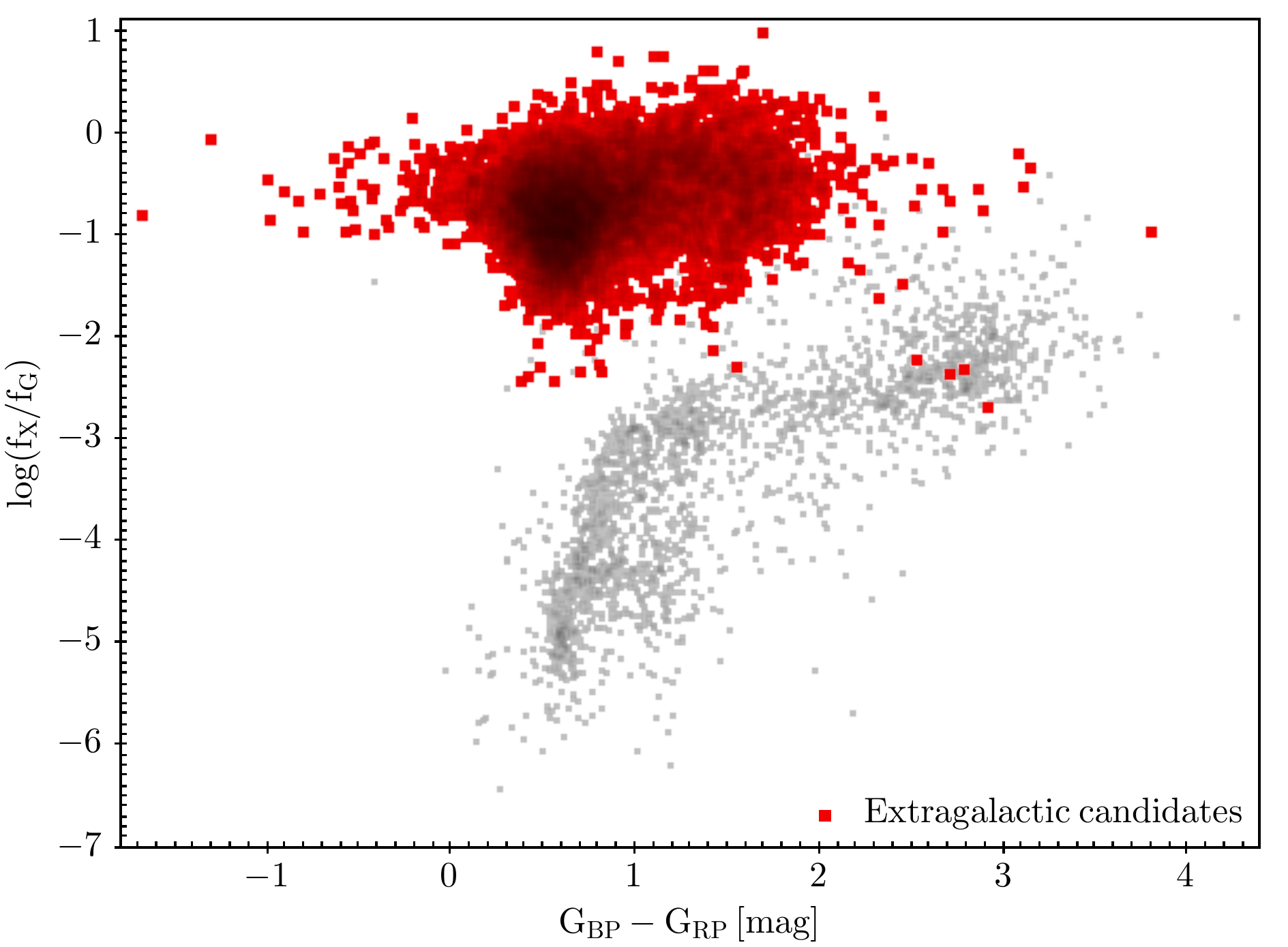}
}
\parbox{8.5cm}{
\includegraphics[width=8.5cm]{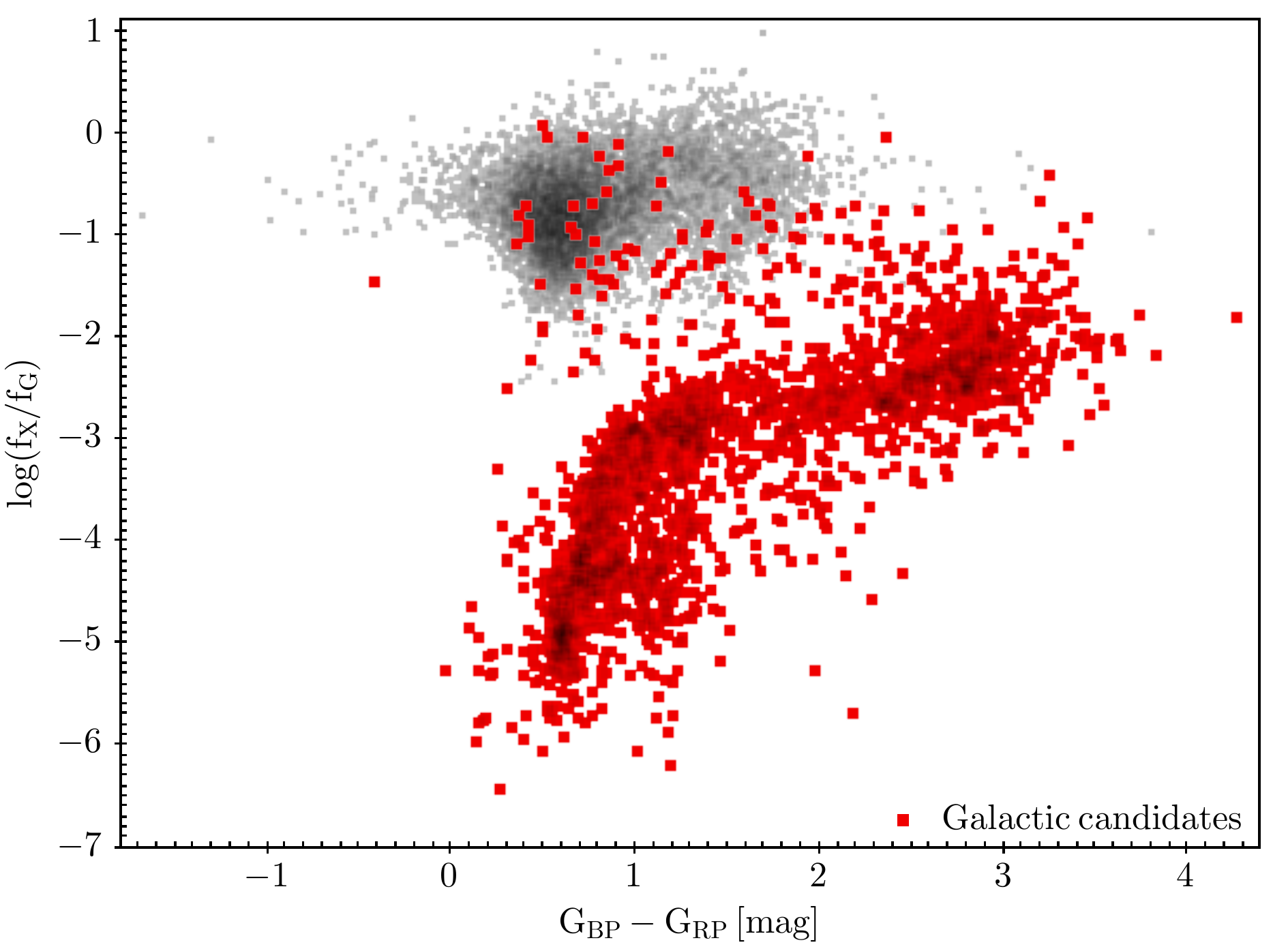}
}
}
\caption{Logarithm of the X-ray over $G$ band flux versus {\em Gaia} 
blue and red photometer color for the full eFEDS sample. Two panels with the same
data are shown to visualize the overlap of the extragalactic and galactic  
candidates selected from Fig.~\ref{fig:eFEDS_W1_logfx}. The extragalactic sample is 
highlighted in red in the left panel and the galactic sample is red in the right panel. 
}
\label{fig:eFEDS_logfxfoptG_BPminRP}
\end{center}
\end{figure*}

We finally show the X-ray over optical flux ratio versus {\em Gaia}
color for the 
eFEDS sample in Fig.~\ref{fig:eFEDS_logfxfoptG_BPminRP}
where with few exceptions the `galactic' and `extragalactic' objects are
clearly separated. The small group in the right panel that is above the main locus of 
the `galactic' sources shows strong overlap with the galactic objects above the separator
in the $W1$ vs $f_{\rm x}$ diagram,
with $155$ of $176$ galactic sources ($\sim ~88$\,\%) 
that are upwards outliers in terms of $f_{\rm x}/f_{\rm G}$ ratio 
being located above the line in the $W1 - f_{\rm x}$ diagram. These objects could be
flaring stars or compact objects.  
%
%
%

To summarize, these figures support the classification scheme of Salvato et al. (this
volume) in terms of galactic versus extragalactic sources, 
but a small subsample of them is overlapping each other. 
The 
extragalactic sample 
reaches down to the upper 
stellar main-sequence.  
The upwards outliers in the sample
of 
`galactic' 
objects 
(with $\log{(f_{\rm x}/f_{\rm G})} \geq -2$) could be AGN or flaring stars. 

\section{Non-UCD eRASS1 counterparts}\label{app:nonucds} 

In Table~\ref{tab:app_nonUCDs} we list the properties of the `plausible' 
counterparts to eRASS1 X-ray sources that are not UCDs or UCD candidates. 
Table~\ref{tab:app_nonUCDs} provides their photometry ($G$, $J$ and 
$W1$ in cols.~3-5), spectral type estimated from $G-J$ assuming a main-sequence star and adopting
the same calibration as in Sect.~\ref{subsect:identifications_eRASS1_bonafide}
(col.~6).
Clearly, the spectral type is meaningful only if the object is a star. 
Furthermore, Table~\ref{tab:app_nonUCDs} holds 
the {\em Gaia}\,DR2 distance from \cite{BailerJones18.0} (col.~7), X-ray fluxes from \ecat for the
$0.6-2.3$\,keV band and the $2.3-5.0$\,keV band (cols.~8 and~9).
The latter energy band was included because is relevant for extragalactic objects 
which tend to be harder X-ray sources than coronally active stars. Finally, in cols.~10 and~11 
the classification from the literature and the corresponding reference are given.

The objects in Table~\ref{tab:app_nonUCDs} belong to two groups selected with different criteria
(see Sect.~\ref{subsect:identifications_eRASS1_plausible}). 
The final classification of these objects is not within the scope of our work. 
However, here we briefly discuss the two groups and the possible nature of its members. 

The first group comprises eRASS1 sources  
that have a UCD or UCD candidate 
as the {\em Gaia} object with the smallest X-ray to optical separation 
but where another {\em Gaia} source in the 
match radius is 
brighter
suggesting the association of the eRASS1 emission with this object 
rather than the UCD. All but one of these $7$ objects have been classified as a star
in the literature. 
In contrast to the objects from Appendix~\ref{app:cpmpairs} these
stars do not share the proper motion with the UCD. 

The other group are {\em Gaia} sources that have the smallest value
for the X-ray to optical separation among all {\em Gaia} objects in the 
match radius of $3 \times$ the {\sc radec\_err} around the eRASS1 position. 
One of these objects, {\em Gaia}\,DR2\,3902395843353397248, was identified as a bonafide 
eRASS1 X-ray source on the basis of an archival {\em XMM-Newton} observation. 
This is 
a known quasar, 2XMM\,J123337.5+073133, from the catalog of \cite{Miller11.0}. 
The {\em XMM-Newton} data was analyzed by \cite{Stelzer20.0} in a study of serendipitous
X-ray detections of UCDs from the {\em XMM-Newton} archive. In that work the source was
recognized as an extragalactic object on the basis of its X-ray spectrum which is 
consistent with a power-law of typical index ($\Gamma \sim 2$) but not with thermal
emission at temperatures typical for M-type dwarfs. 

The position of the other non-UCDs that are closest {\em Gaia} counterparts to 
eRASS1 sources in the 
multiwavelength parameter space can be seen in the right panels of 
Fig.~\ref{fig:eRASS1_selectiondiagrams}, where they are represented as gray circles surrounded
by an annulus. Note that some sources without blue and red photometer, and WISE data
are missing in these diagrams. From comparison  
with Fig.~\ref{fig:eFEDS_logfxfoptG_BPminRP}  
these objects are likely a mixture of extragalactic objects and stars, and this is
consistent with the information compiled from the literature.

\begin{sidewaystable*}\small
\begin{center}
\caption{Multi-wavelength properties of non-UCDs considered to be more plausible counterparts to eRASS1 sources than the UCD in its vicinity.}
\label{tab:app_nonUCDs}
\begin{tabular}{llccclrrrcc}
\hline
\hline
\noalign{\smallskip}
  \multicolumn{1}{c}{Gaia DR2 designation} &
  \multicolumn{1}{c}{Gaia DR2 designation} &
  \multicolumn{1}{c}{$G$} &
  \multicolumn{1}{c}{$J$} &
  \multicolumn{1}{c}{$W1$} &
  \multicolumn{1}{c}{SpType} &
  \multicolumn{1}{c}{Dist} &
  \multicolumn{1}{c}{$f_{x,~band~2}$} &
  \multicolumn{1}{c}{$f_{x,~band~3}$} &
  \multicolumn{1}{c}{Object type} &
  \multicolumn{1}{c}{Ref.} \\
\noalign{\smallskip}
  \multicolumn{1}{c}{of UCD} &
  \multicolumn{1}{c}{of alternative counterpart} &
  \multicolumn{1}{c}{[mag]} &
  \multicolumn{1}{c}{[mag]} &
  \multicolumn{1}{c}{[mag]} &
  \multicolumn{1}{c}{} &
  \multicolumn{1}{c}{[pc]} &
  \multicolumn{1}{c}{[erg/cm$^{2}$/s]} &
  \multicolumn{1}{c}{[erg/cm$^{2}$/s]} &
  \multicolumn{1}{c}{} &
  \multicolumn{1}{c}{} \\
\noalign{\smallskip}
\hline
\noalign{\smallskip}
\multicolumn{11}{c}{\em bright Gaia source but not closest to X-ray source} \\
\noalign{\smallskip}
\hline
\noalign{\smallskip}
  4842003953209804672 & 4842003953209804928 & 14.63 & 13.64 & 12.78 & G1--G2V & $1737.61_{-58.92}^{+63.10}$ & $(27.78\pm10.79)\cdot 10^{-15}$ & \dots & Star & 4\\
\noalign{\smallskip}
  652944911438186368  & 652944915732919168  & 16.92 & 15.25 & 14.28 & K4--K4.5V & $1095.41_{-140.00}^{+184.69}$ & $(47.52\pm27.08)\cdot 10^{-15}$ & $(36.46\pm33.49)\cdot 10^{-14}$ & Star & 2,10\\
\noalign{\smallskip}
  720165105283287808  & 720165105284271744  & 13.02 & 11.05 & 10.21 & K7--K8V & $134.26_{-7.51}^{+8.44}$ & $(109.63\pm42.96)\cdot 10^{-15}$ & \dots & Star & 2,10\\
\noalign{\smallskip}
  5411645461381996928 & 5411645465686894208 & 13.76 & 12.10 & 11.48 & K4--K4.5V & $3357.67_{-212.13}^{+241.87}$ & $(84.62\pm25.74)\cdot 10^{-15}$ & \dots & Star & 4\\
\noalign{\smallskip}
  5449824525188538112 & 5449824490828691584 & 18.77 & 16.41 & 15.06 & M1V & $1616.61_{-503.95}^{+908.12}$ & \dots & $(62.07\pm30.40)\cdot 10^{-14}$ & Star & 4\\
\noalign{\smallskip}
  5854998824943562240 & 5854998820633388544 & 19.75 & \dots   & \dots   & \dots   & $3380.43_{-1639.10}^{+2929.38}$ & $(26.81\pm8.66)\cdot 10^{-15}$ & $(2.08\pm4.96)\cdot 10^{-14}$ & Star & 4\\
\noalign{\smallskip}
  5854998824943562240 & 5854998824943561728 & 15.96 & 13.25 & 11.73 & M3--M3.5V & $430.05_{-9.35}^{+9.76}$ & $(26.81\pm8.66)\cdot 10^{-15}$ & $(2.08\pm4.96)\cdot 10^{-14}$ & Galaxy & 4\\
\noalign{\smallskip}
  5905136482143969280 & 5905136486450575616 & 13.80 & 12.49 & 11.52 & K1--K1.5V & $903.21_{-23.89}^{+25.20}$ & $(43.00\pm20.74)\cdot 10^{-15}$ & $(28.82\pm22.38)\cdot 10^{-14}$ & Star & 4\\
\noalign{\smallskip}
\hline
\noalign{\smallskip}
\multicolumn{11}{c}{\em non-UCD Gaia source closest to X-ray source} \\
\noalign{\smallskip}
\hline
\noalign{\smallskip}
  5985290231327158144 & 5985290231264640000 & 20.29 & \dots   & \dots   & \dots   & $2136.11_{-1251.67}^{+2527.68}$ & $(34.85\pm16.38)\cdot 10^{-15}$ & \dots & \dots & \dots \\
\noalign{\smallskip}
  6235533167875990272 & 6235533163576299520 & 18.14 & 16.14 & \dots   & K7--K8V & $2387.40_{-1011.38}^{+2823.78}$ & $(64.78\pm25.84)\cdot 10^{-15}$ & $(14.81\pm18.61)\cdot 10^{-14}$ & Galaxy & 4\\
\noalign{\smallskip}
  3426062907707797376 & 3426062912004133632 & 15.74 & 12.69 & 11.72 & M4V & \dots & $(63.15\pm30.31)\cdot 10^{-15}$ & \dots & Star & 6,10\\
\noalign{\smallskip}
  3537678013829356032 & 3537678009534395520 & 20.81 & \dots   & 15.99 & \dots & $1011.59_{-473.74}^{+928.31}$ & $(47.71\pm17.37)\cdot 10^{-15}$ & $(6.4\pm7.67)\cdot 10^{-14}$ & Galaxy & 4\\
\noalign{\smallskip}
  3902395843353397248 & 3902395813288871936$^*$ & 19.17 & \dots & 14.10 & \dots & $2301.06_{-651.12}^{+939.98}$ & $(119.02\pm34.17)\cdot 10^{-15}$ & $(11.00\pm12.02)\cdot 10^{-14}$ & QSO & 1\\
\noalign{\smallskip}
  4633181951787913344 & 4633181956082468736 & 20.86 & \dots   & 14.87 & \dots & \dots & $(32.75\pm13.85)\cdot 10^{-15}$ & \dots & AGN & 5,8\\
\noalign{\smallskip}
  4682521921904033408 & 4682521921904033024 & 19.69 & \dots   & 14.67 & \dots & $1179.81_{-419.12}^{+754.84}$ & $(29.13\pm8.86)\cdot 10^{-15}$ & \dots & AGN & 5,8\\
\noalign{\smallskip}
  4820214342870659968 & 4820214347166884096 & 17.05 & 15.98 & 14.89 & G4--G5V & $3625.70_{-567.14}^{+745.52}$ & $(29.31\pm11.73)\cdot 10^{-15}$ & \dots & Star & 4\\
\noalign{\smallskip}
  4907225329405739904 & 4907225329405740672 & 19.79 & \dots   & \dots   & \dots  & $1900.08_{-613.83}^{+953.67}$ & $(28.66\pm14.92)\cdot 10^{-15}$ & \dots & Star & 4\\
\noalign{\smallskip}
  4948028068473626368 & 4948028068473523200 & 20.71 & \dots   & 16.02 & \dots & \dots & $(29.12\pm13.13)\cdot 10^{-15}$ & \dots & Galaxy & 4\\
\noalign{\smallskip}
  5263797568772960256 & 5263797641784398080 & 21.08 & \dots   & \dots   & \dots & \dots & $(9.46\pm3.73)\cdot 10^{-15}$ & $(2.74\pm2.68)\cdot 10^{-14}$ & Galaxy & 4\\
\noalign{\smallskip}
  5317092619649544960 & 5317092550945911552$^\dagger$ & 10.86 & 10.06 & 9.50 & F6--F7V & $370.06_{-3.35}^{+3.41}$ & $(28.11\pm14.56)\cdot 10^{-15}$ & $(3.85\pm13.49)\cdot 10^{-14}$ & Star & 7,9\\
\noalign{\smallskip}
  5387954181261184896 & 5387953979397582336 & 20.92 & \dots   & 15.47 & \dots & \dots & $(46.72\pm18.39)\cdot 10^{-15}$ & \dots & AGN & 5,8\\
\noalign{\smallskip}
  5479448529535828992 & 5479448529538083200 & 20.58 & \dots   & 14.59 & \dots & $1782.32_{-791.88}^{+1312.52}$ & $(16.46\pm6.95)\cdot 10^{-15}$ & $(3.76\pm4.51)\cdot 10^{-14}$ & AGN & 5,8\\
\noalign{\smallskip}
  5790470927042226816 & 5790470927034553984 & 19.83 & \dots   & \dots   & \dots  & $1731.19_{-849.79}^{+1917.51}$ & $(8.32\pm6.27)\cdot 10^{-15}$ & \dots & \dots & \dots \\
\noalign{\smallskip}
  5800676697044928640 & 5800676697044928000 & 12.58 & 10.47 & 9.84   & K9--M0V & $201.28_{-10.17}^{+11.29}$ & $(14.92\pm10.04)\cdot 10^{-15}$ & $(6.54\pm8.89)\cdot 10^{-14}$ & VB & 11\\
\noalign{\smallskip}
  5800676697044928640 & 5800676697042535296 & 12.71 & 10.47 & 9.84   & M0--M0.5V & $206.97_{-8.20}^{+8.89}$ & $(14.92\pm10.04)\cdot 10^{-15}$ & $(6.54\pm8.89)\cdot 10^{-14}$ & VB & 11\\
\noalign{\smallskip}
  5878773221280041472 & 5878773216960788096 & 12.14 & 10.92 & 10.39 & G9--K0V & $314.34_{-4.09}^{+4.19}$ & $(72.80\pm26.28)\cdot 10^{-15}$ & \dots & Star & 3\\
\noalign{\smallskip}
  5991061121183402880 & 5991061121134832896 & 20.41 & \dots   & \dots   & \dots   & $3448.57_{-1706.88}^{+2794.23}$ & $(96.52\pm28.33)\cdot 10^{-15}$ & \dots & \dots & \dots \\
\noalign{\smallskip}
\hline
\multicolumn{11}{l}{Other name from SIMBAD: $^*$ 2XMM J123337.5+073133, $^\dagger$ TYC 8573-3247-1} \\
\multicolumn{11}{l}{References: (1) \cite{Schneider07.0}, (2) \cite{Vasconcellos11.0}, (3) \cite{Saito12.0}, (4) \cite{McMahon13.0}, (5) \cite{Secrest15.0}, (6) \cite{Cook16.0}, (7) \cite{Stevens17.0},} \\
\multicolumn{11}{l}{(8) \cite{Assef18.0} , (9) \cite{CantatGaudin18.0}, (10) \cite{Xiang19.0}, (11) this work.} \\
\end{tabular}
\end{center}
\end{sidewaystable*}

\begin{acknowledgements}
We thank the anonymous referee for careful reading and constructive comments. 
AK is supported by the Deutsche Forschungsgemeinschaft (DFG) project number 413113723. 
MC and EM acknowledge funding by the Bundesministerium für Wirtschaft und
Energie through the Deutsches Zentrum für Luft- und Raumfahrt e.V. (DLR)
under grant numbers FKZ 50 OR 2008 and FKZ 50 OR 1808.
This work is based on data from {\em eROSITA}, the primary instrument aboard SRG, a joint 
Russian-German science mission supported by the Russian Space Agency (Roskosmos), in the 
interests of the Russian Academy of Sciences represented by its Space Research Institute 
(IKI), and the Deutsches Zentrum für Luft- und Raumfahrt (DLR). The SRG spacecraft was built 
by Lavochkin Association (NPOL) and its subcontractors, and is operated by NPOL with support 
from the Max-Planck Institute for Extraterrestrial Physics (MPE).

The development and construction of the {\em eROSITA} X-ray instrument was led by MPE, with 
contributions from the Dr. Karl Remeis Observatory Bamberg, the University of Hamburg 
Observatory, the Leibniz Institute for Astrophysics Potsdam (AIP), and the Institute for 
Astronomy and Astrophysics of the University of Tübingen, with the support of DLR and the 
Max Planck Society. The Argelander Institute for Astronomy of the University of Bonn and 
the Ludwig Maximilians Universität Munich also participated in the science preparation for 
{\em eROSITA}.

The {\em eROSITA} data shown here were processed using the eSASS/NRTA software system developed 
by the German {\em eROSITA} consortium.

This publication makes use of data products from the Two Micron All Sky Survey, which is a 
joint project of the University of Massachusetts and the Infrared Processing and Analysis 
Center/California Institute of Technology, funded by the National Aeronautics and Space 
Administration and the National Science Foundation and 
of data products from the Wide-field Infrared Survey Explorer, 
which is a joint project of the University of California, Los Angeles, and the Jet 
Propulsion Laboratory/California Institute of Technology, funded by the National 
Aeronautics and Space Administration.
We present results from the European Space Agency (ESA) space mission {\em Gaia}. {\em Gaia} 
data are being processed by the {\em Gaia} Data Processing and Analysis Consortium (DPAC). 
Funding for the DPAC is provided by national institutions, in particular the institutions 
participating in the {\em Gaia} MultiLateral Agreement (MLA). The {\em Gaia} mission website 
is https://www.cosmos.esa.int/gaia. 

We acknowledge ESASky, developed by the ESAC Science Data Centre (ESDC) team 
and maintained alongside other ESA science mission's archives at ESA's European Space 
Astronomy Centre (ESAC, Madrid, Spain) and the SVO Filter Profile Service 
(http://svo2.cab.inta-csic.es/theory/fps/) supported from the Spanish MINECO through 
grant AYA2017-84089. 

\end{acknowledgements}

\bibliographystyle{aa} 
\bibliography{eRASS1ucds}

\end{document}